\documentclass{emulateapj}
\usepackage{graphicx}
\usepackage{amssymb}
\usepackage{lscape}
\usepackage{rotating}
\usepackage{natbib} 
\begin{document}

\title{A Faint Flux-Limited Lyman Alpha Emitter Sample at $z\sim0.3$\altaffilmark{1}}

\author{Isak G. B. Wold\altaffilmark{2}, Steven L. Finkelstein\altaffilmark{2},
Amy J. Barger\altaffilmark{3,4,5}, Lennox L. Cowie\altaffilmark{5},
and Benjamin Rosenwasser\altaffilmark{3}}

\altaffiltext{1}{Some of the data presented herein were obtained at the W.M. Keck Observatory, which is operated as a scientific partnership among the California Institute of Technology, the University of California and the National Aeronautics and Space Administration. The Observatory was made possible by the generous financial support of the W.M. Keck Foundation.}
\altaffiltext{2}{Department of Astronomy, The University of Texas at Austin, 2515 Speedway, Stop C1400, Austin, Texas 78712, USA; wold@astro.as.utexas.edu}
\altaffiltext{3}{Department of Astronomy, University of Wisconsin-Madison, 475 North Charter Street, Madison, WI 53706, USA} 
\altaffiltext{4}{Department of Physics and Astronomy, University of Hawaii, 2505 Correa Road, Honolulu, HI 96822, USA} 
\altaffiltext{5}{Institute for Astronomy, University of Hawaii, 2680 Woodlawn Drive, Honolulu, HI 96822, USA}  
\begin{abstract}
We present a flux-limited sample of $z\sim0.3$ Ly$\alpha$ emitters
(LAEs) from \emph{Galaxy Evolution Explorer} (\emph{GALEX}) grism
spectroscopic data. The published \emph{GALEX} $z\sim0.3$ LAE sample
is pre-selected from continuum-bright objects and thus is biased against
high equivalent width (EW) LAEs. We remove this continuum pre-selection
and compute the EW distribution and the luminosity function of the
Ly$\alpha$ emission line directly from our sample. We examine the
evolution of these quantities from $z\sim0.3$ to $2.2$ and find
that the EW distribution shows little evidence for evolution over
this redshift range. As shown by previous studies, the Ly$\alpha$
luminosity density from star-forming galaxies declines rapidly with
declining redshift. However, we find that the decline in Ly$\alpha$
luminosity density from $z=2.2$ to $z=0.3$ may simply mirror the
decline seen in the H$\alpha$ luminosity density from $z=2.2$ to
$z=0.4$, implying little change in the volumetric Ly$\alpha$ escape
fraction. Finally, we show that the observed Ly$\alpha$ luminosity
density from AGNs is comparable to the observed Ly$\alpha$ luminosity
density from star-forming galaxies at $z=0.3$. We suggest that this
significant contribution from AGNs to the total observed Ly$\alpha$
luminosity density persists out to $z\sim2.2$. 
\end{abstract}


\section{Introduction }

Observational surveys of Ly$\alpha$ emitters (LAEs) have proven to
be an efficient method to identify and study large numbers of galaxies
over a wide redshift range. To understand what types of galaxies are
selected in surveys -- and how this evolves with redshift -- it
is important to establish a low-redshift reference sample that can
be directly compared to high-redshift samples. While $z\sim0$ LAE
studies have provided insight into the physical conditions that facilitate
strong Ly$\alpha$ emission \citep[e.g.,][]{hayes13,ostlin14,rivera-thorsen15,alexandroff15,henry15,izotov16},
it is very difficult to make statistical comparisons to high-redshift
LAE populations because -- unlike the high-redshift samples -- the
$z\sim0$ studies have not been selected based solely on their Ly$\alpha$
emission. There is not currently a survey instrument capable of observing
a large number of $z\sim0$ LAEs. Thus, local LAEs are typically pre-selected
from identified high equivalent width H$\alpha$ emitters, compact
{[}O{\small{}III}{]} emitters, or ultraviolet-luminous galaxies and
subsequently observed with the \emph{Hubble Space Telescope (HST)}
to investigate the existence of Ly$\alpha$ emission. 

The lowest redshift where a direct LAE survey is presently possible
is at a redshift of $z\sim0.3$ via the \emph{Galaxy Evolution Explorer}
(\emph{GALEX}) Far Ultraviolet (FUV) ($1344-1786$~\AA) grism data.
By examining the \emph{GALEX} pipeline spectra for emission line objects,
a sample of about $50$ $z\sim0.3$ LAE galaxies was discovered \citep{deharveng08,cowie10}.
The advent of this low-redshift LAE sample has been very exciting,
and many follow-up papers have been written \citep[e.g.,][]{finkelstein09a,finkelstein09b,finkelstein11,atek09,scarlata09,cowie11}.
Furthermore, using this $z\sim0.3$ sample as an anchor point, studies
of the evolution of LAE samples have suggested that at low redshifts
high equivalent width (EW) LAEs become less prevalent and that the
amount of escaping Ly$\alpha$ emission declines rapidly \citep[e.g.,][]{hayes11,blanc11,zheng14,konno16}.
A number of explanations for these trends have been suggested including
increasing dust content, increasing neutral gas column density, and/or
increasing metallicity of star-forming galaxies at lower redshifts.
However, the \emph{GALEX} pipeline sample is biased against continuum-faint
objects. It is therefore of interest to determine the effect of this
bias on the evolutionary trends listed above. 

The \textit{GALEX} pipeline only extracts sources with a bright Near
Ultraviolet continuum counterpart (NUV $<22$). Thus, the LAE pipeline
sample is analogous to locating LAEs in the high-redshift Lyman break
galaxy (LBG) population (which is continuum selected) via spectroscopy
\citep[e.g.,][]{shapley03}. This results in a sample that is biased
against high-EW LAEs - objects with detectable emission lines but
continuum magnitudes that fall below the pipeline's threshold. In
the pipeline sample, no LAE galaxies are found with a rest-frame EW(Ly$\alpha$)$>$120
\AA \citep[][Section 5.4]{cowie10}. Beyond having an unbiased LAE
sample, searching for these extreme EW LAEs is of interest given the
recent studies suggesting that high-EW LAEs may be efficient emitters
of ionizing photons and potential analogs of reionization-era galaxies
\citep[e.g.,][]{jaskot14,erb16,trainor16}.\begin{deluxetable*}{cccccccc} 
\tablecolumns{8} 
\tablewidth{0pc} 

\tablecaption{\textit{GALEX} field exposure time, $50\%$ completeness flux threshold, number of LAE candidates, number of confirmed LAE galaxies, and number of star-forming LAEs in our final sample} 
\tablehead{ \colhead{\textit{GALEX}} & \colhead{$\alpha$} & \colhead{$\delta$} & \colhead{Exposure} & \colhead{$f_{\rm{Ly}\alpha}$ } & \colhead{Number of}& \colhead{Number of} & \colhead{Number of}\\ 
\colhead{Field} & \colhead{(J2000)} & \colhead{(J2000)} & \colhead{time} & \colhead{(erg cm$^{-2}$ s$^{-1}$)} & \colhead{candidate LAEs}& \colhead{confirmed LAEs} & \colhead{final SF LAEs}\\
\colhead{(1)} & \colhead{(2)} & \colhead{(3)} & \colhead{(4)} & \colhead{(5)} & \colhead{(6)}& \colhead{(7)} & \colhead{(8)}} 
\startdata 
CDFS & 3$^h$30$^m$40$^s$ & -27$^\circ$27$'$43$''$ & 353 ks & 1.2$\times$10$^{-15}$ & 62 & 57 & 33\\ 
GROTH & 14$^{h}$19$^{m}$58$^{s}$ & 52$^{\circ}$46$'$54$''$ & 291 ks & 1.2$\times$10$^{-15}$ & 51 & 43 & 27\\ 
NGPDWS & 14$^{h}$36$^{m}$37$^{s}$ & 35$^{\circ}$10$'$17$''$ & 165 ks & 1.5$\times$10$^{-15}$ & 22 & 16 & 6\\ 
COSMOS & 10$^{h}$00$^{m}$29$^{s}$& +2$^{\circ}$12$'$21$''$ & 140 ks & 1.6$\times$10$^{-15}$ & 38 & 28 & 17 
\enddata 
\label{fields}
\end{deluxetable*}

In this paper, we apply our data cube reduction technique \citep{barger12,wold14}
on the deepest archival \emph{GALEX} FUV grism data to remove the
continuum pre-selection and investigate whether high-EW LAEs exist
in the low-redshift universe. While previous studies have attempted
to account for these missing LAEs when computing the $z\sim0.3$ luminosity
function (LF), these corrections rely on ad-hoc assumptions and the
two independently computed pipeline LFs are offset by an overall multiplicative
factor of $\sim5$ \citep{deharveng08,cowie10}. By removing the continuum
selection and obtaining a sample that is limited by Ly$\alpha$ emission
line flux, we avoid these problems and increase the sample size of
known $z\sim0.3$ LAEs to better measure the Ly$\alpha$ EW distribution
and LF. Unless otherwise noted, we give all magnitudes in the AB
magnitude system ($m_{\mbox{AB}}=23.9-2.5$log$_{10}f_{\nu}$ with
$f_{\nu}$ in units of $\mu$Jy) and EWs are given in the rest-frame.
We use a standard $H_{0}=70$ km s$^{-1}$ Mpc$^{-1}$, $\Omega_{M}=0.3$,
and $\Omega_{\Lambda}=0.7$ cosmology.

\section{Choice of Fields and Existing Ancillary Data}

Our data cube reconstruction of the \emph{GALEX} grism data requires
fields observed with hundreds of rotation angles \citep[see Section 3.1 and ][for details]{barger12}.
This limits our study to the four deepest FUV grism observations:
Chandra Deep Field South, Groth, the North Galactic Pole Deep Wide
Survey, and the Cosmic Evolution Survey (archival tilename: CDFS-00,
GROTH-00, NGPDWS-00, and COSMOS-00). These fields are some of the
most heavily studied extra-galactic fields and contain ancillary data
which has greatly aided this work. We note that the \emph{GALEX} fields
are large ($\sim1$~deg$^2$), and with the exception of the archival
ground-based imaging, the existing ancillary surveys only cover subregions
of the fields.

These ancillary data include archival optical spectra and redshifts
which were used to verify the redshifts derived from the candidate
Ly$\alpha$ emission. We used cataloged redshifts in CDFS \citep{cooper12,cardamone10,mao12,lefevre13},
GROTH \citep{matthews13,flesch15}, NGPDWS \citep{kochanek12}, and
COSMOS \citep{prescott06,lilly07,adelmanmccarthy09,adams11,knobel12},
and we used the CDFS and COSMOS optical spectra published by \citet{lefevre13,lilly07},
respectively. The 7~Ms {\em Chandra\/} image \citep{luo17} of
the CDFS \citep[][]{giacconi02,luo08} region, along with shallower
X-ray observations in the Extended CDFS \citep[][]{lehmer05, virani06},
COSMOS \citep[][]{civano16,elvis09}, GROTH \citep[][]{laird09},
and NGPDWS \citep[][]{kenter05} fields, were used to identify AGNs.
We also used data from the \textit{Wide-field Infrared Survey Explorer}
(\textit{WISE}) to identify AGNs via the color cut prescribed by
\citet{assef13}.

\section{\emph{GALEX} FUV LAEs}

\subsection{Data Cube Catalog Extraction}

\begin{figure*}[!t]
\includegraphics[bb=80bp 70bp 750bp 520bp,clip,scale=0.29]{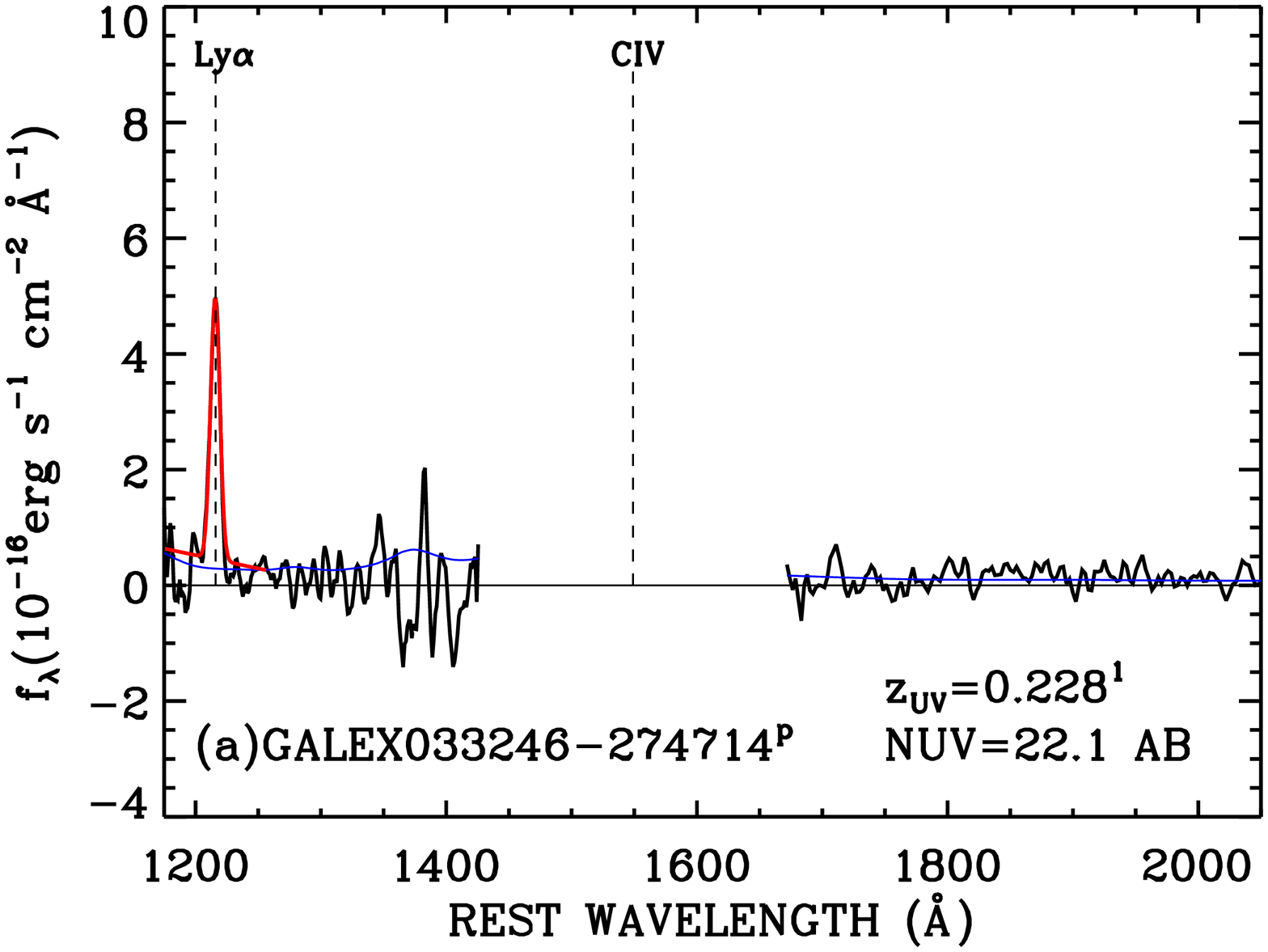}\includegraphics[bb=175bp 70bp 750bp 520bp,clip,scale=0.29]{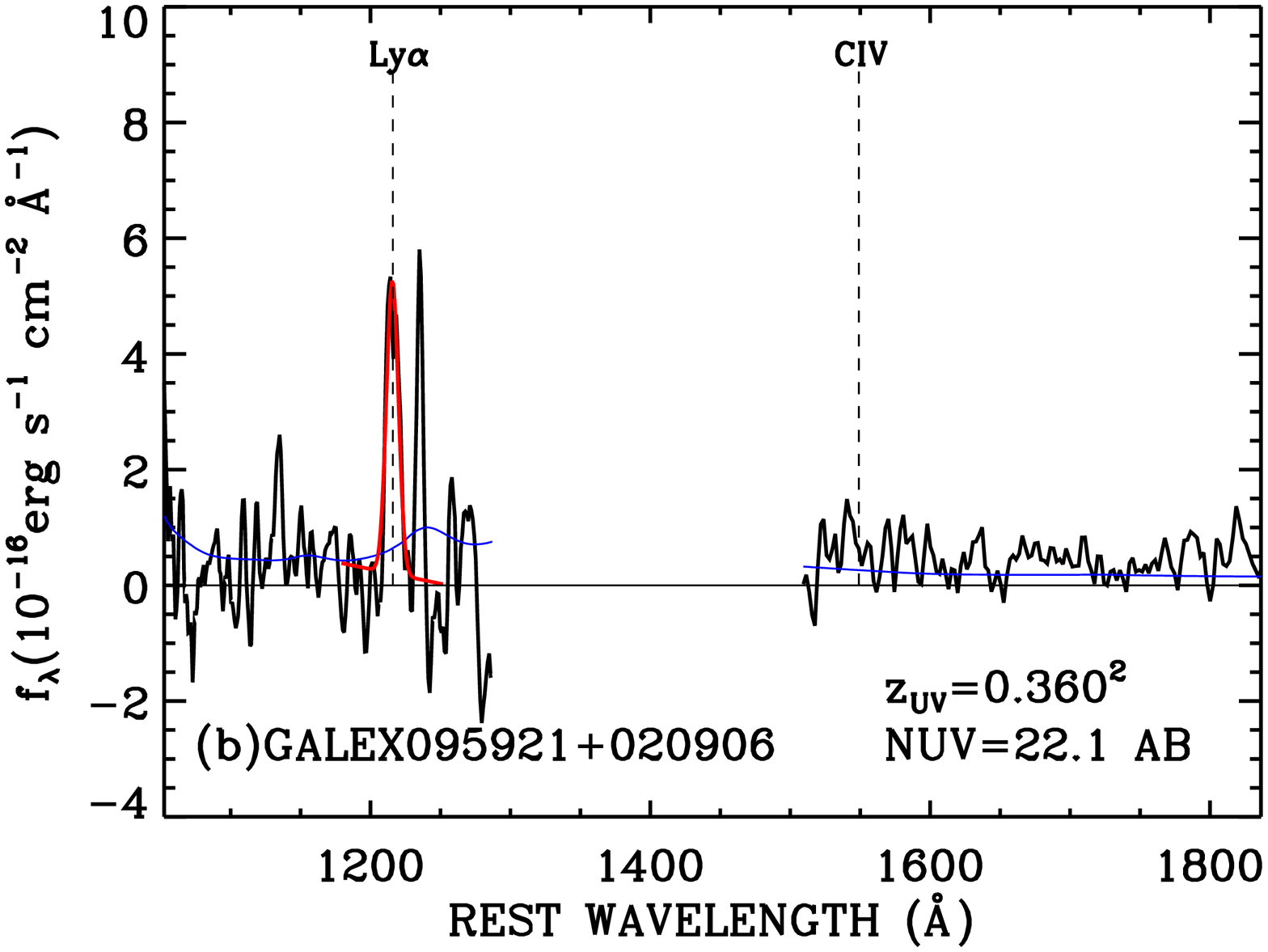}\includegraphics[bb=175bp 70bp 710bp 520bp,clip,scale=0.29]{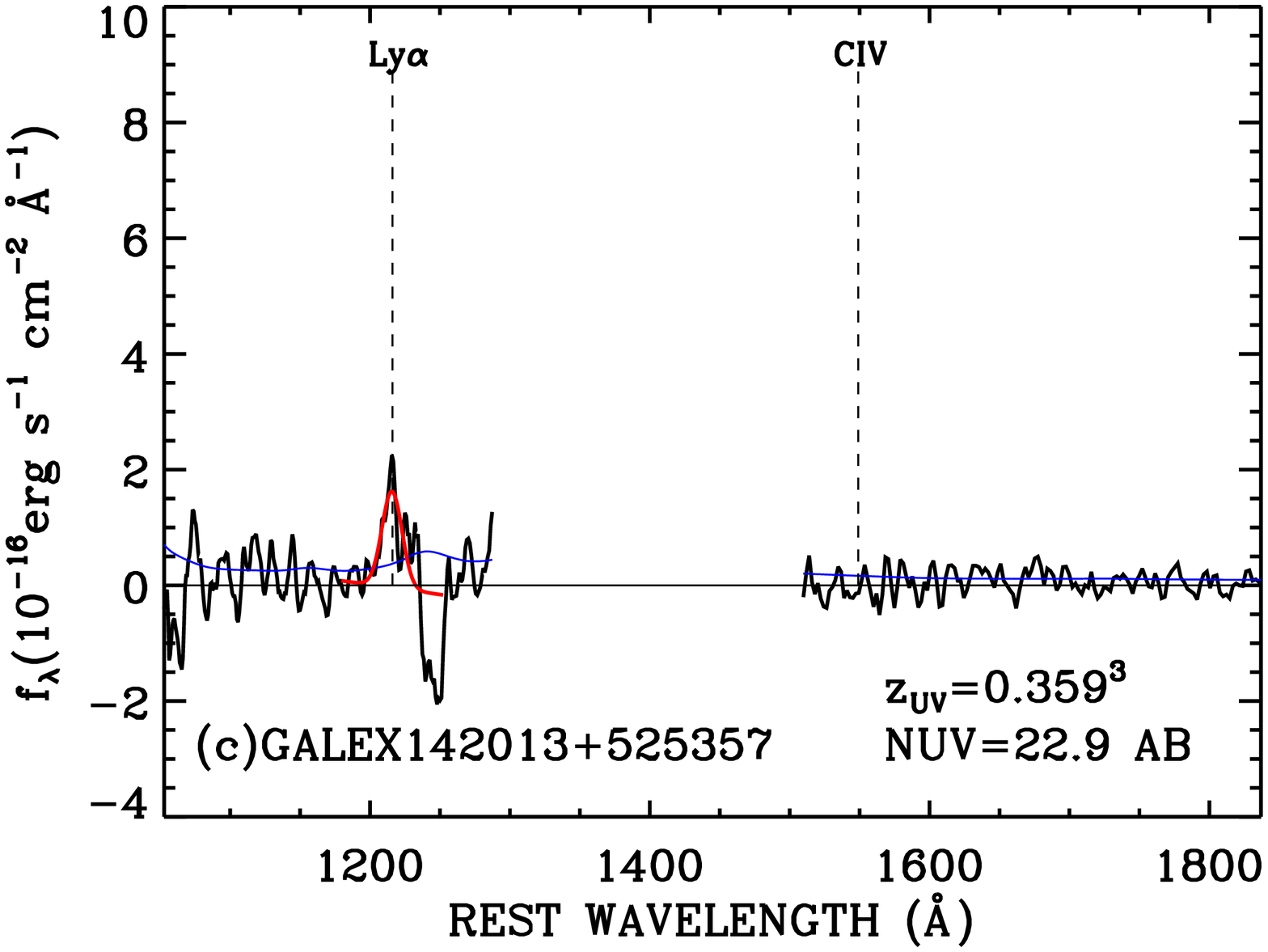}

\includegraphics[bb=80bp 70bp 750bp 520bp,clip,scale=0.29]{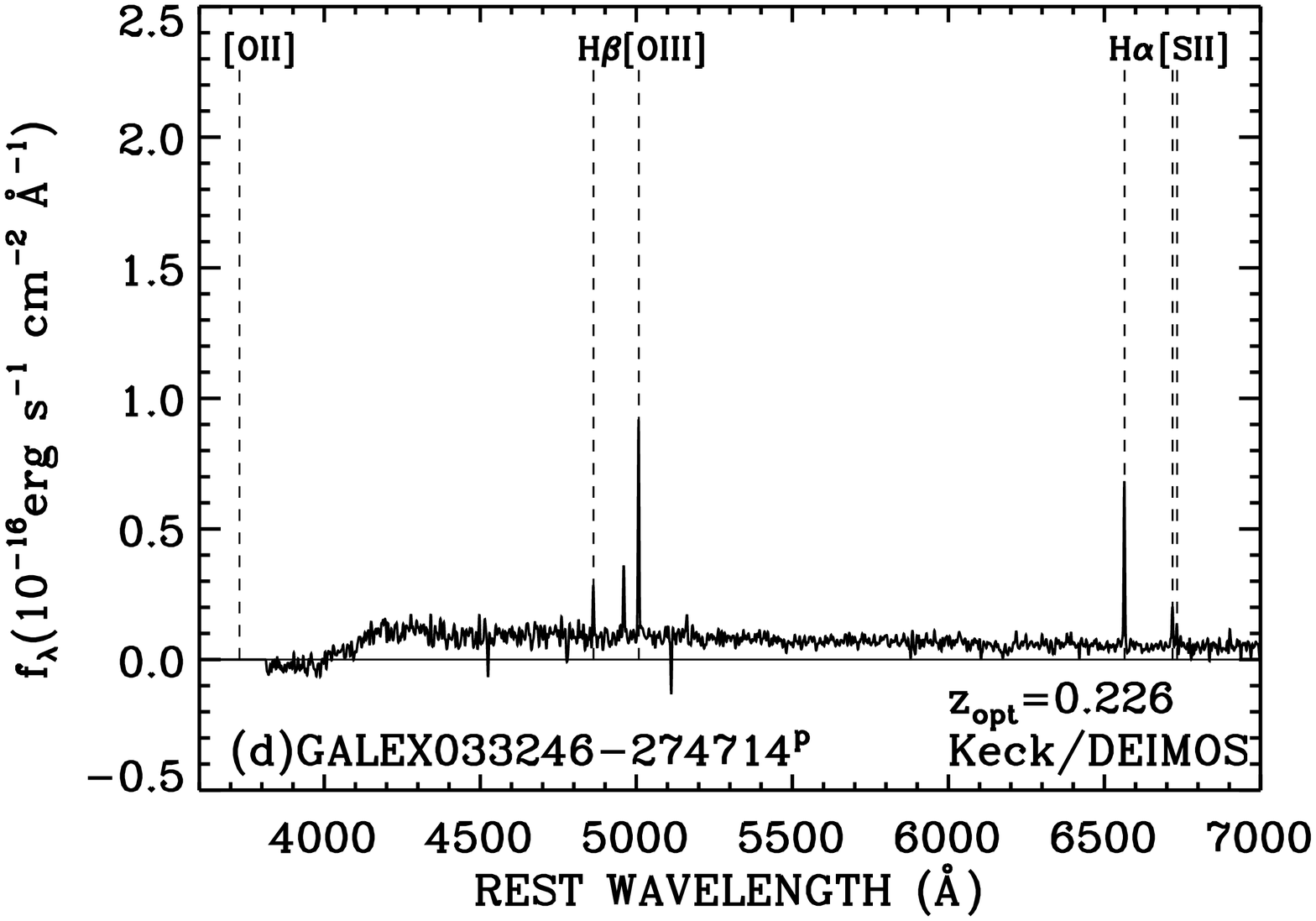}\includegraphics[bb=175bp 70bp 750bp 520bp,clip,scale=0.29]{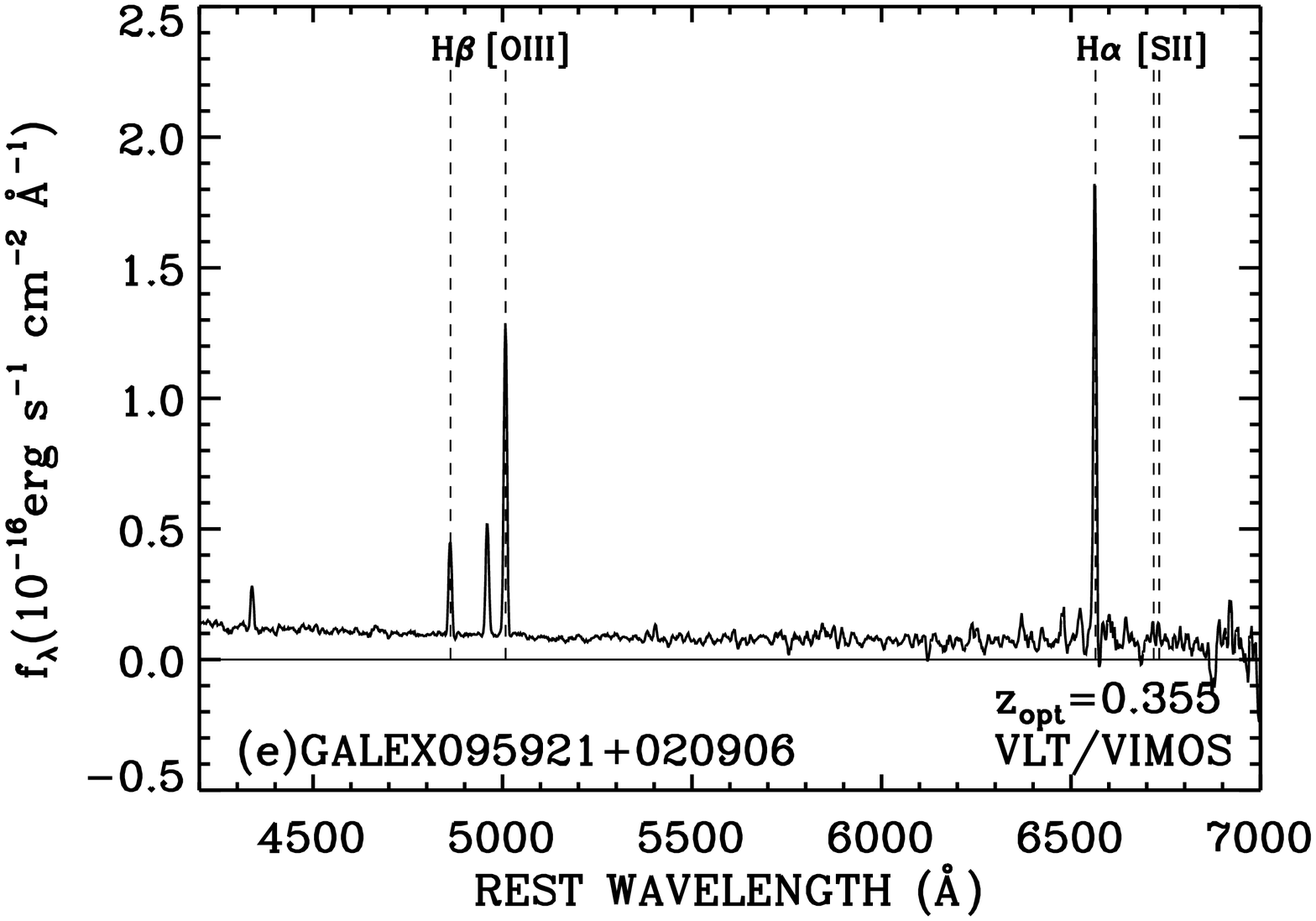}\includegraphics[bb=175bp 70bp 750bp 520bp,clip,scale=0.29]{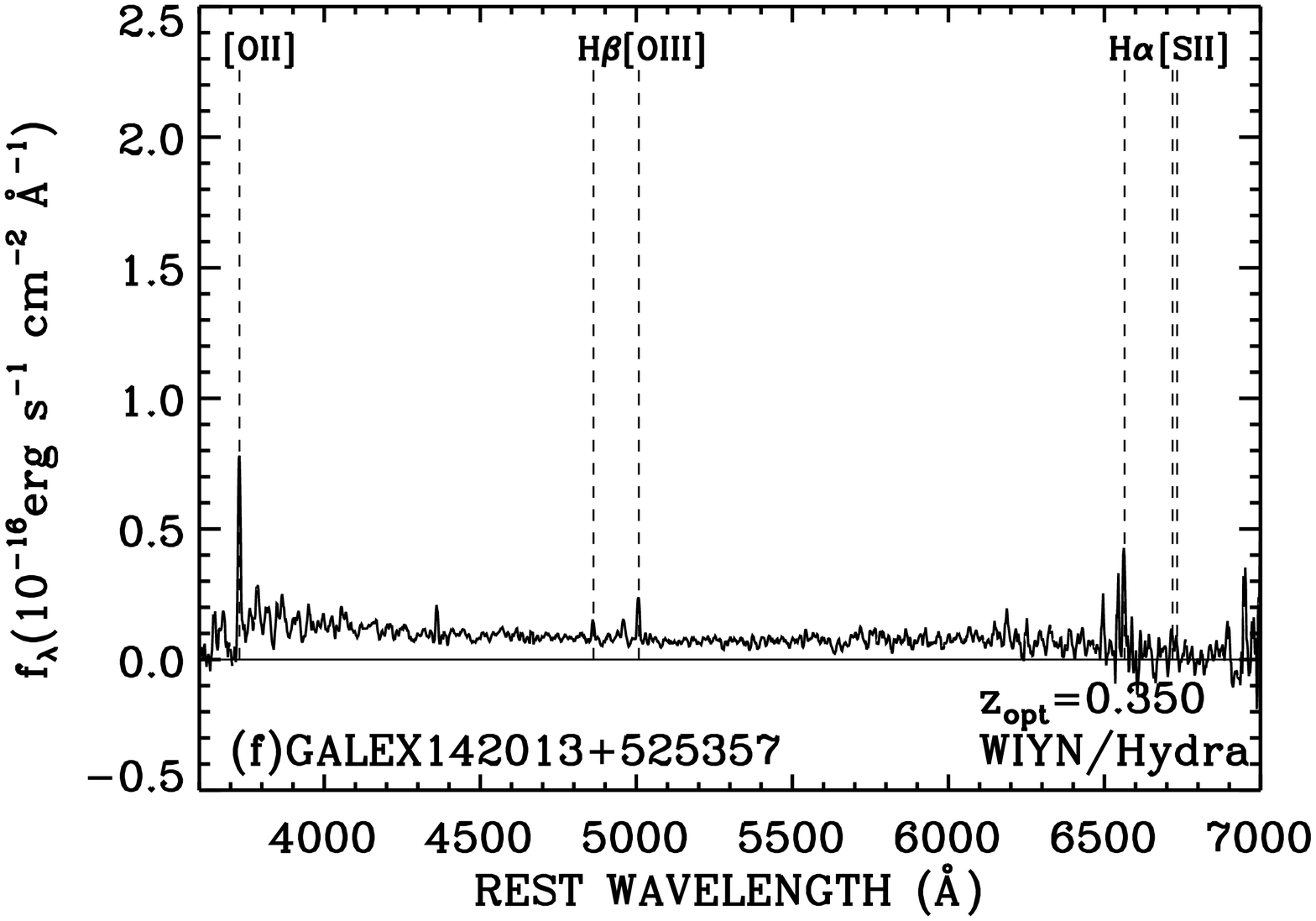}

\caption{\textbf{(Top row)} Examples of LAE UV spectra with the full range
of assigned confidence classes (1=good, 2=marginal, 3=poor) indicated
by superscripts to the listed Ly$\alpha$ redshifts. The Ly$\alpha$
redshift and the Ly$\alpha$ flux are measured from the Gaussian fit
(red profile). We indicate the 1$\sigma$ noise array with the blue
line. Each UV spectrum consists of the \emph{GALEX} FUV spectral band,
a band gap, and the\emph{ GALEX} NUV spectral band. We examine the
UV spectra for high-excitation lines like C{\scriptsize{}IV} to help
identify AGNs. \textbf{(Bottom row)} Below each UV spectrum, we show
the corresponding optical spectrum. The three example LAEs were also
selected to illustrate the quality of the optical spectra obtained
from different telescopes/instruments. The archival VLT/VIMOS spectrum
is from zCOSMOS \citep{lilly07}. We use these optical spectra to
confirm our LAE candidates and to help identify AGNs via the BPT diagnostic
diagram. }

\label{example}

{\footnotesize (A color version of this figure is available in the online journal.)}
\end{figure*}

In \citet{barger12}, we describe in detail our method to convert
multiple \emph{GALEX} low-resolution slitless spectroscopic images
into a three-dimensional (two spatial axes and one wavelength axis)
data cube. Here, we provide a brief overview of this process. For
each of our four fields, we begin our data cube construction with
archival 1.25 degree diameter FUV grism intensity maps. For each intensity
map, we know the wavelength dispersion and the dispersion direction,
and this allows us to extract a spectrum for each spatial position
thus forming an initial data cube. A data cube constructed from a
single slitless spectroscopic image will suffer from overlapping spectra
caused by neighboring objects that are oriented in-line with the dispersion
direction. However, the spectral dispersion direction can be altered
from one exposure to the next by changing the grism rotation angle,
and objects that overlap in one rotation angle are unlikely to overlap
in another rotation angle. Thus, we are able to disentangle overlaps
by requiring our selected fields to have hundreds of exposures with
a corresponding number of rotation angles. 

For each field, we construct hundreds of data cubes - one for each
exposure - and then combine these initial data cubes applying a 5$\sigma$
cut to remove contamination from overlapping sources. This results
in an intermediate data cube that has a wavelength step of 2.5 \AA\
and a wavelength range of $1345$ to $1795$ \AA. We resample this
intermediate cube to form wavelength slices with a 10 \AA\ wavelength
extent sampled every 5 \AA. We designed the wavelength slices to
have a wavelength extent that matches the spectral resolution of \emph{GALEX}.
To account for emission line objects that would otherwise be split
into two adjacent wavelength slices, we decided to make wavelength
slices every 5 \AA\ interval. For each slice, we subtracted the average
of independent slices on either side of the primary slice, N. We used
slices N-10, N-8, N-6 and slices N+6, N+8, and N+10 to form this average.
This procedure subtracts most of the background residual structure
and most of the continuum from objects within the data cube. 

The final background subtracted FUV data cubes have a 50$'$ diameter
field of view and cover a wavelength range of 1395 to 1745 \AA\ or
a Ly$\alpha$ redshift range of $z=$ 0.15 to 0.44. For each wavelength
slice, we used SExtractor \citep{bertin96} to identify all 4$\sigma$
sources within the cube and then visually inspected each source and
its spectrum (1-D and 2-D) to eliminate objects that were artifacts.
During this visual inspection, we assigned a confidence category ($1=$
good, $2=$ fair, $3=$ uncertain) reflecting our confidence that
the identified candidate is real and not an artifact. We applied this
data cube search method and found 62 CDFS, 51 GROTH, 22 NGPDWS, and
38 COSMOS candidate LAEs (see Table \ref{fields}). In Figure \ref{example}(a),
(b), and (c), we show extracted 1D spectra for all confidence categories
to illustrate the quality of our \emph{GALEX} spectra. We estimate
spectral noise by examining regions above and below the object's two-dimensional
spectrum. In general, spectral noise will increase as contamination
from neighboring sources increases and as the spectral response falls
off toward the edges of the spectral window.  As in \citet{barger12}
and \citet{wold14}, we use our modified version of the \emph{GALEX}
pipeline software to extract two and one-dimensional spectra rather
than extracting spectra directly from our data cubes. Our modified
method uses profile-weighted spectral extraction \citep{horne86}
which provides modest improvement to the spectral signal to noise.
Additionally, our modified extraction method, which is optimized to
extract a single spectrum, provides a check on our LAE candidate sample
which is selected based on a search of our four data cubes.

\subsection{Optical Spectroscopic Follow-up}

\noindent \label{opt_spec}

We used optical spectroscopic follow-up to confirm the veracity of
our candidate LAEs and to identify optical AGNs. For our sample of
173 candidate LAEs, we obtained optical spectroscopic information
for 171. To populate our optical spectroscopic target list, we visually
identified the closest optical counterpart to the LAE candidate's
position in the FUV image which has a spatial resolution of $\sim5''$.
Follow-up spectroscopic observations were primarily obtained with
the Hydra fiber spectrograph on the Wisconsin\textendash Indiana\textendash Yale\textendash NOAO
(WIYN) telescope. Each WIYN target was observed for a total of $\sim3$
hours in a series of runs from January to March 2016. We configured
the spectrograph using the \textquotedblleft red\textquotedblright{}
fiber bundle and the 316@7.0 grating at first order with the GG-420
filter to provide a spectral window of \ensuremath{\sim}4500\textendash 9500
\AA\ with a pixel scale of 2.6 \AA\ per pixel. The Hydra \textquotedblleft red\textquotedblright{}
fibers are 2$''$ diameter and have a positional accuracy of 0.3$''$,
which ensured that the majority of light from our target galaxies
was observed with little contamination from the sky and neighboring
sources. We employed the IRAF task \emph{dohydra} in the reduction
of our spectra. This task is specifically designed for reduction of
data from the Hydra spectrograph and includes steps for dark and bias
subtraction, flat fielding, dispersion calibration, and sky subtraction.
In Figure \ref{example}(f), we show an example of a WIYN/HYDRA obtained
spectrum.

We also targeted a subset of LAE candidates with the DEep Imaging
Multi-Object Spectrograph \citep[DEIMOS;][]{faber03} on Keck II.
The observations were made with the ZD600 line mm$^{-1}$ grating
blazed at 7500 \AA. This gives a resolution of $\sim$ 5 \AA~ with
a 1$''$ slit and a wavelength coverage of 5300 \AA.  Each $\sim$30
minute exposure was broken into three subsets, with the objects stepped
along the slit by 1.5$''$ in each direction. The raw two-dimensional
spectra were reduced and extracted using the procedure described in
\citet{cowie96}. In Figure \ref{example}(d), we show an example
of a Keck/DEIMOS spectrum.

Our optical follow-up with WIYN and Keck was designed to have sufficient
signal-to-noise to place our sources on the BPT diagnostic diagram
\citep[][see Section \ref{agn}]{baldwin81}. Thus, our optical spectra
typically displayed easily identifiable H$\alpha$ and {[}O{\small{}III}{]}
emission lines. In all cases, at least two spectroscopic lines were
required to measure the optical redshift. From our observed Keck spectra,
we find that our LAEs have a median H$\alpha$ line flux of $9\times10^{-16}$
erg s$^{-1}$cm$^{-2}$ giving an uncorrected-for-dust SFR of 2 M$_{\odot}$
yr$^{-1}$ at $z=0.3$. From our shallower WIYN spectra, we estimate
any H$\alpha$ line with a flux greater than $2\times10^{-16}$ erg
s$^{-1}$cm$^{-2}$ will be detected at the 5$\sigma$ level. Thus,
we do not expect any significant confirmation bias to be introduced
by optically following-up our candidates with two different telescopes.
In Section \ref{spur}, we show that the vast majority of LAE candidates
without a recovered optical redshift are assigned the lowest confidence
category, and we argue that as a general rule increasing the depth
of our optical spectra would only serve to further follow-up spurious
LAE candidates.

When a LAE candidate's optical counterpart was found to have an existing
archival optical redshift, we relied on the archival data to confirm
our proposed Ly$\alpha$ based redshift. This practice reduces the
need for telescope time and should not impose a significant sample
selection bias since (in all but two cases) we have targeted the non-archival
sources with WIYN or Keck. We used archival VLT/VIMOS redshifts and
spectra from the zCOSMOS survey \citep[for COSMOS;][]{lilly07} and
from the VIMOS VLT Deep Survey \citep[for CDFS;][]{lefevre13}. In
Figure \ref{example}(e), we show an example of the archival VLT/VIMOS
spectra. For 23 LAE candidates, their published optical redshifts
lack accompanying optical spectra. These optical redshifts allow us
to confirm the veracity of 19 candidates and falsify 4 candidates,
but we are unable to examine their optical spectra for AGN features.
In one case, GALEX033150-280811, there is a published optical spectroscopic
AGN classification \citep{mao12}, and we use this classification
in our study. They find that this object has emission line ratios
typical of AGN activity, which is broadly consistent with one of our
optical AGN classes. In Section \ref{agn}, we discuss our AGN classification
scheme in more detail.

\subsection{Catalog Completeness}

\noindent 
\begin{figure}
\includegraphics[bb=120bp 80bp 690bp 550bp,clip,angle=180,scale=0.4]{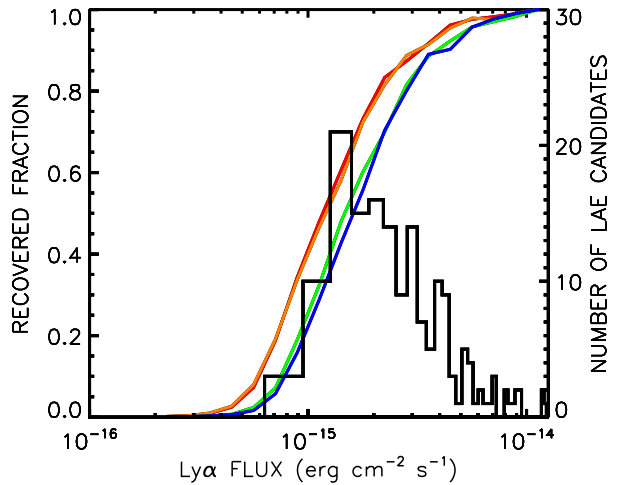}

\caption{Fraction of simulated Ly$\alpha$ emitters recovered as a function
of the emission-line flux. The red, orange, green, and blue curves
show recovered fractions from the CDFS, GROTH, NGPDWS, and COSMOS
fields, respectively (also see Table \ref{fields}). The histogram
shows the number of LAE candidates as a function of flux for all four
\textit{GALEX} fields.}

\label{compl}

{\footnotesize (A color version of this figure is available in the online journal.)}
\end{figure}
To determine the limitations of our multi-field catalogs and to compute
the LAE galaxy LF, we measured our ability to recover fake emitters
as a function of flux. For each field, we added 1000 simulated emitters
uniformly within the field's data cube. We did not model morphology
or size difference, since nearly all emitters are unresolved at the
spatial ($\sim5''$) and spectral resolution ($\sim10$ \AA) of the
\textit{\small{}GALEX} grism data. We then ran our standard selection
procedure and found the number of recovered objects. We independently
performed the above procedure ten times, giving a total of 10,000
input sources. In Table \ref{fields}, we list the flux threshold
above which each field is greater than 50$\%$ complete. As expected,
the completeness limit scales as the inverse square root of the exposure
time. In Figure \ref{compl}, we show the completeness as a function
of the emission-line flux. The black histogram displays the Ly$\alpha$
flux distribution of our 173 LAE candidates.

\subsection{Catalogs of LAE Candidates by Field}

In Tables \ref{cdfs}-\ref{cosmos}, we list all of the LAE candidates
in the CDFS (Table \ref{cdfs}), GROTH (Table \ref{groth}), NGPDWS
(Table \ref{ngpdws}), and COSMOS (Table \ref{cosmos}) fields ordered
by right ascension. We measured FUV and NUV AB magnitudes from the
archival \textit{GALEX} background subtracted intensity maps \citep[][]{morrissey07}.
We first determined the magnitudes within 8$''$ diameter apertures
centered on each of the emitter positions. To correct for flux that
falls outside our apertures, we measured the offset between 8$''$
aperture magnitudes and \textit{GALEX} pipeline total magnitudes for
all bright cataloged objects (20-23 mag range) within our fields.
We determined the median offset for each field (typically $\sim0.5$
mag) and applied these to our aperture magnitudes.  For extended
sources we adopt the \emph{GALEX} cataloged magnitude which uses SExtractor's
AUTO aperture.  We list these magnitudes in  Tables \ref{cdfs}-\ref{cosmos}. 

We corrected our one-dimensional FUV spectra for Galactic extinction
assuming a \citet[][]{fitzpatrick99} reddening law with R$_{V}$=3.1.
We obtained $A_{V}$ values from the \citet[][]{schlafly11} recalibration
of the \citet[][]{cardelli89} extinction map as listed in the NASA/IPAC
Extragalactic Database (NED). Galactic extinction increases the Ly$\alpha$
flux by $\sim$11\% for the COSMOS LAEs, $\sim$4\% for the GROTH
LAEs, and $\sim$6\% for the CDFS and NGPDWS LAEs.

From these extinction corrected spectra, we measured the redshifts,
the Ly$\alpha$ fluxes, and the line widths using a two step process.
First, we fit a 140 \AA~ rest-frame region around the Ly$\alpha$
line with a Gaussian and a sloped continuum (e.g., see Figure \ref{example}(a),
(b), and (c)). A downhill simplex optimization routine was used to
$\chi^{2}$ fit the five free parameters (continuum level and slope
plus Gaussian center, width, and area). We used the results of this
fitting process to subtract the continuum and as a starting point
for the second step. In the second step, we used the IDL MPFIT procedures
of \citet[][]{markwardt09} to $\chi^{2}$ fit the remaining three
Gaussian parameters. We found that this two step procedure rather
than a 5 parameter MPFIT solution resulted better $\chi^{2}$ fits.
With the best-fit redshifts and Ly$\alpha$ fluxes, we calculated
Ly$\alpha$ luminosities. When available, we used the more precise
optical redshift rather than the Ly$\alpha$ redshift to calculate
the Ly$\alpha$ luminosities. We list the Ly$\alpha$ redshifts and
luminosities in Tables \ref{cdfs}-\ref{cosmos}. During the initial
visual inspection of the 1-D and 2-D spectra, we classified our LAE
candidates into three qualitative categories ($1=$good, $2=$fair,
$3=$uncertain) reflecting our confidence that the identified candidate
is real and not an artifact. Our LAE detection confidence is given
in Tables \ref{cdfs}-\ref{cosmos} as superscripts to the Ly$\alpha$
redshift. 

The rest-frame EW$_{{\rm r}}$(Ly$\alpha$) measured on the spectra
are quite uncertain due to the very faint UV continuum. We obtained
a more accurate rest-frame EW by dividing the measured Ly$\alpha$
flux by the continuum flux measured from the broadband FUV image (corrected
for the emission-line contribution). We computed the EW uncertainty
by propagating the 1$\sigma$ error from our FUV and Ly$\alpha$ flux
measurements. It is these rest-frame EWs with 1$\sigma$ errors that
are listed in Tables \ref{cdfs}-\ref{cosmos}. In Section \ref{sec_ew},
we use these measurements to construct the $z\sim0.3$ rest-frame
EW distribution for star-forming LAEs. 

Candidate X-ray counterparts were identified by matching all X-ray
sources within a $6''$ radius from the data cube position. We then
manually inspected the matches to reject false counterparts caused
by X-ray sources with an optical counterpart neighboring but not associated
with the LAE in question. We list the \emph{Chandra} X-ray luminosity
of each identified counterpart in Tables \ref{cdfs}-\ref{cosmos}.
LAE candidates within the X-ray footprint that lack detections were
given X-ray luminosities of `-999'. At our survey's redshift of $z\sim0.3$,
the X-ray imaging depth ($f_{2-8,10,10{\rm \,{keV}}}\sim6.7$, $3.8$,
$8.9\times10^{-16}$ erg cm$^{-2}$ s$^{-1}$) corresponds to an X-ray
luminosity of $\sim10^{41}$ erg s$^{-1}$ for our CDFS, GROTH and
COSMOS fields \citep[][]{lehmer05,laird09,civano16}. The X-ray imaging
depth for the NGPDWS field ($f_{2-7\,{\rm {keV}}}\sim1.5\times10^{-16}$
erg cm$^{-2}$ s$^{-1}$) corresponds to an X-ray luminosity of $\sim10^{42}$
erg s$^{-1}$ \citep{kenter05}. For the central 484.2 arcmin$^{2}$
of the CDFS, we use a deeper X-ray imaging survey that has a sensitivity
limit ($f_{2-7\,{\rm {keV}}}\sim2.7\times10^{-17}$ erg cm$^{-2}$
s$^{-1}$) that corresponds to an X-ray luminosity of $\sim10^{40}$
erg s$^{-1}$ \citep{luo17}.

In the final column of Tables \ref{cdfs}-\ref{cosmos}, we give an
AGN classification. Our AGN classification scheme is described in
Section \ref{agn}.

\subsection{Spurious LAE Candidates}

\noindent \label{spur}

We obtained optical redshifts and spectra from archival sources and
combined this with our own optical spectra from Keck-DEIMOS and WIYN-Hydra
(see Section \ref{opt_spec}). For our sample of 173 candidate LAEs,
we have optical spectroscopic information for 171. Using these data,
we found that 27 LAE candidates are spurious. These spurious sources
have optical redshifts that are not consistent with the redshifts
derived from the candidate Ly$\alpha$ emission line ($z_{UV}$) or
have no viable optical counterpart. Specifically, we consider any
source with an optical redshift outside of $z_{UV}\pm0.03$ to be
spurious. We found that two of our spurious LAE candidates are known
X-ray bright stars. Both stars are relatively high confidence candidates
(given 1 and 2 confidence classifications) and were selected based
on emission at an observed wavelength of $1550$ \AA\ indicating
C {\small{}IV} $\lambda1549$ emission. We are confident that these
are C {\small{}IV} $\lambda1549$ selected because both stars display
Mg {\small{}II }$\lambda2798$ emission in their \emph{GALEX} NUV
spectrum. We also found two O {\small{}VI }$\lambda1035$ selected
AGNs (GALEX142010+524029 and 143554+351910) at $z\sim0.55$. For these
two high-redshift interlopers, Ly$\alpha$ emission falls in the gap
between the \emph{GALEX} FUV and NUV bands. For both sources, strong
C {\small{}IV} $\lambda1549$ emission is observed in the NUV spectrum.
Overall, our optical spectroscopic follow up indicates that our data
cube search selects real emission line objects (Ly$\alpha$, C {\small{}IV},
and O {\small{}VI} emitters) 87\% of the time or 148 confirmed sources
out of a total of 171 candidates. For the purposes of this study,
we consider any non-Ly$\alpha$ selected source to be spurious. 

 In Tables \ref{cdfs}-\ref{cosmos}, we indicate spurious objects
with optical redshifts not consistent with their Ly$\alpha$ redshifts
by showing their optical redshift in parentheses. We indicate stars
by setting their optical redshift to `star' in Column 14. Additionally,
we targeted 8 candidate LAEs with WIYN but did not recover an optical
redshift. We indicate these objects by setting their optical redshifts
to `no z' in Column 14. All spurious LAE candidates are given blank
entries for the Ly$\alpha$ luminosity and the rest-frame EW$_{{\rm r}}$(Ly$\alpha$)
fields in Columns 8 and 9. 

Given the $\sim5''$ spatial resolution of \emph{GALEX}, it is possible
that some of the spurious LAE candidates result from closely paired
systems in which we have inadvertently targeted the wrong optical
counterpart. To investigate this possibility, we examine the available
optical images and find that the majority ($55\%$) of candidates
have alternative optical counterparts within $5''$. However, the
centroid of the \emph{GALEX} source can typically be determined with
an accuracy much less than $5''$, and we know that $87\%$ of our
candidates are confirmed with optical spectra. Thus, we suspect that
the importance of inadvertently targeting the wrong optical counterpart
can be better assessed by computing our confirmation rate of high
confidence candidates. As discussed in Section 3.1, during our initial
data cube search we visually inspected each \emph{GALEX} spectrum
(1-D and 2-D) and assigned a confidence category ($1=$ good, $2=$
fair, $3=$ uncertain) reflecting our confidence that the identified
candidate is real and not an artifact. Candidates with higher confidence
measures (1 or 2) are optically confirmed $98\%$ percent of the time,
or 116 out of a total of 118. On the other hand, candidates with low
confidence measures (3) are optically confirmed $60\%$ percent of
the time, or 32 out of a total of 53. Applying the high confidence
percentage to our total sample size, we estimate that $\sim3$ spurious
LAE candidates could result from closely paired systems in which we
have inadvertently targeted the wrong optical counterpart. Given this
low estimate, we make no attempt to correct for this effect, and we
simply exclude all spurious candidates from further analysis.
\begin{figure}
\includegraphics[bb=120bp 75bp 695bp 510bp,clip,scale=0.4]{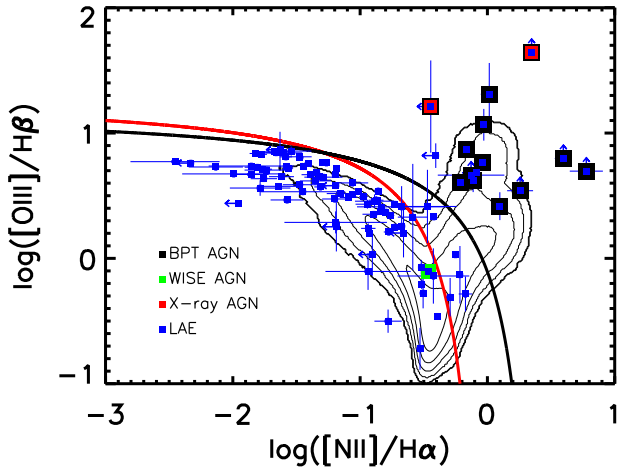}

\caption{BPT diagram for our LAE sample with narrow emission line optical spectra.
The black curve shows the theoretical separation between AGNs and
star-forming galaxies proposed by \citet{kewley01}. The red curve
shows the empirical separation between SDSS AGNs and star-forming
galaxies proposed by \citet{kauffmann03}. For the purposes of our
study we require BPT AGNs to lie to the upper-right of both curves
(see Section \ref{agn} for details). We show narrow-line X-ray AGNs
which are LAEs with X-ray luminosities greater than $10^{42}$ erg
s$^{-1}$ (red squares). We also indicate a \emph{WISE} AGN which
is identified via the color cut prescribed by \citet[][green square]{assef13}.
The black contours show the distribution of SDSS sources on a log
scale.}

\label{bpt}

{\footnotesize (A color version of this figure is available in the online journal.)}
\end{figure}
\begin{figure}
\includegraphics[bb=120bp 75bp 695bp 510bp,clip,scale=0.4]{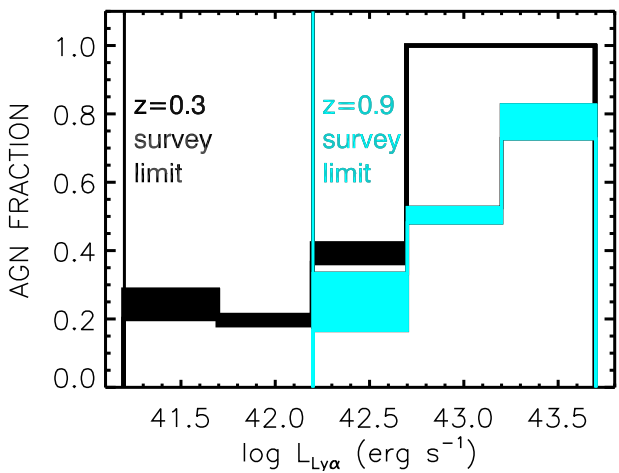}

\caption{AGN fraction per Ly$\alpha$ luminosity bin for both the $z=0.3$
(black histogram) and $z=0.9$ \citep[cyan histogram;][]{wold14}
EW $>20$ \AA\ LAE samples. The filled histogram regions show the
boost in the AGN fractions caused by restricting the survey to areas
with deep X-ray data. This restriction ensures a robust AGN classification
at the expense of sample size. We show the approximate luminosity
limit for each LAE survey with vertical black and cyan lines. In all
observed $0.5$ dex Ly$\alpha$ luminosity bins, we find AGN fractions
that are $\sim20\%$ or greater. }

\label{agnhisto}

{\footnotesize (A color version of this figure is available in the online journal.)}
\end{figure}

\subsection{AGN - Galaxy Identification}

\noindent \label{agn}We made a classification of whether an emitter
was an AGN based on: X-ray imaging, UV spectra, optical spectra, and
infrared imaging. We classified objects as X-ray AGNs (denoted by
`x' in Tables \ref{cdfs}-\ref{cosmos} Column 15) if their X-ray
luminosity exceeded $10^{42}$ erg s$^{-1}$ \citep[e.g., see][]{hornschemeier01,barger02,szokoly04}.
We note that archival X-ray imaging is available for $75\%$ percent
of our survey area, and deep X-ray imaging that has a depth better
than $1\times10^{-15}$ erg cm$^{-2}$ s$^{-1}$ is available for
$49\%$ percent of our survey area. We classified objects as UV AGNs
(denoted by `u' in Tables \ref{cdfs}-\ref{cosmos} Column 15) by
examining the \emph{GALEX} spectra for high-excitation lines such
as C {\small{}IV} $\lambda1549$ \citep[for details on this procedure see][]{cowie10,cowie11}.
We note that this does not provide a uniform AGN diagnostic because
in some cases the gap between the FUV and NUV bands prevents the observation
of potential high excitation UV lines (e.g., see Figure \ref{example}(a)).
 We classified objects as \emph{WISE} AGNs (denoted by `w' in Tables
\ref{cdfs}-\ref{cosmos} Column 15) via the color cut as prescribed
by \citet{assef13}. Finally, we classified objects as optical AGNs
based on emission line ratios via the BPT diagnostic diagram \citep{baldwin81}
or the presence of broad emission lines ($\gtrsim1000$ km s$^{-1}$
denoted by `n' or `b', respectively in Tables \ref{cdfs}-\ref{cosmos}
Column 15). 

In Figure \ref{bpt}, we show a BPT diagram of {[}O{\small{}III}{]}$\lambda5007/$H$\beta$
versus {[}N{\small{}II}{]}$\lambda6584$/H$\alpha$ for our sample
of LAEs with narrow emission line optical spectra. The BPT diagram
uses the ratio of neighboring emission lines which are insensitive
to flux calibration and reddening effects to separate star-forming
(SF) galaxies from AGNs. The red curve shows the empirical separation
between SDSS AGNs and SFs proposed by \citet{kauffmann03}. The black
curve shows the theoretical separation between AGNs and SFs proposed
by \citet{kewley01}. Objects that lie in between these two curves
are generally classified as intermediate objects with both AGN and
SF contributions. For our study, we require a BPT AGN to be positioned
above or to the right of both curves. As a reference we show contours
representing the distribution of sources from the Sloan Digital Sky
Survey \citep[SDSS;][]{york00} on a log scale. We have taken emission
line measurements from the MPA-JHU catalog for SDSS DR7.

We restricted our BPT sample to sources with either H$\alpha$ or
{[}O{\small{}III}{]}$\lambda5007$ detected with a signal-to-noise
above 4. Objects with {[}N{\small{}II}{]}$\lambda6584$ or H$\beta$
detected with a signal-to-noise below 1, have their flux values set
to 1$\sigma$ and are displayed as upper or lower limits, respectively.
In Figure \ref{bpt}, the 13 LAEs identified as BPT AGNs are outlined
in black. Two of these BPT AGNs are also identified as X-ray AGNs
(red outlined symbols). We find one \emph{WISE} AGN that is not identified
as a BPT AGN (green outlined symbol). For our sample of LAEs, all
UV AGNs are also found to be broad-line AGNs (BLAGNs; note only narrow-emission
line objects are shown in Figure \ref{bpt}). Overall, we find that
37 out of our 146 non-spurious LAEs are classified as AGNs by some
means. As described in the next section, we classify optical absorber
LAEs as AGNs, and we include these objects in our total AGN count
of 37.
\begin{figure*}[t]
\includegraphics[width=18cm]{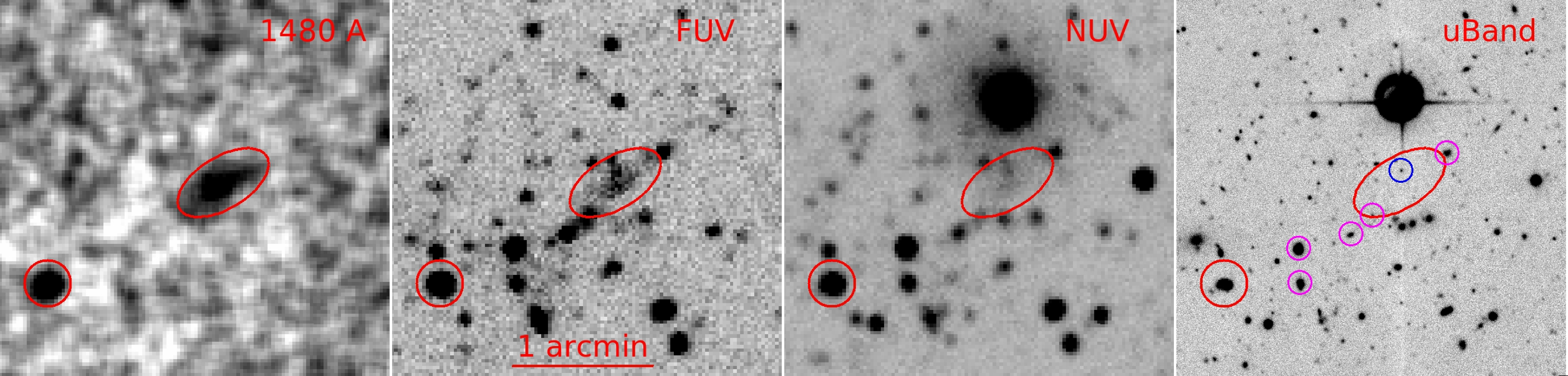}

\caption{Images of the extended LAE candidate GALEX033145-281038 with $z_{UV}=0.218$
(red ellipse). From left to right, we show a 1480\AA\ data cube slice
with a width of 10\AA, the \emph{GALEX} FUV band ($\Delta\lambda=$
1344-1786 \AA), the \emph{GALEX} NUV band ($\Delta\lambda=$ 1771-2831
\AA), and a CFHT u-band ($\Delta\lambda=$ 3400-4100 \AA) image
with a 5$\sigma$ depth of $\sim26.5$ AB magnitude. The extended
LAE has a major axis of about 23$''$ based on the FUV broad band
image or $\sim$80 kpc at $z=0.218$. It has a $\sim$22 AB FUV counterpart
but is very faint in the NUV and u-band. We highlight a nearby LAE,
GALEX033150-281120, with $z_{UV}\sim z_{opt}=0.213$ (red $r=10''$
circle), five sources with known optical redshifts at $z\sim0.215$
(magenta $r=5''$ circles), and a background source with an optical
redshift of $z=0.387$ (blue $r=5''$ circle). We note that the wavelength
range of FUV broadband image encompasses the wavelength of the proposed
extended Ly$\alpha$ emission. In the FUV image, we indicate the angular
size of 1$'$.}

\label{lab}

{\footnotesize (A color version of this figure is available in the online journal.)}
\end{figure*}
\begin{figure}[!t]
\includegraphics[bb=90bp 70bp 705bp 520bp,clip,width=8.5cm]{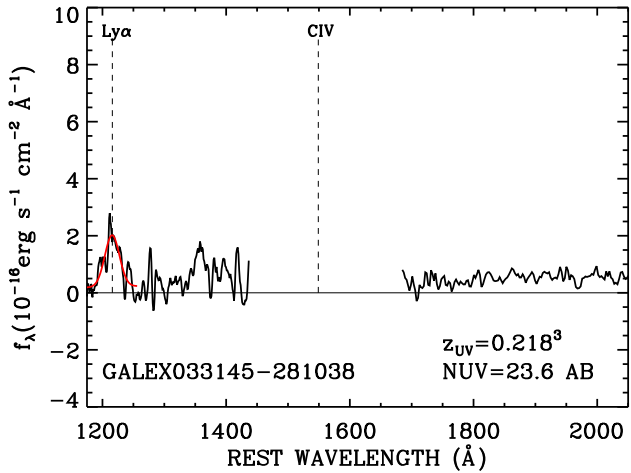}

\caption{The \emph{GALEX} spectrum of the extended LAE candidate GALEX033145-281038.
As in Figure \ref{example}, we show our Gaussian fit from which we
measure the Ly$\alpha$ redshift and the Ly$\alpha$ flux (red profile).
This source was not discovered in previous \emph{GALEX} studies because
of their requirement for all objects to have a bright NUV continuum
(NUV$<22$). }
\label{lab2}
{\footnotesize (A color version of this figure is available in the online journal.)}
\end{figure}
\begin{figure}[!t]
\includegraphics[bb=80bp 130bp 750bp 520bp,clip,width=8.5cm]{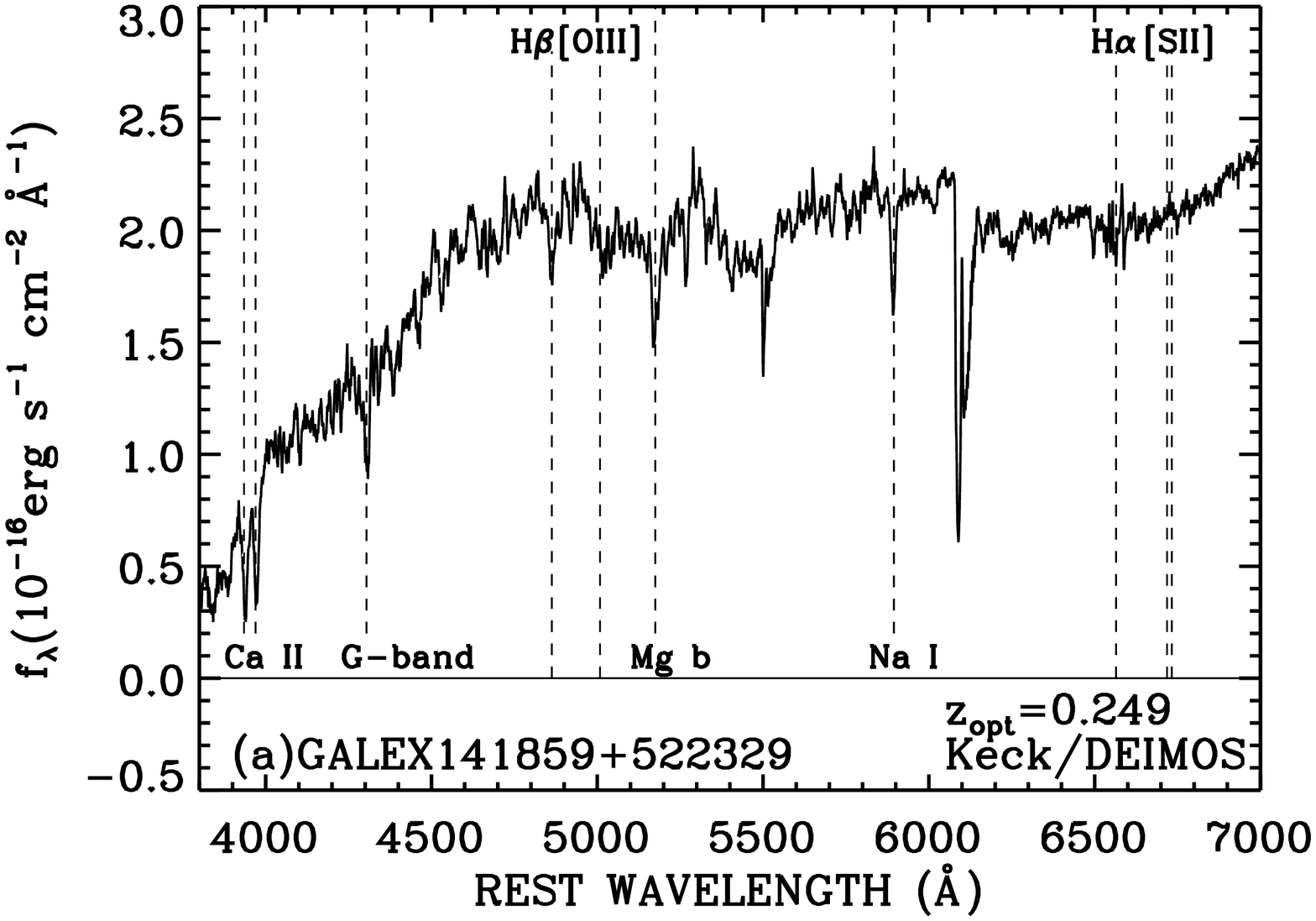}

\includegraphics[bb=80bp 130bp 750bp 520bp,clip,width=8.5cm]{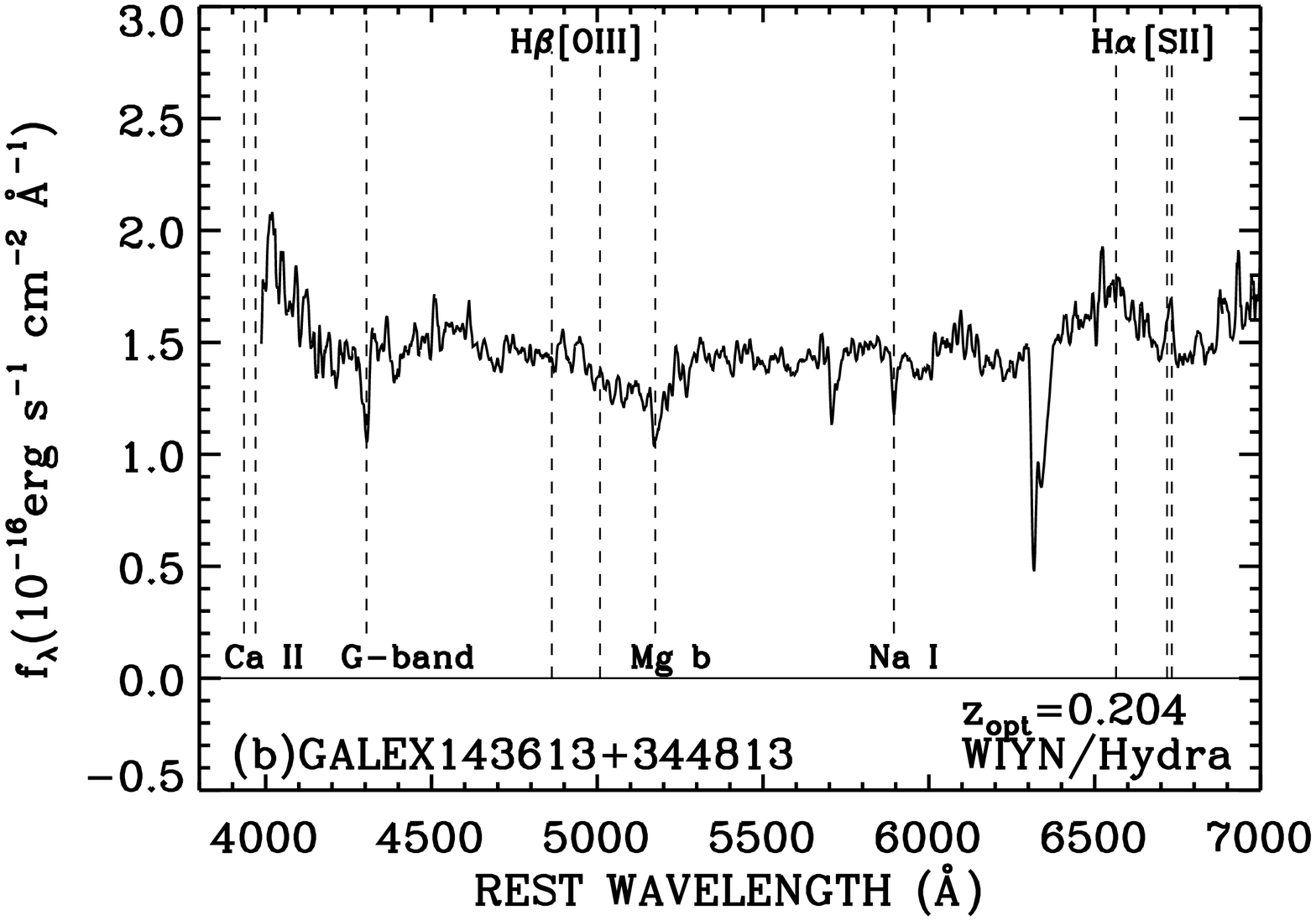}

\includegraphics[bb=80bp 70bp 750bp 520bp,clip,width=8.5cm]{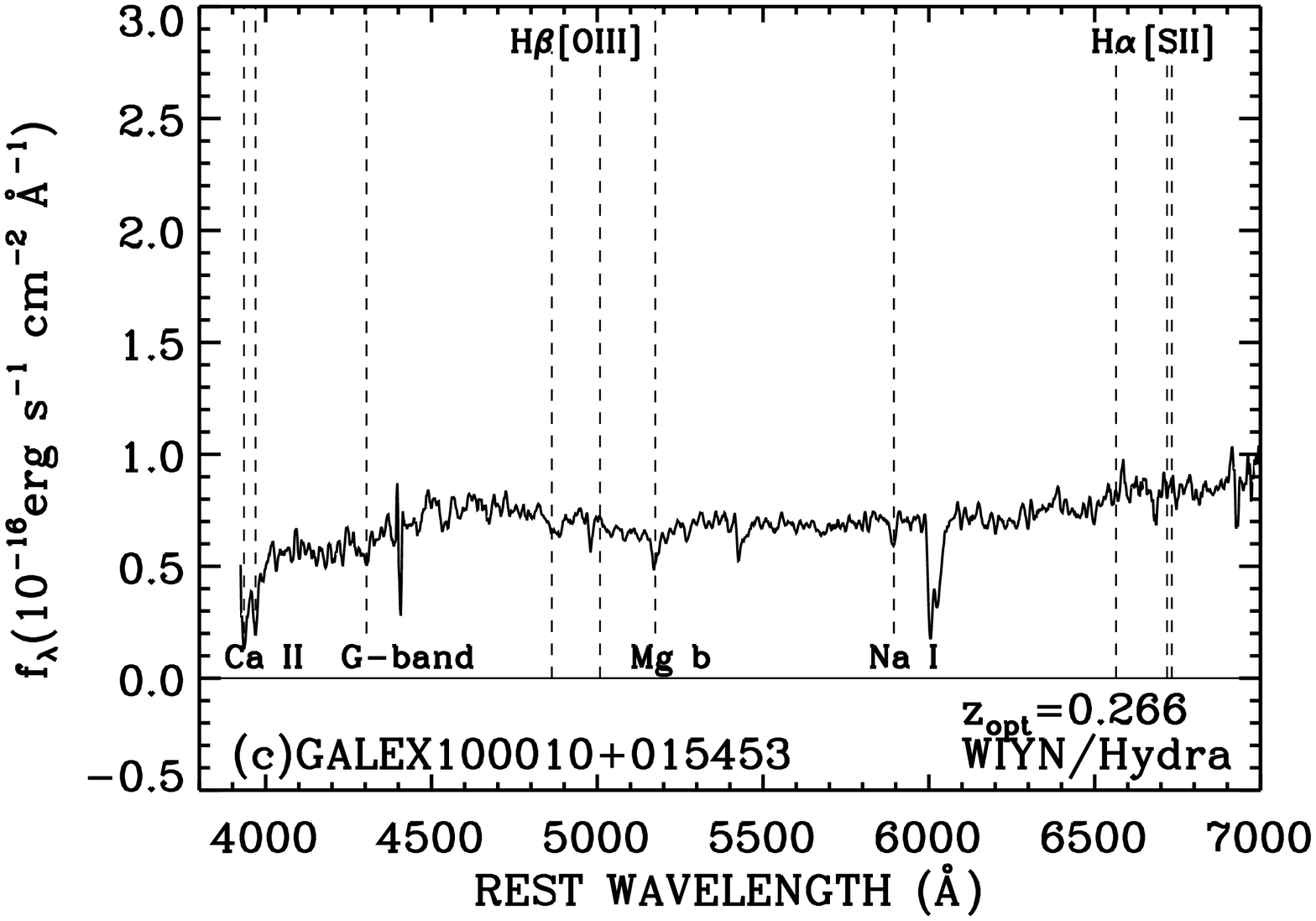}

\caption{Optical spectra for three LAEs with weak optical emission lines, referred
to as absorber LAEs. Ly$\alpha$ emission requires a source of relatively
hard ionizing radiation and, as discussed in Section \ref{odd}, we
suspect that these objects are obscured AGNs with favorable geometry
and/or kinematics that allows for the escape of Ly$\alpha$ photons. }
\label{abs1}
\end{figure}
\begin{figure*}[!t]
\includegraphics[width=18cm]{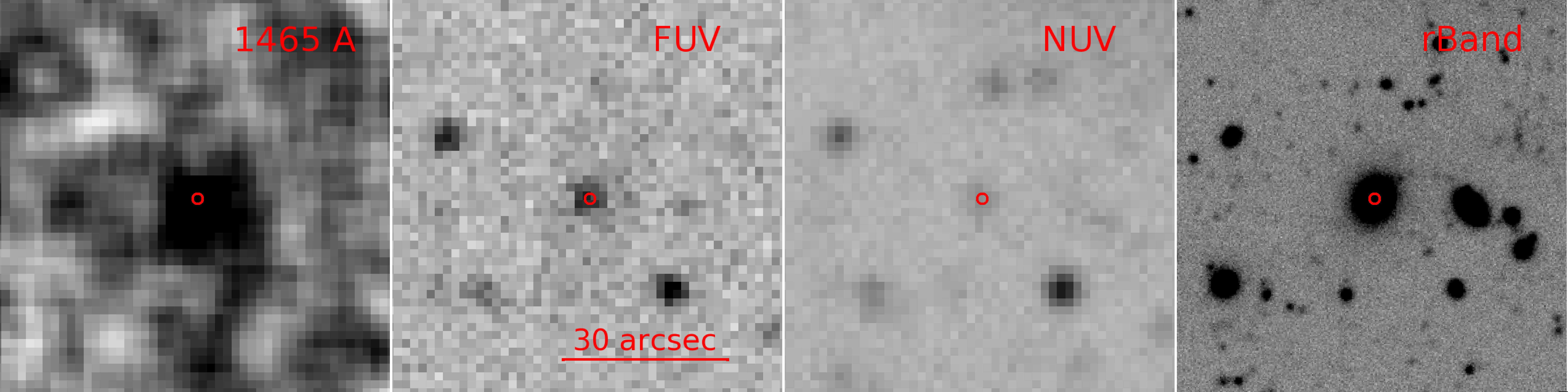}

\caption{Images of the absorber LAE GALEX143613+344813 with $z_{UV}\sim z_{opt}=0.204$.
From left to right, we show a 1465\AA\ data cube slice with a width
of 10\AA, the \emph{GALEX} FUV band ($\Delta\lambda=$ 1344-1786
\AA), the \emph{GALEX} NUV band ($\Delta\lambda=$ 1771-2831 \AA),
and a KPNO Mayall 4m telescope/MOSAIC r-band ($\Delta\lambda=$ 5700-7000
\AA) image with a 5$\sigma$ depth of $\sim26$ AB magnitude. We
obtained an optical spectrum with WIYN/HYDRA. We indicate the $r=1''$
WIYN/HYDRA fiber position with a red circle. The resulting spectrum
is shown in Figure \ref{abs1}(b). We argue that the source targeted
by WIYN is the only viable target and this suggests that - at least
in some cases - our absorber LAEs are not closely paired systems in
which we have inadvertently targeted the wrong optical counterpart.
In the FUV image, we indicate the angular size of 30$''$.}

\label{abs2}

{\footnotesize (A color version of this figure is available in the online journal.)}
\end{figure*}

Previous $z=0.3$ studies based on the \emph{GALEX} pipeline reductions
have estimated a wide range AGN contribution to the FUV LAE sample.
These AGN fraction estimates have ranged from approximately 15 to
45$\%$ \citep{finkelstein09b,cowie11}. Our AGN contribution estimate
for our EW $>20$ \AA\ LAE sample is 26 $\pm$ 5$\%$, or 22 $\pm$
5 $\%$ if absorber LAEs are not included in the AGN count. We apply
the EW cut to be consistent with high-redshift LAEs samples which
typically use this constraint to remove low-redshift interlopers such
as {[}O{\small{}II}{]} emitters. If we limit our sample to EW $>20$
\AA\ LAEs previously discovered in the \emph{GALEX} pipeline reductions,
then we find an AGN fraction of $\sim34\%$. We note that the AGN
fraction is not an invariant property of LAE samples. As previously
pointed out by \citet{nilsson11}, \citet{wold14}, and discussed
in Section \ref{sec_lf}, the AGN fraction is strongly dependent on
the sample's Ly$\alpha$ luminosity range, such that - holding everything
else constant - samples probing more luminous LAEs will have higher
AGN fractions. In Figure \ref{agnhisto}, we show how the AGN fraction
increases with Ly$\alpha$ luminosity in both the $z=0.3$ and $z=0.9$
EW $>20$ \AA\ LAE samples. For this Figure, we have counted the
$z=0.3$ absorber LAEs as AGNs. In all observed $0.5$ dex Ly$\alpha$
luminosity bins, we find AGN fractions that are $\sim20\%$ or greater.
At a given Ly$\alpha$ luminosity the $z=0.3$ sample has a higher
AGN fraction. This can be attributed to the strong luminosity boost
from $z=0.3$ to $0.9$ observed in the typical LAE galaxy (discussed
further in Section \ref{evo}). 

Previous studies have shown that in order to achieve a complete consensus
of AGNs, multi-wavelength datasets are required \citep[e.g.,][]{hickox09}.
Thus, our primary reason for using X-ray imaging, UV spectra, optical
spectra, and infrared imaging to identify AGNs is to increase the
completeness of our AGN sample. The other reason we use multiple identification
methods is because we lack uniform coverage for any one method. We
lack deep X-ray imaging for $49\%$ of our survey. Depending on the
LAE's redshift, the \emph{GALEX} band gap between FUV and NUV bands
may prevent us from observing high-excitation lines in the UV spectrum.
While we have optical redshifts for all but two of our LAEs, we only
have optical spectra for $86\%$ of our LAE sample (archival redshifts
are more readily available than archival spectra). We have infrared
imaging for all fields via the all sky \emph{WISE }survey, but the
depth of this survey depends strongly on ecliptic latitude. By using
our multi-wavelength data, we ensure that every LAE is classified
by at least two methods. While utilizing all methods clearly provides
advantages, we may also be reducing the reliability of our AGN sample.
For example, \citet{assef13} estimate that their prescribed \emph{WISE
}color selection reliably identifies AGNs 90\% of the time. 

We assess the importance of these completeness and purity concerns
by limiting our survey to regions with deep X-ray imaging. X-ray selection
provides a robust AGN identification which is often used as the base-line
truth in studies that compare AGN classification methods \citep[e.g.,][]{trouille10}.
Furthermore, by limiting our survey to regions with deep X-ray imaging,
we ensure that every LAE is classified by at least three methods.
We compare results derived from our full sample to results computed
from our X-ray covered sample to assess any significant incompleteness
in our AGN sample. Additionally, within the deep X-ray fields, we
find that all UV and \emph{WISE} AGNs are independently classified
as X-ray AGNs. Falsely identified AGN in one method are unlikely to
be falsely identified in another method. Thus, concerns about the
purity of the \emph{WISE} and UV selected AGNs should be eased. We
note that within the deep X-ray fields 8 optically identified AGN
are not identified as X-ray AGN. Three of these eight are $10^{41}$erg
s$^{-1}$ X-ray sources, perhaps indicating that our straight $10^{42}$
erg s$^{-1}$ luminosity cut is missing some X-ray faint AGN. The
remaining 5 optical only AGNs are composed of two `absorbers' (see
\ref{odd}) and three BPT AGNs. These could indicate falsely identified
optical AGNs or represent a population of heavily obscured AGNs. 

While the comparison of our full sample to our X-ray deep sample does
not completely alleviate all completeness and purity concerns, it
does significantly improve our AGN classification and allows us to
assess any effect on our main results. Furthermore, in Section \ref{sec_lf},
we consider a method to measure the LAE luminosity function without
AGN identification. Here we simultaneously fit the combined SF+AGN
LF with a Schechter + power-law function. This bypasses AGN identification
concerns at the expense of having to assume a functional form to the
AGN luminosity function. In Section \ref{sec_lf}, we show that restricting
the survey's area to deep X-ray fields or simultaneously fitting the
combined SF+AGN LF does not significantly change our LAE luminosity
function results.

\subsection{Extended and Absorber LAE Candidates }

\noindent \label{odd}

We found one highly extended Ly$\alpha$ source, LAE candidate GALEX033145-281038
with $z_{UV}=0.218$. In Figure \ref{lab}, we show this candidate
in a 1480A data cube slice with a width of 10A, the \emph{GALEX} FUV
band ($\Delta\lambda=$ 1344-1786 \AA), the \emph{GALEX} NUV band
($\Delta\lambda=$ 1771-2831 \AA), and a CFHT u-band ($\Delta\lambda=$
3400-4100 \AA) image with a 5$\sigma$ depth of $\sim$26.5 AB magnitude.
In Figure \ref{lab2}, we show our exacted 1-D \emph{GALEX} spectrum
for this object. With a measured FUV major axis of 0.37$'$ or 80
kpc at $z=0.218$ and a Ly$\alpha$ luminosity of $7.8\times10^{41}$
erg s$^{-1}$, this extended source falls below the typical high-redshift
Ly$\alpha$ blob physical extent ($\sim100$ kpc) and Ly$\alpha$
luminosity ($\sim10^{43}$ erg s$^{-1}$). It has a 22 AB FUV counterpart
but is very faint in the NUV and u-band. In Figure \ref{lab}, we
highlight a nearby LAE, GALEX033150-281120, with $z_{UV}\sim z_{opt}=0.213$
(red $r=10''$ circle), five sources with known optical redshifts
at $z\sim0.215$ (magenta $r=5''$ circles), and a background source
with an optical redshift of $z=0.387$ (blue $r=5''$ circle). The
closest object with known matching redshift ($z=0.216$) is about
60 kpc away from the centroid of the extended source (magenta circle
to the south-east or lower-left relative to the extended LAE). The
ECDFS X-ray field lies to the north of this source, and we lack X-ray
data for this source or any of the potential counterparts. 

While more data is needed to study this extended object, we note some
similarities with more extensively studied spatially extended LAEs.
In particular, the lack of a clear optical counterpart and the apparent
over-density of nearby $z\sim0.215$ sources is consistent with the
properties of the \citet{nilsson06} Ly$\alpha$ nebula at $z=3.157$.
The Nilsson et al.\ Ly$\alpha$ nebula has recently been re-examined
by \citet{prescott15} with data from the \emph{Hubble Space Telescope}
and the \emph{Herschel Space Observatory}. This object exists within
a local over-density of galaxies and has no continuum source located
within the nebula. Prescott et al.\ conclude that the Ly$\alpha$
nebula is likely powered by an obscured AGN located $\sim$30 kpc
away. We also note that the only probable candidate found for the
\citet{barger12} low-redshift ($z=0.977$) Ly$\alpha$ nebula was
an AGN located 170 kpc away \citep{barger12}. Further advancing an
AGN power source, \citet{schirmer16} have suggested that SDSS galaxies
selected for their strong {[}O{\small{}III}{]} emission lines and
for their large spatial extent (Green Beans) are likely ionized by
AGNs. Green Beans are estimated to be extremely rare ($\sim3.3$ Gpc$^{-3}$)
and to have very high Ly$\alpha$ luminosities ($\sim10^{43}$ erg
s$^{-1}$), so it is not clear that these objects are directly related
to our relatively faint extended object found in a survey volume of
$\sim0.90\times10^{6}$ Mpc$^{3}$.

In Figure \ref{abs1}, we show 3 of the 6 LAEs with very weak optical
emission lines. We refer to these LAEs as absorbers, and they are
denoted with an `a' in Column 15 in Tables \ref{cdfs}-\ref{cosmos}.
Given the poor resolution of \emph{GALEX}, it is possible that absorbers
are closely paired systems in which we have inadvertently targeted
the wrong optical counterpart. In this scenario, the real LAE counterpart
could still have an emission line optical spectrum. However, in Figure
\ref{abs2}, we present our strongest case against this interpretation
being true for all cases. For this LAE, the Ly$\alpha$ emission seen
in the data cube slice has only one viable FUV counterpart which we
targeted with WIYN/HYDRA (red circle indicates HYDRA's $r=1''$ fiber
location). The resulting optical spectra is shown in Figure \ref{abs1}(b).
As might be expected from an absorber spectrum, the $r$-band morphology
appears to be spheroidal. This LAE is within an X-ray imaging survey
\citep{kenter05} but is not detected. Based on the hard X-ray band
detection limit, this absorber has an X-ray luminosity upper limit
of $\sim2\times10^{42}$ erg s$^{-1}$. Four of the other absorbers,
GALEX033145-274615, 033213-280405, 033251-280305, and 100010+015453,
are also within X-ray surveys and are not detected in the hard X-ray
band. This places an upper limit on their X-ray luminosities of $5\times10^{39}$,
$2\times10^{41}$, $1\times10^{41}$, and $2\times10^{41}$ erg s$^{-1}$,
respectively.  We find that absorber GALEX033251-280305 is a soft
X-ray source with a luminosity of $1\times10^{41}$ erg s$^{-1}$
\citep{lehmer05}. 

Another plausible explanation for an absorber LAE is that a faint
star-forming galaxy responsible for the Ly$\alpha$ emission is out-shined
in the optical by a superimposed absorber galaxy at the same redshift.
In this scenario, even if we followup the correct optical counterpart,
we would not recover the LAE's uncontaminated optical spectrum. 

We note that 4 out of 6 absorbers have blue FUV to NUV colors (FUV-NUV$<0$
) and very large rest-frame EWs ($>200$ \AA). While these objects
lack a clear SF or AGN signature, we suggest that these objects are
likely obscured AGNs with favorable geometry and/or kinematics that
allows for the escape of Ly$\alpha$ photons.  

Further evidence that heavily obscured AGNs can be strong Ly$\alpha$
emitters is demonstrated by our newly discovered data cube LAE GALEX095910+020732.
This object was the focus of a multi-wavelength study that concluded
that its nuclear emission must be suppressed by a N$_{{\rm {H}}}$
$\gtrsim10^{25}$ cm$^{-2}$ column density \citep{lanzuisi15}. Unlike
our absorber LAEs, this obscured object has strong optical emission
lines including H$\alpha$. With our data cube search, we now know
that this object is also a very luminous LAE with $L_{{\rm {Ly}\alpha}}=10^{42.7}$
erg s$^{-1}$. While this object did not meet our X-ray or infrared
AGN criteria, we classified this object as a UV AGN based on a strong
C {\small{}IV} emission line. For all six of our absorber LAEs, we
find that C {\small{}IV} is not observable with the \emph{GALEX} grism
data because the emission line feature falls between the FUV and NUV
bandpasses. 

Throughout our subsequent analysis we classify these 6 absorber LAEs
as AGNs. Furthermore, we exclude the extended LAE GALEX033145-281038
from both SF and AGN categories. 
\begin{figure}
\includegraphics[bb=70bp 150bp 695bp 520bp,clip,width=8.5cm]{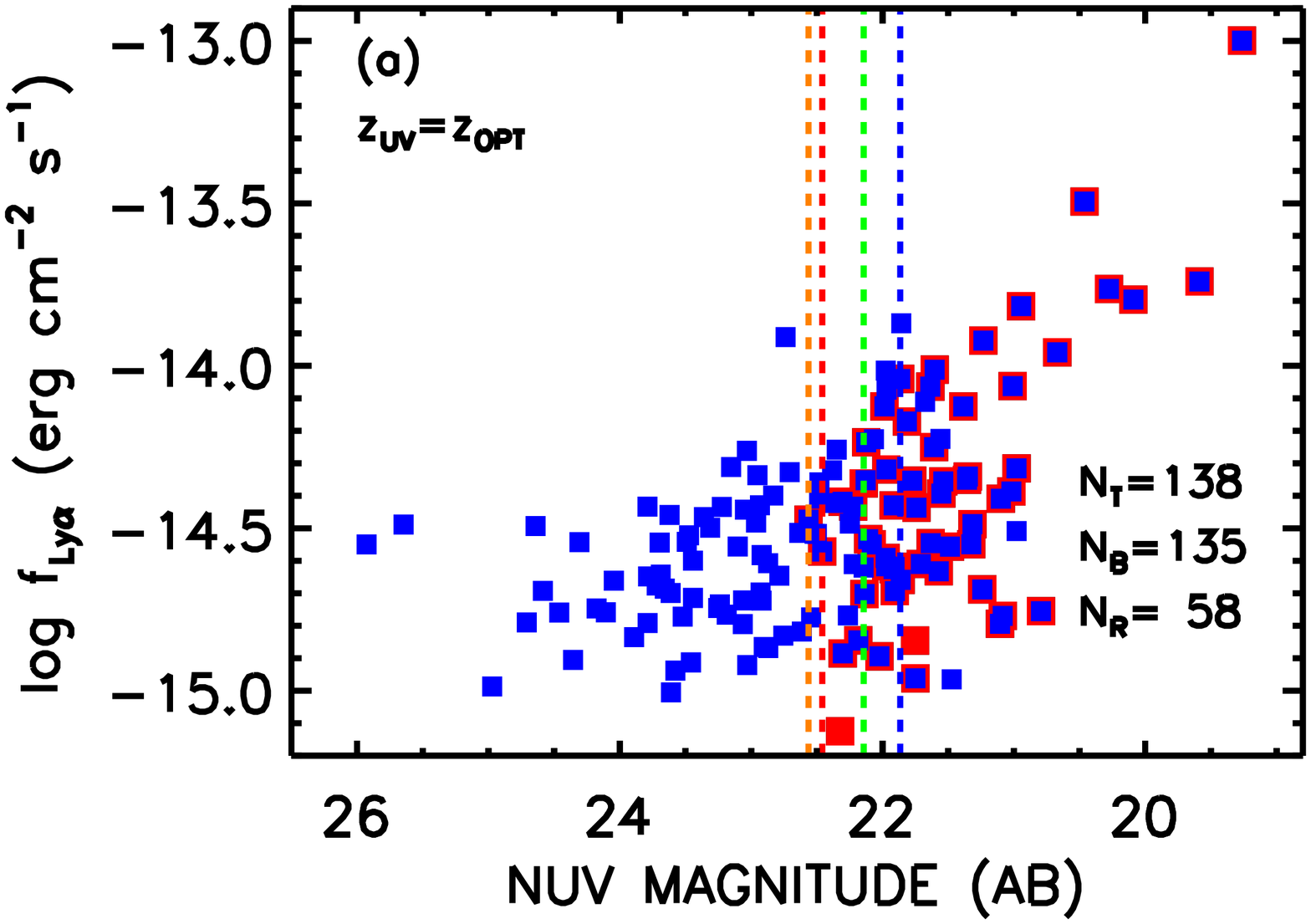}

\includegraphics[bb=70bp 150bp 695bp 520bp,clip,width=8.5cm]{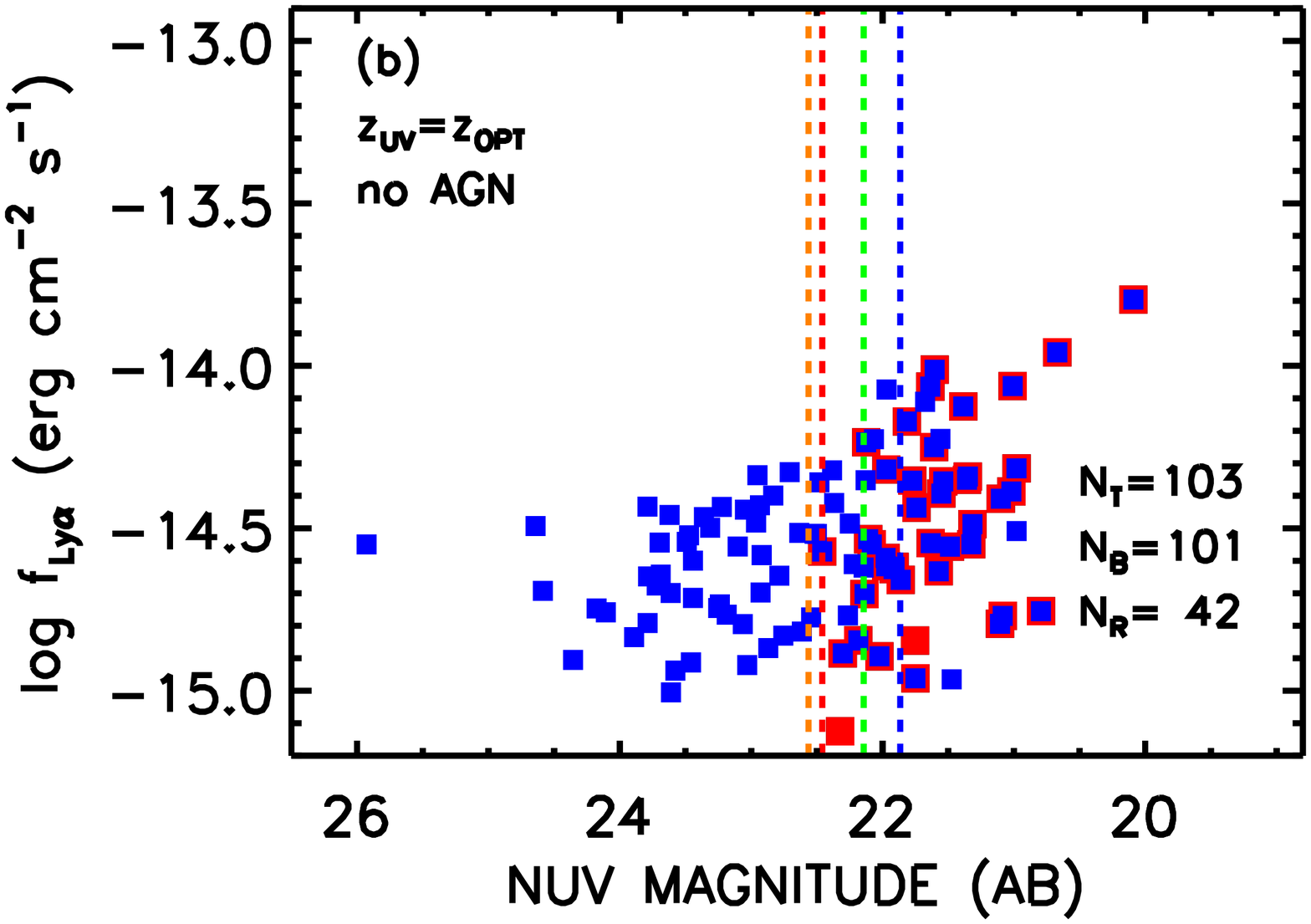}

\includegraphics[bb=70bp 75bp 695bp 520bp,clip,width=8.5cm]{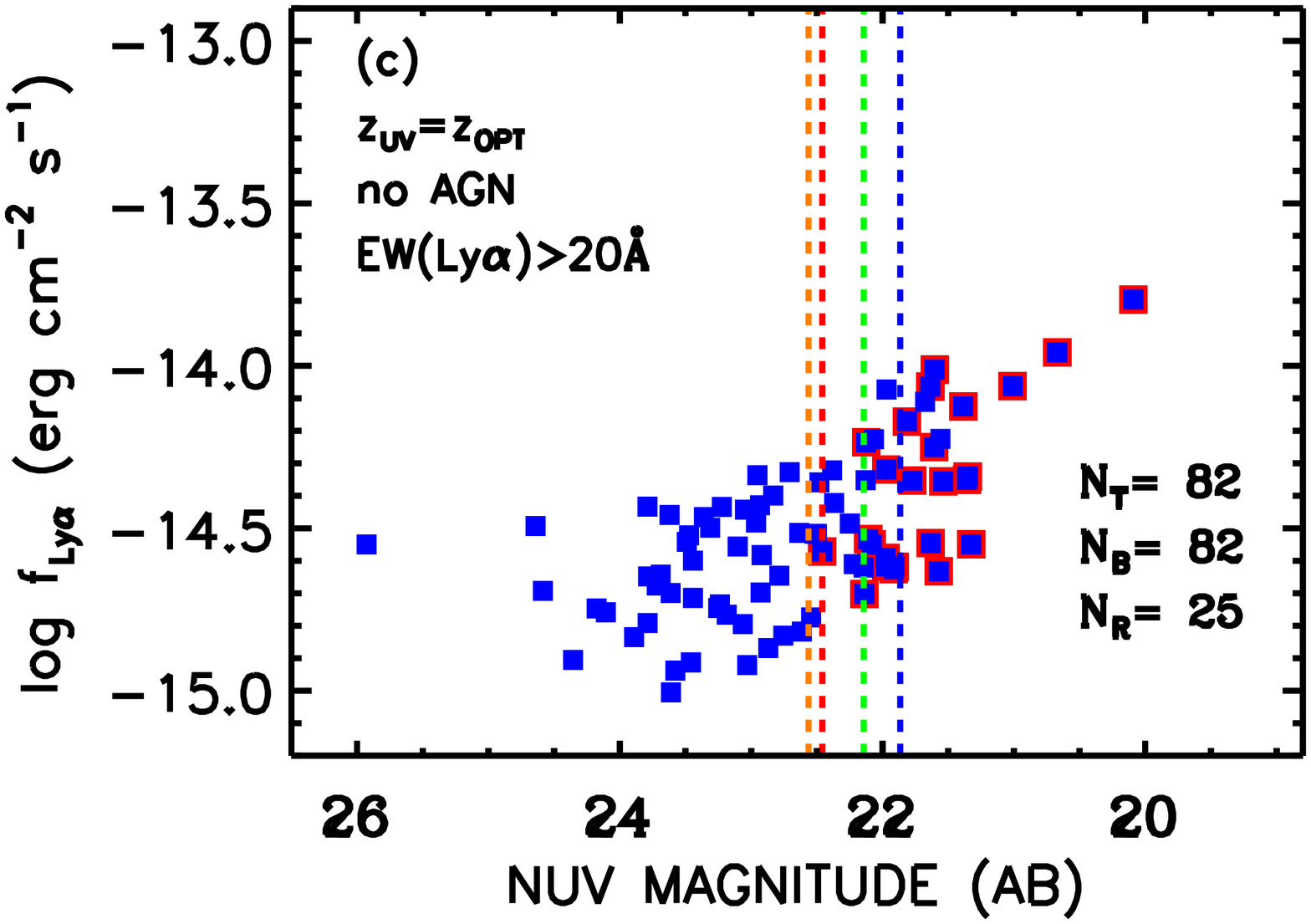}

\caption{\textbf{(a)} LAE NUV magnitude vs.\ Ly$\alpha$ flux for LAEs with
optical redshifts that confirm their Ly$\alpha$ redshifts based on
Ly$\alpha$ emission. Red squares show the pipeline LAEs found by
\citet{cowie10,cowie11} constrained to the data cubes' 50$'$ diameter
FOVs. Blue squares show our data cube LAEs constrained to the pipeline's
redshift range ($z=0.195-0.44$). The dashed red, orange, green, and
blue line shows the maximum NUV magnitude found in the pipeline sample
for the CDFS, GROTH, NGPDWS, and COSMOS field, respectively. This
roughly corresponds to the \textit{GALEX} pipeline's magnitude limit
of NUV$\sim22$.\textbf{\textit{ }}\textbf{(b) }The same as Figure
\ref{pipeVcube}(a), but with all AGNs removed. We note that many
of the most luminous sources are removed by this cut. \textbf{(c)
}The same as Figure \ref{pipeVcube}(b), but with all EW$_{{\rm r}}$(Ly$\alpha$)$<20$\AA~
objects removed. We note that the final sample contains $25$ pipeline
LAEs. The data cube sample recovers all 25 of these objects plus 57
previously unidentified LAEs. In all panels, we indicate the number
of data cube LAEs or blue squares (N$_{{\rm {B}}}$), the number of
pipeline LAEs or red squares (N$_{{\rm {R}}}$), and the total number
of LAEs (N$_{{\rm {T}}}$).}

\label{pipeVcube}

{\footnotesize (A color version of this figure is available in the online journal.)}
\end{figure}
\begin{figure*}[!t]
\includegraphics[bb=70bp 55bp 720bp 520bp,clip,width=8.5cm]{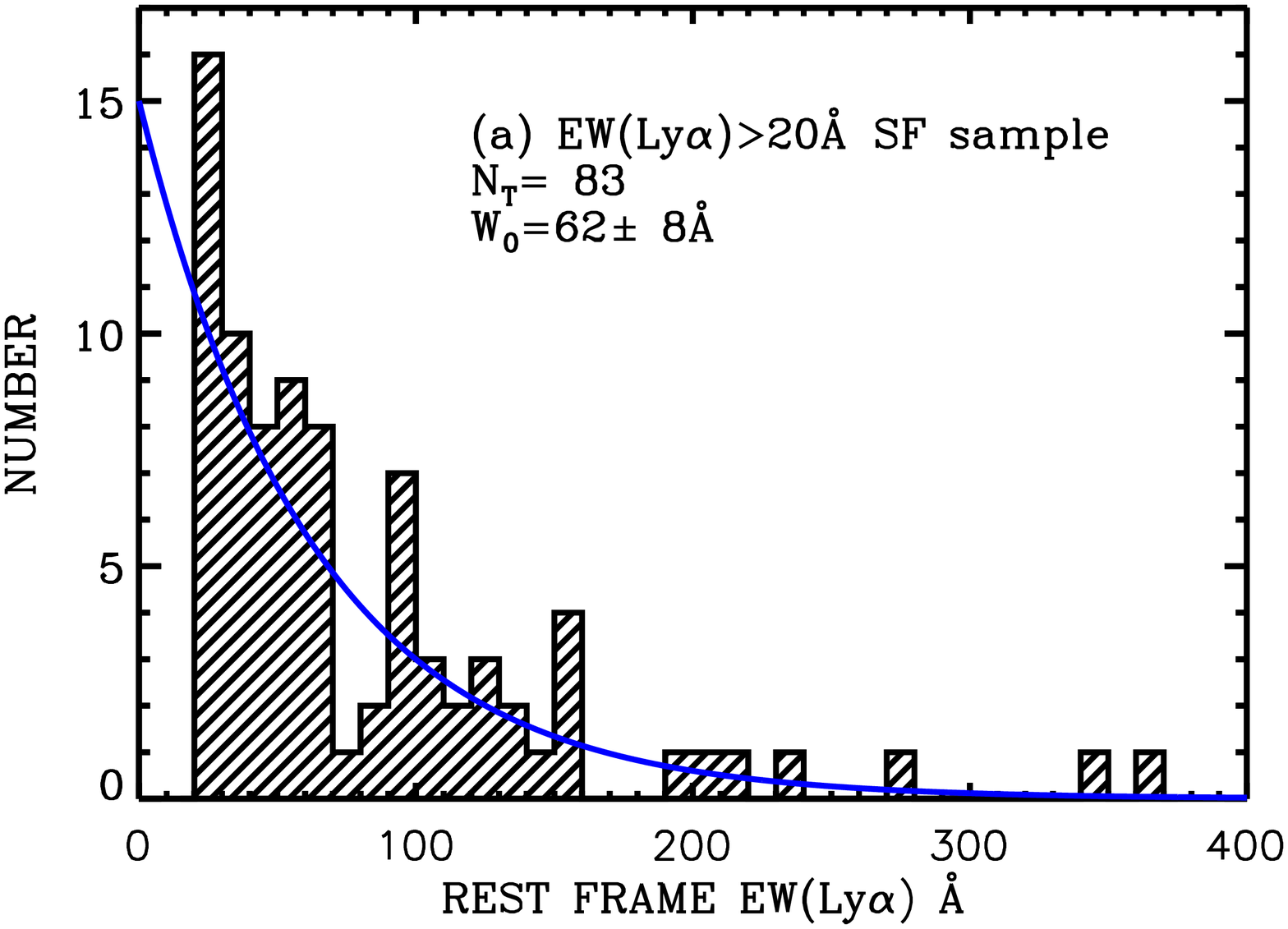}\includegraphics[bb=70bp 55bp 720bp 520bp,clip,width=8.5cm]{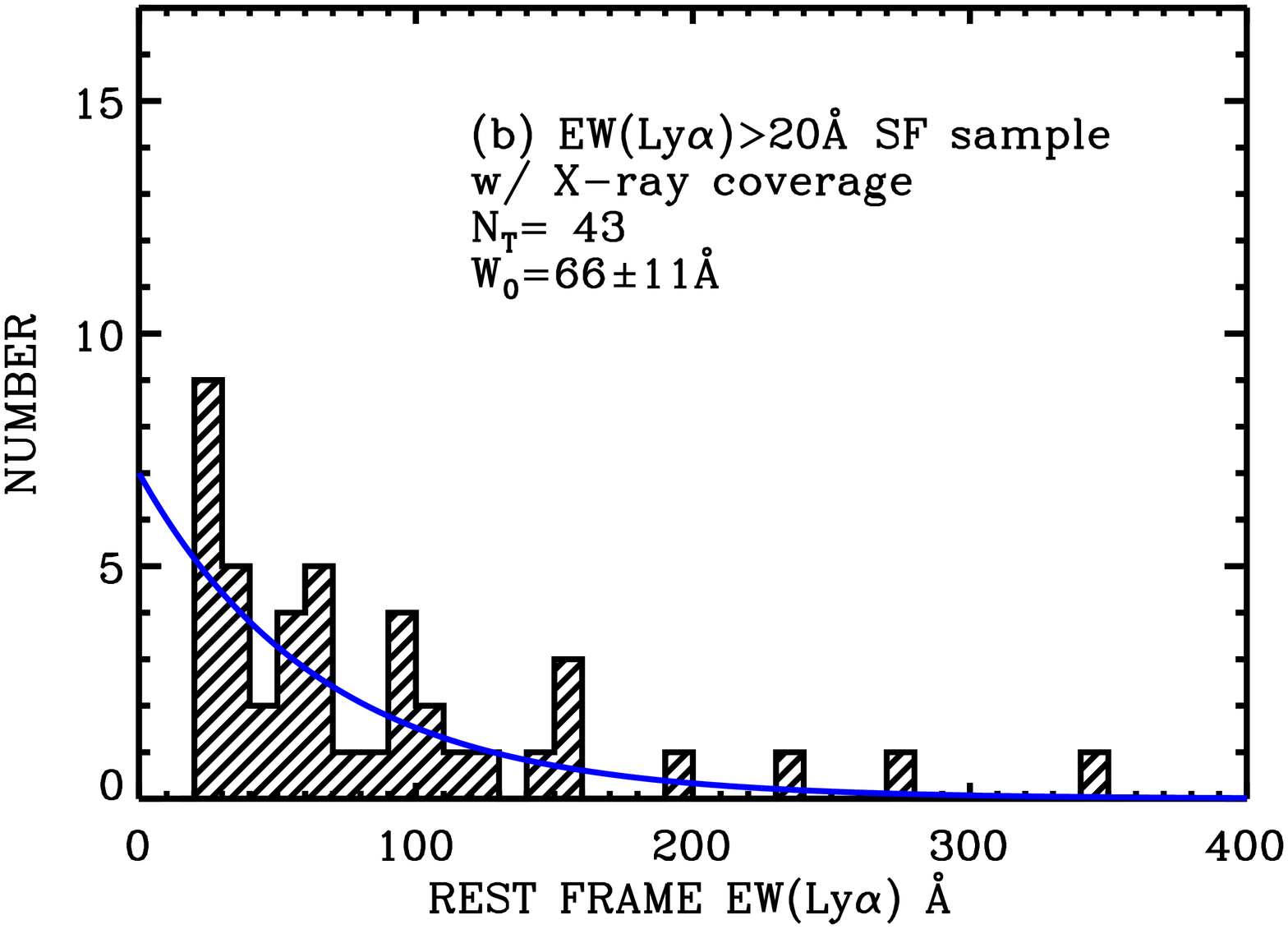}

\caption{\textbf{(a)} Ly$\alpha$ galaxy EW distribution.\textbf{\textit{ }}\textbf{(b)
}Same as Figure \ref{EW}(a), but with the LAE survey limited to regions
with deep X-ray data to ensure a robust AGN classification. In each
Figure, we indicate the best fit exponential (blue curve), the total
number (N$_{{\rm {T}}}$) of star-forming LAEs with EW $>20$ \AA~
used to compute the distribution and the EW distribution scale length
(W$_{{\rm {0}}}$). }

\label{EW}

{\footnotesize (A color version of this figure is available in the online journal.)}
\end{figure*}
\begin{figure}[!h]
\includegraphics[bb=70bp 55bp 720bp 520bp,clip,width=8.5cm]{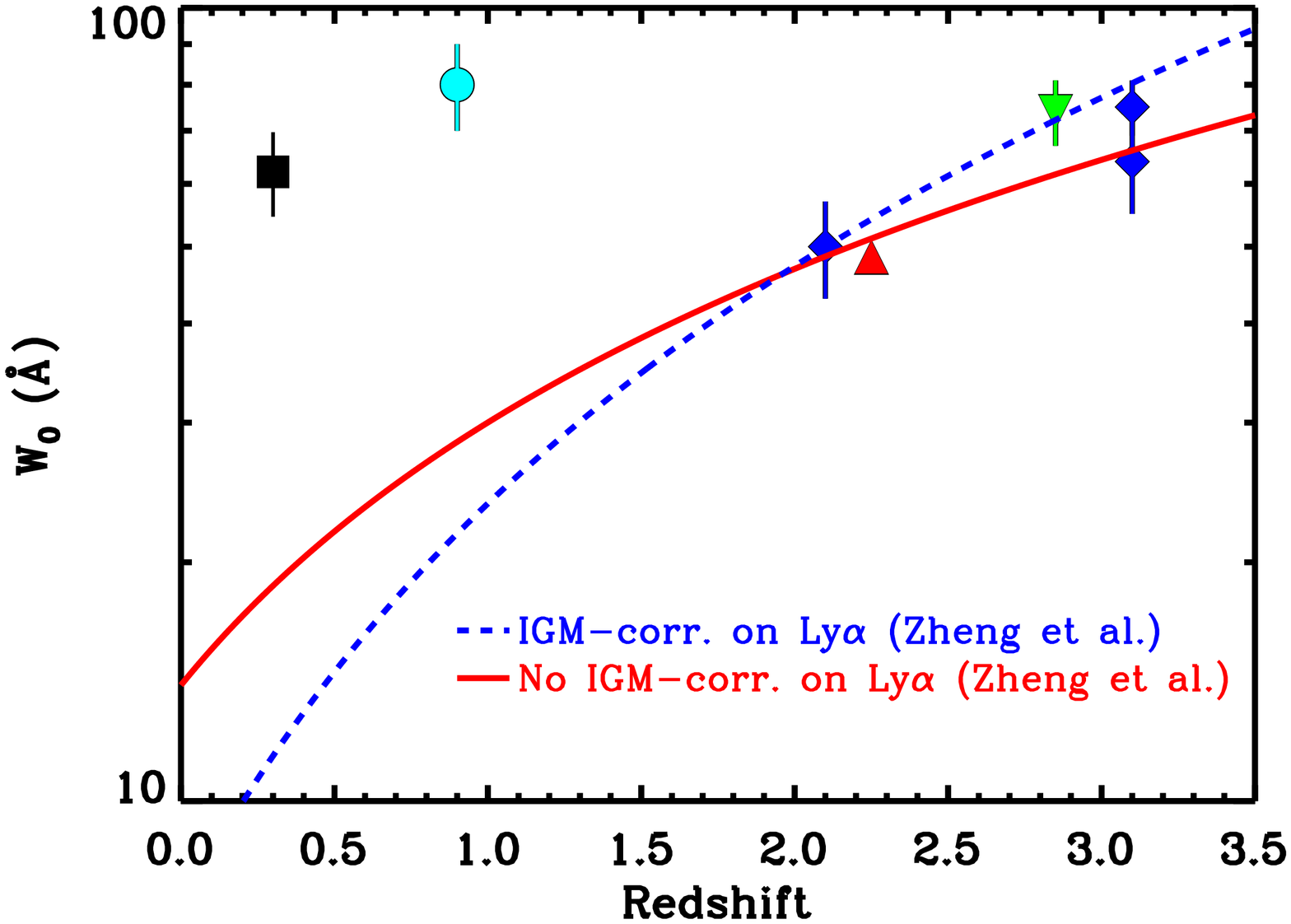}

\caption{Rest-frame EW scale lengths from our $z\sim0.3$ sample (black square),
the $z\sim0.9$ \citet{wold14} sample (cyan circle), the $z=2.25$
\citet{nilsson09} sample (red upward triangle), the $z=2.1$ \citet{guaita10},
$z=3.1$ \citet{gronwall07}, and $z=3.1$ \citet{ciardullo12} samples
(blue diamonds; values quoted from Ciardullo et al.), and $z\sim2.85$
\citet{blanc11} sample (green downward triangle) vs.\ redshift.
The dashed blue and solid red curves indicate the empirical EW scale
length evolution with and without IGM absorption to the Ly$\alpha$
line flux, respectively, proposed by \citet{zheng14}. In contrast
to these previously suggested evolutionary trends, our new $z=0.3$
result plus our recent $z=0.9$ result favors a relatively constant
EW scale length from $z=0.3$ - $3$.}

\label{EW_evo}

{\footnotesize (A color version of this figure is available in the online journal.)}
\end{figure}

\subsection{Comparison of the \emph{GALEX} Data Cube Sample with the \emph{GALEX}
Pipeline Sample}

In Figure \ref{pipeVcube}, we compare our data cube sample (blue
squares) to the pipeline sample (red squares) as presented in \citet{cowie10}
and \citet{cowie11} for \emph{GALEX} fields CDFS, GROTH, NGPDWS,
and COSMOS constrained to the data cubes' 50$'$ diameter FOVs. We
require the sources to have optical redshifts in agreement with their
Ly$\alpha$ based redshifts, and we limit our sample in this comparison
to LAEs with $z>0.195$ to be consistent with the pipeline sample.
The dashed vertical lines indicate the pipeline\textquoteright s NUV
continuum thresholds. As expected, the pipeline begins to miss objects
fainter than the pipeline's extraction threshold of $\sim$22 AB magnitude.
In Figure \ref{pipeVcube}(a), we show that within the same FOV our
sample contains 135 LAEs, while the pipeline sample contains 58 LAEs.
We note that there are three low-EW LAEs detected in the pipeline
but not found with our data cube search. These objects have relatively
low Ly$\alpha$ flux measurements (ranging from 8$\times10^{-16}$
to 4 $\times$ 10$^{-15}$ erg cm$^{-2}$ s$^{-1}$ ). We find that
our recovered fraction of fake sources falls below $\sim$90\% at
4$\times$10$^{-15}$ erg cm$^{-2}$ s$^{-1}$ and then quickly declines
with a $\sim$30$\%$ recovery at 1 $\times$ 10$^{-15}$ erg cm$^{-2}$
s$^{-1}$. Thus, we suspect that the missed pipeline LAE are accounted
for by the data cube\textquoteright s flux limit. Regardless, the
missed sources have low EWs ($<20$\AA) and are thus excluded from
our final sample used to compute the Ly$\alpha$ EW distribution and
LF. In Figure \ref{pipeVcube}(b), we remove all AGNs (see Section
\ref{agn} for AGN classification) and show that our sample contains
101 SF LAEs, while the pipeline sample contains 42 SF LAEs. In Figure
\ref{pipeVcube}(c), we show that the data cube sample finds all pipeline
EW(Ly$\alpha$)$>$20 \AA\  star-forming LAEs plus an additional
57 LAEs that fall below the pipeline's continuum detection threshold.

\subsection{EW and LF Sample Definition}

For the 146 non-spurious sources, we have 144 (125) optical redshifts
(spectra) that agree with our Ly$\alpha$ redshifts. We note that
19 LAEs with optical redshifts were obtained from archival sources
(see Tables \ref{cdfs}-\ref{cosmos}) that lacked published spectra.
For our final SF LAE sample, we start from these 146 LAEs and require
sources to not be identified as an AGN in any way, have EW(Ly$\alpha$)
$\geq$ 20\AA, have $z>0.195$, and be detected above the $50\%$
flux completeness threshold as determined from our Monte Carlo simulations.
 We require our LAEs to have $z>0.195$ to be consistent with previous
studies \citep{cowie10,cowie11}. This removes 6 LAEs with $0.15<z<0.195$
from our final sample. Our final SF sample which is used to derive
the Ly$\alpha$ LF and EW distribution has a size of 83 objects. Of
these 83 LAEs, we have optical redshifts (spectra) for 81 (71) objects.
The two LAEs in our final sample without optical followup (GALEX033108-274214
and 033346-274736) are assigned a high LAE confidence classification
of 2 and 1, respectively. We include these optically un-targeted LAEs
because targeted high confidence (1 and 2) candidates are optically
confirmed $98\%$ percent of the time.

We emphasize that previous $z\sim0.3$ samples used to compute the
Ly$\alpha$ LF were biased against high-EW objects and had a smaller
sample size of SF LAEs \citep[][]{deharveng08,cowie10}. For example,
\citet{cowie10} derived the $z\sim0.3$ LF from 41 star-forming LAEs
with EW(Ly$\alpha$)$>$20 \AA\ in nine \emph{GALEX} fields. With
only four fields we have obtained a sample of 83 SF LAEs. Most importantly,
our sample is not pre-selected from continuum bright objects, which
facilitates the comparison of our LAE sample to high-redshift samples.

\section{Equivalent Width Distribution}

\label{sec_ew}

The Ly$\alpha$ EW in the rest frame is the ratio of Ly$\alpha$ flux
relative to the continuum flux density divided by ($1+z$). Galaxies
with extremely high Ly$\alpha$ EWs are proposed sites of low metallicity
starbursts \citep{schaerer03,tumlinson03}, and these extreme emitters
may play an increasingly dominant role at higher redshifts. In line
with these expectations, there are studies that find a shift toward
lower EW objects at lower redshifts \citep[e.g.,][]{ciardullo12,zheng14}.
However, these studies lack an unbiased low-redshift constraint. Previously,
the $z\sim0.3$ EW distribution was derived from the \emph{GALEX}
pipeline reductions and thus was biased against high-EW objects \citep[as described in Section 5.4 of][]{cowie10}.
With our data cube sample we remove this bias and make a valid comparison
to high-redshift EW distributions. 

In Figure \ref{EW}(a), we show our $z\sim0.3$ rest-frame EW distribution
for all LAEs in our SF sample. To compare our EW distribution to previous
studies, we fit it with an exponential and find a scale length of
62 $\pm$ $8$ \AA. We compute a maximum likelihood estimate of the
scale length and compute the 1$\sigma$ error using the parameterized
bootstrap method. In Figure \ref{EW}(b), we show our $z\sim0.3$
rest-frame EW distribution for all LAEs in our SF sample that have
available deep X-ray data. The deep X-ray data have a depth of $\sim10^{41}$erg
s$^{-1}$ at $z\sim0.3$ and provide a uniform AGN diagnostic. We
find that the computed scale length is not significantly altered by
the requirement of a more strict AGN diagnostic.

In Figure \ref{EW_evo}, we show the redshift evolution of the EW
scale length. The dashed blue and solid red curves indicate the empirical
EW scale length evolution with and without IGM absorption to the Ly$\alpha$
line flux, respectively, proposed by \citet{zheng14}.  In contrast
to these previously suggested evolutionary trends, our new $z=0.3$
result plus our recent $z=0.9$ result \citep{wold14} favors a relatively
constant EW scale length from $z=0.3$ - $3$, or roughly 8 Gyrs.
 Our measured large scale length is in sharp contrast to the biased
$z\sim0.3$ \emph{GALEX} pipeline LAE sample, which has an EW scale
length of 23.7 \AA\ \citep{cowie10}. We note that \citet{cowie10}
corrected their EW distribution for the pipeline sample's incompleteness
and found their corrected distribution to be well described by a EW
scale length of 75 \AA. This estimate is within 2$\sigma$ of our
result.

In the pipeline sample, no LAE galaxies are found with an EW(Ly$\alpha$)$>$120
\AA . In our final SF sample, we find 16 of these extreme EW LAEs.
These extreme EW LAEs are of interest given the recent studies suggesting
that high-EW LAEs are efficient emitters of ionizing photons and potential
analogs of reionization-era galaxies \citep[e.g.,][]{erb16,trainor16}.
\begin{figure*}[!t]
\begin{centering}
\includegraphics[scale=0.45]{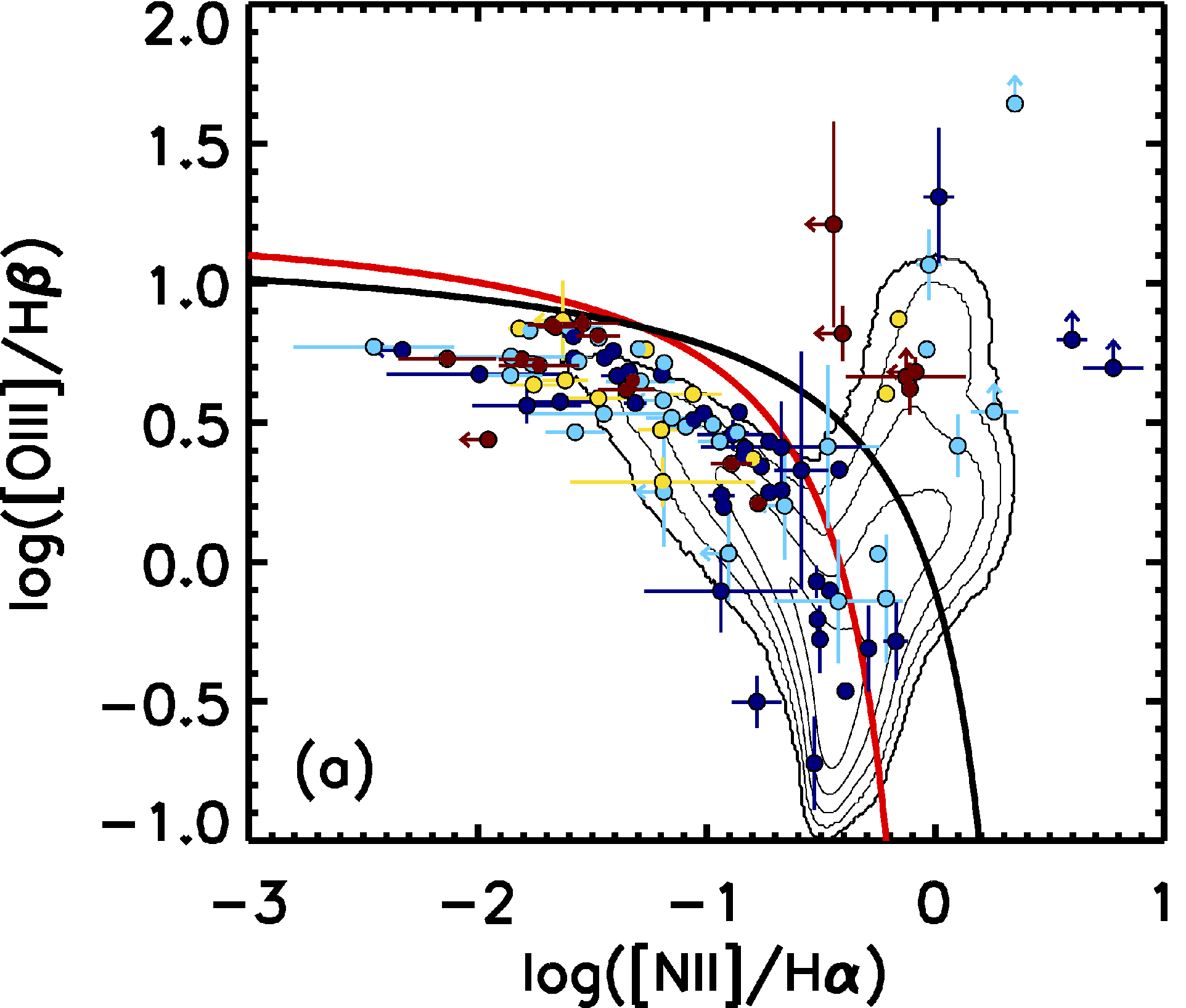}\includegraphics[scale=0.45]{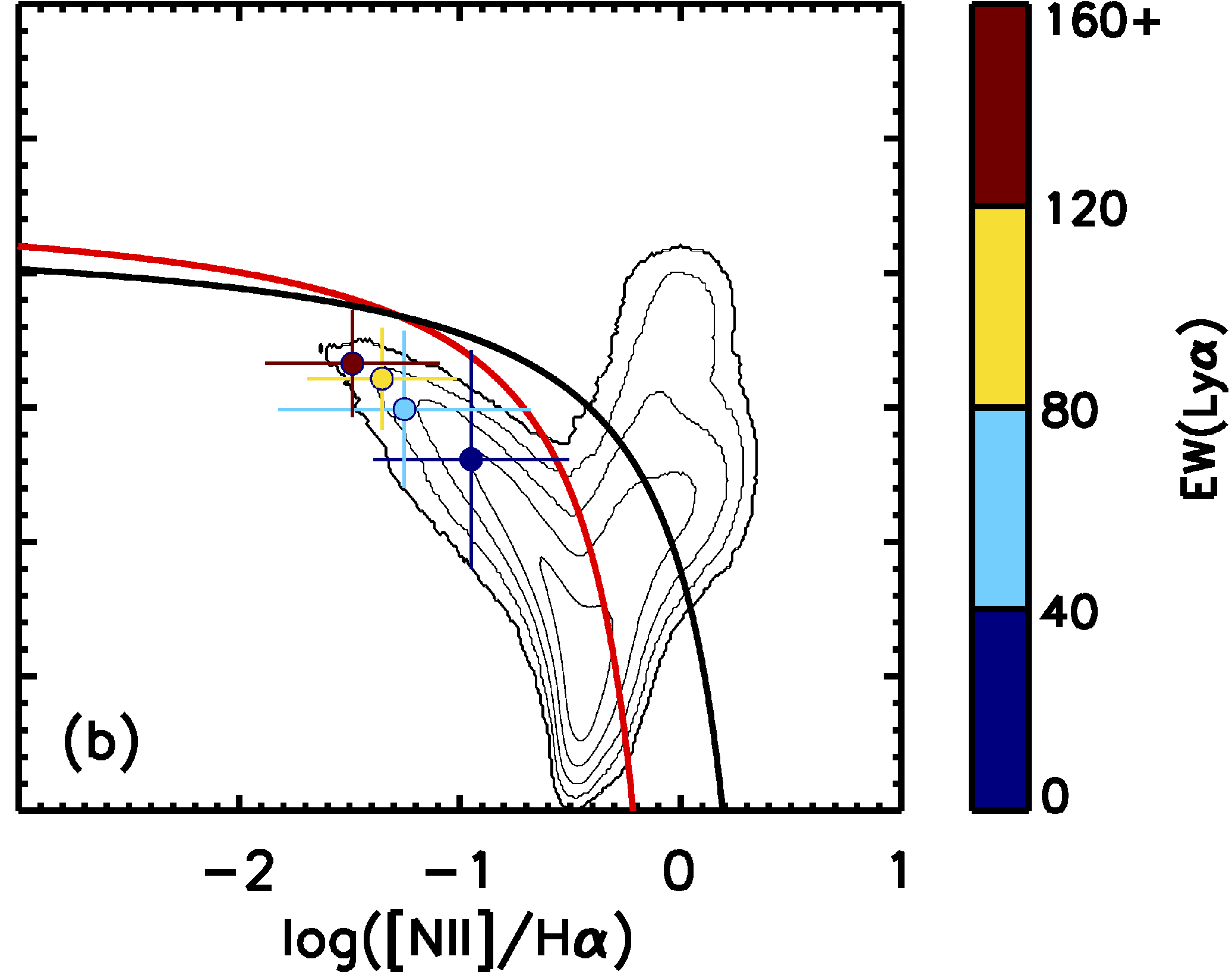}
\par\end{centering}

\caption{\textbf{(a)} BPT diagram as presented in Figure \ref{bpt} but with
data points color-coded to indicate Ly$\alpha$ EWs. We note that
star-forming LAEs with EW $>80$ \AA~ are only found in the upper
left corner of the BPT diagram. \textbf{(b) }The average star-former
{[}N{\scriptsize{}II}{]} to H$\alpha$ and {[}O{\scriptsize{}III}{]}
to H$\beta$ line ratios for each EW bin with error bars that indicate
the standard deviation of the component data points. We find that
the average high-EW LAE probes galaxies with more extreme ISM properties.}

\label{bpt_ew}

{\footnotesize (A color version of this figure is available in the online journal.)}
\end{figure*}

In Figure \ref{bpt_ew}(a), we color-code our BPT diagram data (Figure
\ref{bpt}) to show our Ly$\alpha$ EW measurements. We find that
star-forming LAEs have a wide range of {[}N{\small{}II}{]} to H$\alpha$
and {[}O{\small{}III}{]} to H$\beta$ line ratios, most likely indicating
a wide range of ISM conditions. However, we note that LAEs with EW
$>80$ \AA~ are only found in the upper left corner of the BPT diagram.
This region is thought to be dominated by galaxies with lower metallicities,
higher ionization parameters, and higher electron densities. To illustrate
this trend more clearly, in Figure \ref{bpt_ew}(b), we show the average
star-former {[}N{\small{}II}{]} to H$\alpha$ and {[}O{\small{}III}{]}
to H$\beta$ line ratios for each of our adopted EW bins. The error
bars show the standard deviation of the data points. We note that
our observed trend, where high-EW LAEs preferentially occupy the upper
left corner of the BPT diagram, is consistent with earlier $z=0.3$
results (\citealt{cowie11}; and see \citealt{trainor16} for $z=2.5$ results).
Cowie et al.\ compared LAEs to UV-selected galaxies (Ly$\alpha$
EW $\sim0$ \AA) and found that their $z\sim0.3$ UV-selected galaxies
were preferentially located in the lower right of the BPT diagram,
roughly corresponding to our highest SDSS density contour shown in
Figure \ref{bpt_ew}, while their $z\sim0.3$ LAEs were preferentially
found in the upper left of the BPT diagram. Our results indicate that
higher EW LAEs at $z=0.3$ on average probe galaxies with more extreme
ISM properties and may offer promise as local analogs to high-redshift
galaxies. We will further investigate the emission properties of the
$z\sim0.3$ LAE sample in a follow-up paper.

\begin{figure*}[!t]
\includegraphics[bb=95bp 75bp 720bp 520bp,clip,width=8.5cm]{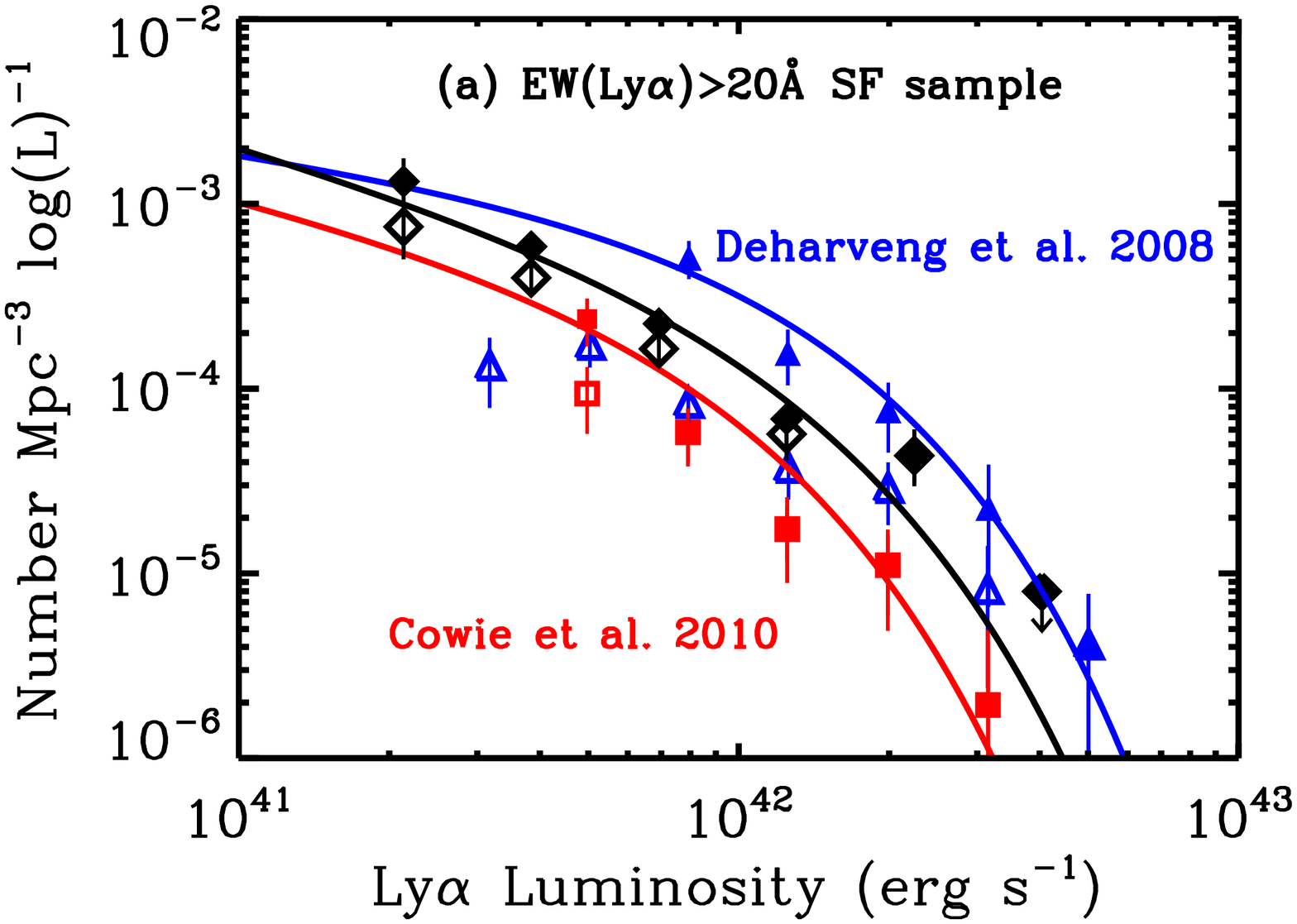}\includegraphics[bb=95bp 75bp 720bp 520bp,clip,width=8.5cm]{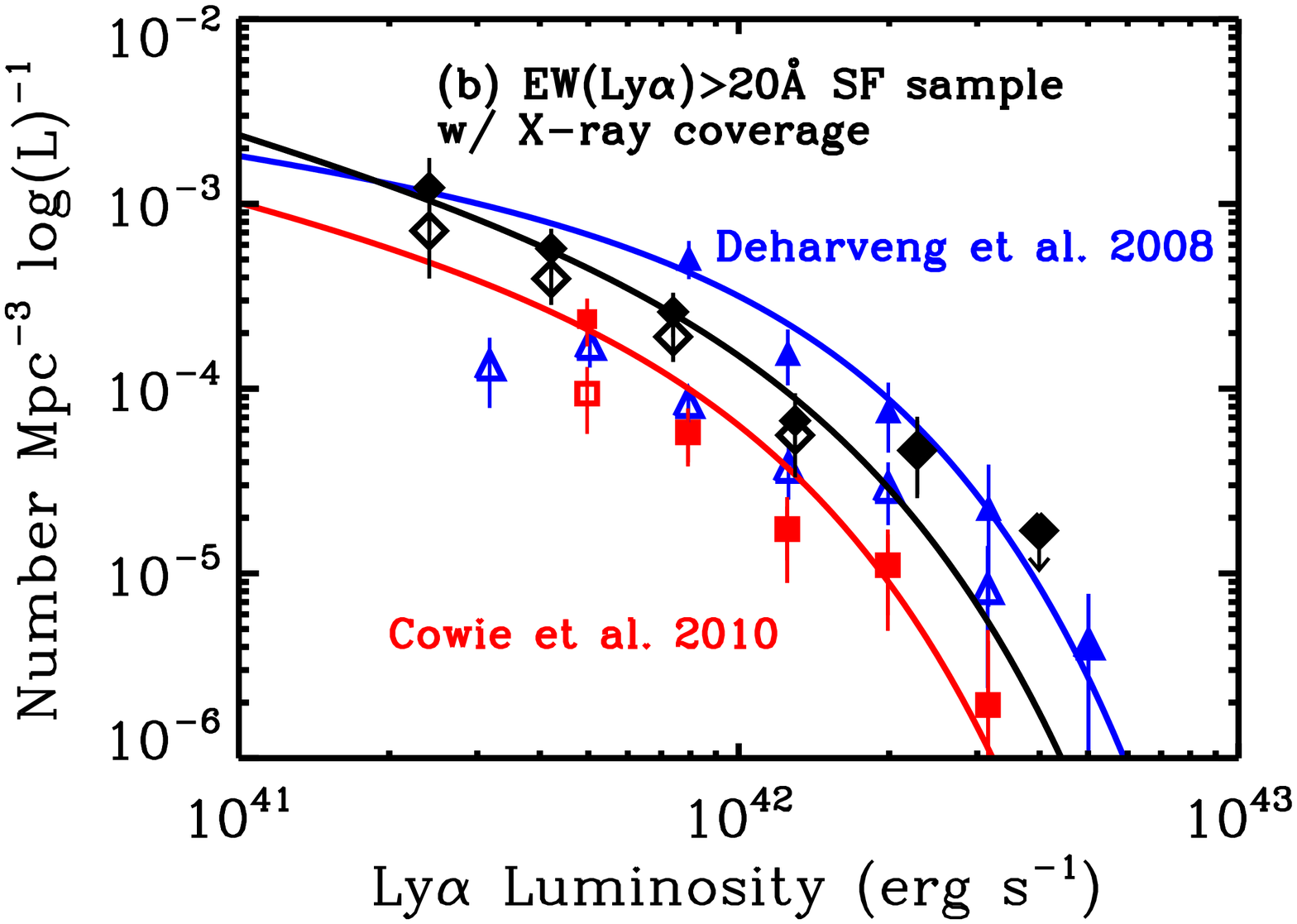}

\caption{\textbf{(a)} The SF Ly$\alpha$ LF at $z$ = 0.195-0.44 in deep \emph{GALEX
}grism fields with $EW(Ly\alpha)\geqslant20$ \AA (open diamonds--raw
data; solid diamonds--corrected for the effects of incompleteness
using the results from our Monte Carlo simulations). The black curve
indicates the best-fit Schechter function to the SF data assuming
a fixed slope of $\alpha=-1.75$. For comparison, we show raw and
completeness corrected data from \citet[][blue curve and triangle symbols]{deharveng08}
and \citet[][red curve and square symbols]{cowie10}. \textbf{(b)
}Same as Figure \ref{LF}(a), but with the LAE survey limited to regions
with deep X-ray data to ensure a robust AGN classification. We find
that requiring a more robust AGN classification does not significantly
alter our results derived from our full sample. }

\label{LF}

{\footnotesize (A color version of this figure is available in the online journal.)}
\end{figure*}
\begin{figure}[!t]
\includegraphics[bb=95bp 75bp 720bp 520bp,clip,width=8.5cm]{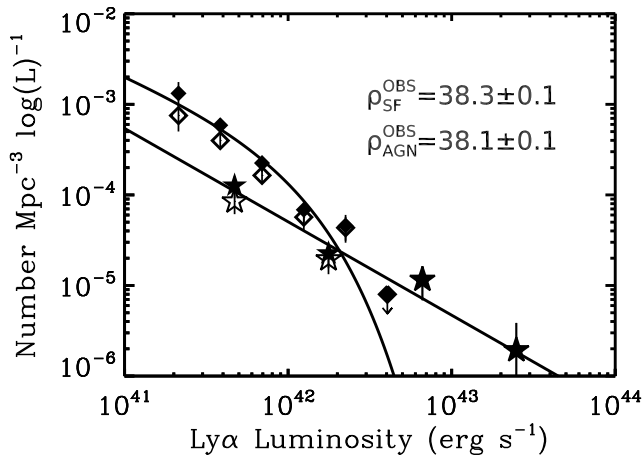}

\caption{The AGN Ly$\alpha$ LF at $z$ = 0.195-0.44 in deep \emph{GALEX }grism
fields with $EW(Ly\alpha)\geqslant20$ \AA (open stars--raw data;
solid stars--corrected for the effects of incompleteness using the
results from our Monte Carlo simulations). The black line indicates
the best-fit power-law to the AGN data. Integrating over the observed
luminosity range, we calculate the observed AGN Ly$\alpha$ luminosity
density and find log $\rho_{Ly\alpha,AGN}^{obs}=38.1\pm0.1$ erg s$^{-1}$Mpc$^{-3}$(from
log $L=41.2$ to $43.7$) which is only 0.2 dex less than the observed
SF Ly$\alpha$ luminosity density. We show our SF Ly$\alpha$ LF from
Figure \ref{LF}(a) for comparison. }

\label{LFagn}
\end{figure}

\noindent \begin{deluxetable*}{ccccccccc}  \tablecolumns{9}  \tablewidth{0pc}  \tablecaption{Best-fit $z\sim0.3$ Ly$\alpha$ Luminosity Function Parameters using a $\rho_{Ly\alpha}$ lower integration limit of log $L=41.2$.}  \tablehead{  \colhead{Reference} & \colhead{$\alpha$} & \colhead{log $L^{\star}$} & \colhead{log $\phi^{\star}$} & \colhead{$\sigma$} & \colhead{log $\rho^{obs}_{Ly\alpha,SF}$} & \colhead{m} & \colhead{b} & \colhead{log $\rho^{obs}_{Ly\alpha,AGN}$} \\  \colhead{ } & \colhead{(fixed)} & \colhead{(erg s$^{-1}$)} & \colhead{(Mpc$^{-3}$)} & \colhead{ } & \colhead{(erg s$^{-1}$ Mpc$^{-3}$)} & \colhead{ } & \colhead{} & \colhead{(erg s$^{-1}$ Mpc$^{-3}$)}\\ \colhead{(1)} & \colhead{(2)} & \colhead{(3)} & \colhead{(4)} & \colhead{(5)} & \colhead{(6)} & \colhead{(7)} & \colhead{(8)} & \colhead{(9)}}   \startdata  Deharveng et al.\ 2008 & -1.35 & 42.0$\pm$0.1 & -3.4$\pm$0.2 & \nodata & 38.6$\pm$0.2 & \nodata & \nodata & \nodata \\ Cowie et al.\ 2010 & -1.60 & 41.8$\pm$0.1 & -3.8$\pm$0.1 & \nodata & 38.0$\pm$0.1 & \nodata & \nodata & \nodata \\ This work (Best estimate) & -1.75 & 42.0$\pm$0.2 & -3.7$\pm$0.3 & \nodata & 38.3$\pm$0.1 & -2.0$\pm$0.3 & 38.6$\pm$12.4 & 38.1$\pm$0.1\tablenotemark{a} \\ This work (X-ray Covered) & -1.75 & 41.9$\pm$0.2 & -3.6$\pm$0.4 & \nodata & 38.3$\pm$0.1 & -2.4$\pm$0.2 & 55.1$\pm$8.6 & 38.1$\pm$0.1\tablenotemark{a} \\ This work (Schechter + fPL)  & -1.75 & 41.9$\pm$0.2 & -3.7$\pm$0.3 & \nodata & 38.3$\pm$0.1 & -2.0(fixed) & 38.8$\pm$0.2 & 38.2$\pm$0.2\tablenotemark{a} \\ This work (Schechter + PL)  & -1.75 & 41.9$\pm$0.2 & -3.7$\pm$0.3 & \nodata & 38.2$\pm$0.2 & -2.1$\pm$0.6 & 43.0$\pm$24.3 & 38.3$\pm$0.2\tablenotemark{a} \\ This work (Saunders + fPL) & -1.75 & 41.2$\pm$1.6 & -3.1$\pm$1.4 & 0.5$\pm$0.4 & 38.4$\pm$0.2 & -2.0(fixed) & 38.7$\pm$0.3 & 38.1$\pm$0.3\tablenotemark{a} \\ This work (Saunders + PL) & -1.75 & 41.3$\pm$1.2 & -3.1$\pm$1.0 & 0.5$\pm$0.4 & 38.4$\pm$0.1 & -1.7$\pm$0.4 & 25.5$\pm$17.0 & 38.0$\pm$0.4\tablenotemark{a} \enddata   \tablecomments{} \tablenotetext{a}{\rm{Upper} integration limit set to the maximum observered $z=0.3$ Ly$\alpha$ luminosity of log $L=43.7$.} \label{schechter} \end{deluxetable*}

\section{Luminosity Function}

\noindent \label{sec_lf}

For the combined CDFS, GROTH, NGPDWS, and COSMOS fields, we compute
the Ly$\alpha$ LF in the redshift range $z=0.195-0.44$ using the
$1/V$ technique \citep{felten76}. The total area covered by our
survey is 7423 arcmin$^{2}$ which indicates a Ly$\alpha$ survey
volume of $0.90\times10^{6}$ Mpc$^{3}$. This is comparable to the
largest LAE survey at $z=2.2$ which has a survey volume of $1.32\times10^{6}$
Mpc$^{3}$ \citep{konno16} and is about 10 times smaller than the
\emph{GALEX} NUV LAE survey at $z=0.67-1.16$ which has a survey volume
of $9.25\times10^{6}$ Mpc$^{3}$ \citep{wold14}.  In Figure \ref{LF}(a),
we show our raw EW$({\rm {Ly}}\alpha)\geqslant20$ \AA\ star-forming
Ly$\alpha$ LF with black open diamonds. We show the LF corrected
for incompleteness using the results from our Monte Carlo simulations
with solid symbols.  Error bars are $\pm1\sigma$ Poisson errors.
We fit a Schechter function \citep{schechter76} to the SF Ly$\alpha$
LF, where

\begin{equation}
\Phi_{SF}(L)dL=\phi^{\star}\left(\frac{L}{L^{\star}}\right)^{\alpha}e^{-(L/L^{\star})}d\left(\frac{L}{L^{\star}}\right).
\end{equation}

\noindent For the Schechter function fit, we assume a fixed faint-end
slope of $\alpha=-1.75$, which is the best-fit $z=2.2$ value found
by \citet{konno16}. This assumption is required because our $z\sim0.3$
data lack the faint luminosity range necessary to constrain $\alpha$.

In Figure \ref{LF}(b), we restrict our fields to regions with deep
X-ray data and re-derive the star-forming Ly$\alpha$ LF. This removes
the NGPDWS field and restricts the area of the remaining \textit{GALEX}
fields but ensures a uniform means of AGN classification (See Section
\ref{agn}). The X-ray imaging depth for our restricted field is $\sim10^{41}$
erg s$^{-1}$ at $z\sim0.3$, which is well below the $10^{42}$ erg
s$^{-1}$ threshold typically used to identify AGN. Comparing this
LF to the LF computed from the full LAE galaxy sample, we find that
all star-forming LF points are consistent within 1$\sigma$ error
bars. We find that requiring a more robust AGN classification does
not significantly alter our results derived from our full sample. 

In both panels of Figure \ref{LF}, we compare our $z\sim0.3$ LF
to the results of two $z\sim0.3$ Ly$\alpha$ LFs derived from \emph{GALEX}
pipeline data \citep[][blue data and red data, respectively]{deharveng08,cowie10}.
\citet{cowie10} pointed out that all of their $z\sim0.3$ raw Ly$\alpha$
LF measurements are comparable to the previously published raw LF
determined by \citet[][]{deharveng08}. It is only after corrections
for incompleteness are applied that the their results differ. Unlike
the previous pipeline samples, our data cube LAE sample is not pre-selected
from continuum bright objects and this greatly simplifies the estimation
of corrections for incompleteness. With our larger and less biased
sample, we find that our LF data points fall in between these two
previous $z\sim0.3$ LFs with best fit Schechter parameters summarized
in Table \ref{schechter}. 

Having identified the AGNs within our LAE sample, we may also compute
the LF for Ly$\alpha$ emitting AGNs. In Figure \ref{LFagn}, we show
the AGN Ly$\alpha$ LF at $z$ = 0.195-0.44 in our survey fields with
EW$({\rm {Ly}}\alpha)\geqslant20$ \AA (open stars--raw data; solid
stars--corrected for the effects of incompleteness using the results
from our Monte Carlo simulations). The black line indicates the best-fit
power-law to the AGN data with the functional form:

\begin{equation}
\Phi_{AGN}(L)dL=10^{b}L^{m}dL.
\end{equation}
The best-fit parameters are listed in Table \ref{schechter}.

To assess the amount of Ly$\alpha$ light emitted by star-formers
and AGNs we calculate the observed Ly$\alpha$ luminosity density:

\begin{equation}
\rho_{Ly\alpha}^{obs}=\int L\Phi(L)dL,
\end{equation}
and find log $\rho_{Ly\alpha,SF}^{obs}=38.3\pm0.1$ (integrating over
the luminosity range of log $L=41.2$ to infinity) and log $\rho_{Ly\alpha,AGN}^{obs}=38.1\pm0.1$
erg s$^{-1}$Mpc$^{-3}$ (integrating over the survey's luminosity
range of log $L=41.2$ to $43.7$). This result indicates that AGNs
are responsible for $\sim39\%$ of the observed Ly$\alpha$ light
at $z\sim0.3$. We emphasize that this result is dependent on our
survey's luminosities limits. To estimate a lower limit to the AGN
contribution, we integrate the SF LF from zero to infinity and compare
this value to the $\rho_{Ly\alpha,AGN}^{obs}$. With these integration
limits the luminosity density is simply 
\begin{equation}
\rho_{{\rm {Ly}\alpha,}SF}^{total}=L^{\star}\phi^{\star}\Gamma(\alpha+2),
\end{equation}

\noindent where $\Gamma$ is the Gamma function. Integrating down
to zero makes our calculation more sensitive to the poorly constrained
faint-end slope, but assuming reasonable $\alpha$ values we do not
expect our total SF luminosity density calculations to be altered
by more than a factor of 2. We find a total SF luminosity density
of log $\rho_{Ly\alpha,SF}^{total}=38.8\pm0.1$ erg s$^{-1}$Mpc$^{-3}$
which gives a lower-limit $z\sim0.3$ AGN contribution of $17\%$.
This significant AGN contribution emphasizes the need for caution
when interpreting higher redshift LAE samples with limited or no AGN
identification. We note that the assumed form of the AGN LF results
in an AGN luminosity density that is very sensitive to the assumed
upper integration limit. The assumed power-law LF is merely the simplest
functional form given the observed Ly$\alpha$ luminosity range. This
is also true for the assumed Schechter function since the SF LF must
turn over at lower luminosities to prevent the total number of galaxies
from diverging (for $\alpha\leq-1$). 
\begin{figure*}[!t]
\includegraphics[bb=95bp 75bp 720bp 520bp,clip,width=8.5cm]{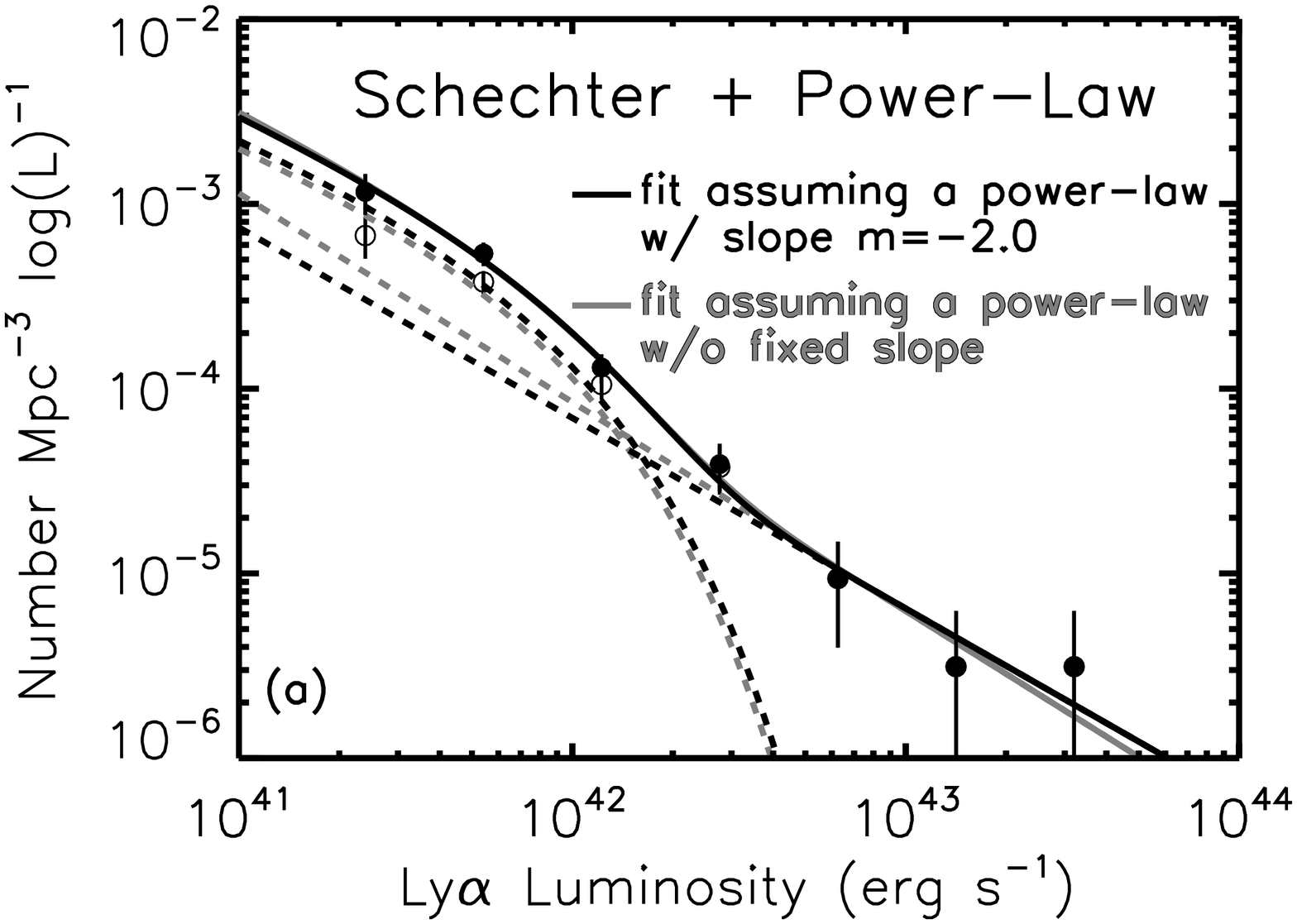}\includegraphics[bb=95bp 75bp 720bp 520bp,clip,width=8.5cm]{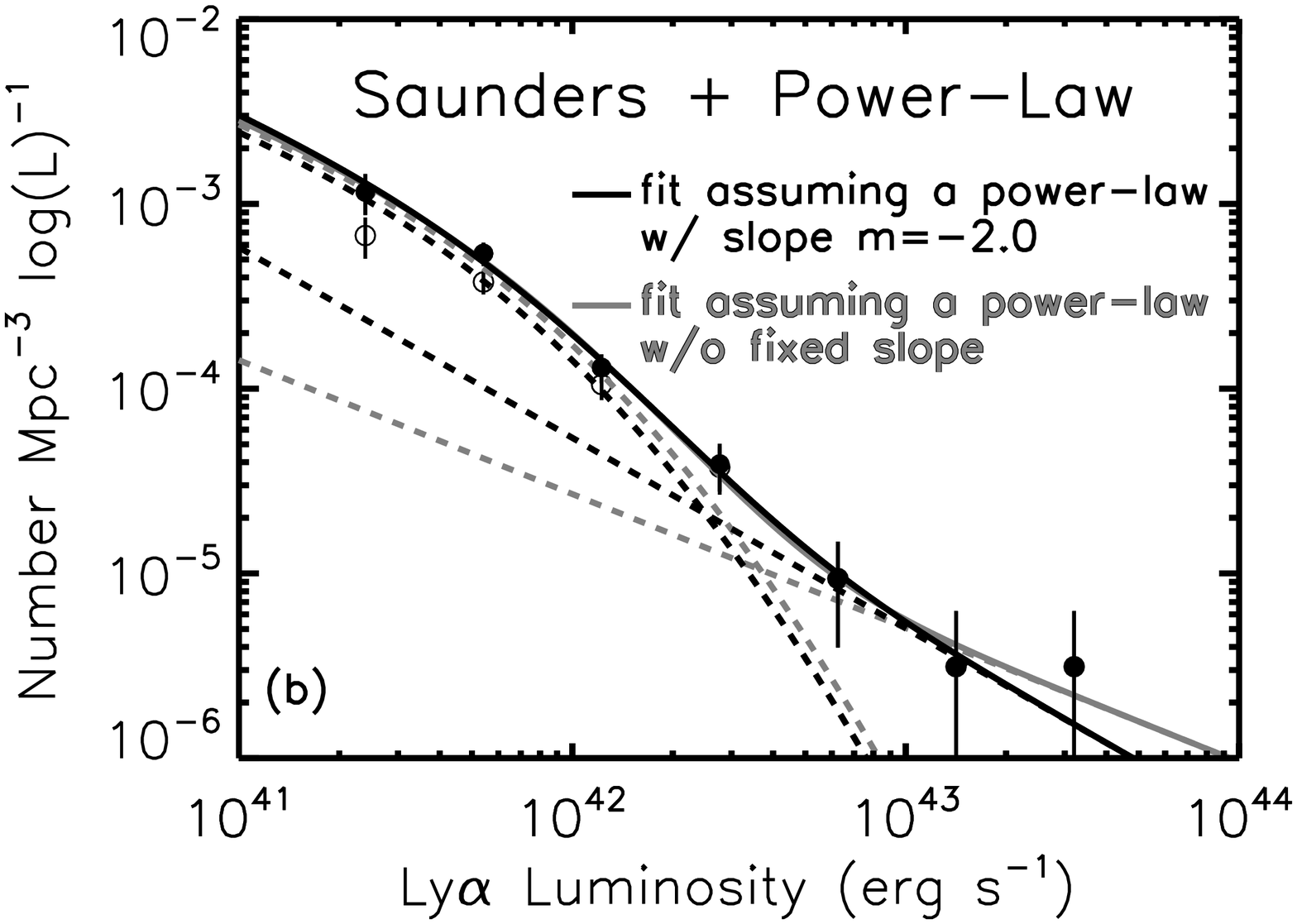}

\caption{\textbf{(a)}\textbf{\emph{ }}Combined SF and AGN Ly$\alpha$ LF at
$z$ = 0.195-0.44 in deep \emph{GALEX }grism fields with $EW(Ly\alpha)\geqslant20$
\AA (open circles--raw data; solid circles--corrected for the effects
of incompleteness using the results from our Monte Carlo simulations).
The black curve indicates the best-fit Schechter function $+$ power-law
to the combined data assuming a Schechter fixed faint-end slope of
$\alpha=-1.75$. The black dashed curve and line indicate the underlying
best-fit Schechter function and power-law, respectively. The slope
of the power-law has been fixed to the isolated best-fit value of
-2.0 (see Figure \ref{LFagn} and Row 3 of Table \ref{schechter}).
 \textbf{(b) }Same as Figure \ref{comboLF}(a), but with the data
fit by a Saunders function $+$  power-law. Computing luminosity densities
for both Schechter and Saunders fits, we find that the results agree
within 1$\sigma$ errors. In both Figures, we also show the effect
of allowing the power law slope to be a free parameter (grey dashed
and solid curves). We find that this alteration does not significantly
alter our computed luminosity densities. All results are summarized
in Table \ref{schechter}.}

\label{comboLF}
\end{figure*}
\begin{deluxetable*}{ccccccccc} \tablecolumns{9}  \tablewidth{0pc}  \tablecaption{Evolution of Ly$\alpha$ Luminosity Function Parameters assuming $\alpha=-1.75$ and a $\rho_{Ly\alpha}$ lower integration limit of log $L=41.41$.}  \tablehead{  \colhead{Redshift} & \colhead{log $L^{\star}$} & \colhead{log $\phi^{\star}$} & \colhead{$\sigma$} & \colhead{log $\rho^{obs}_{Ly\alpha,SF}$} & \colhead{m} & \colhead{b} & \colhead{log $\rho^{obs}_{Ly\alpha,AGN}$} & \colhead{Reference}\\  \colhead{ } & \colhead{(erg s$^{-1}$)} & \colhead{(Mpc$^{-3}$)} & \colhead{ } & \colhead{(erg s$^{-1}$ Mpc$^{-3}$)} & \colhead{ } & \colhead{} & \colhead{(erg s$^{-1}$ Mpc$^{-3}$)} &\colhead{ }\\ \colhead{(1)} & \colhead{(2)} & \colhead{(3)} & \colhead{(4)} & \colhead{(5)} & \colhead{(6)} & \colhead{(7)} & \colhead{(8)} & \colhead{(9)}}   \startdata 0.3 & 41.9$\pm$0.2 & -3.7$\pm$0.3 & \nodata  & 38.2$\pm$0.1 & -2.0(fixed) & 38.8$\pm$0.2 & 38.2$\pm$0.2\tablenotemark{a} & This work (Schechter + fPL)\\ 0.3 & 41.9$\pm$0.2 & -3.7$\pm$0.3 & \nodata  & 38.1$\pm$0.2 & -2.1$\pm$0.5 & 43.0$\pm$23.4 & 38.3$\pm$0.3\tablenotemark{a} & This work (Schechter + PL)\\ 0.3 & 41.2$\pm$1.6 & -3.1$\pm$1.6 & 0.5$\pm$0.4  & 38.2$\pm$1.4 & -2.0(fixed) & 38.7$\pm$0.7 & 38.1$\pm$0.7\tablenotemark{a} & This work (Saunders + fPL)\\ 0.3 & 41.3$\pm$1.6 & -3.1$\pm$1.6 & 0.5$\pm$0.4  & 38.3$\pm$0.9 & -1.7$\pm$0.3 & 25.5$\pm$13.2 & 38.2$\pm$0.6\tablenotemark{a} & This work (Saunders + PL)\\ 0.9 & 43.0$\pm$0.2 & -4.9$\pm$0.3 & \nodata  & 38.4$\pm$0.1 & \nodata & \nodata & \nodata & Wold+2014 (Schechter)\tablenotemark{b}\\ 2.2 & 42.7$\pm$0.1 & -3.2$\pm$0.2 & \nodata  & 39.8$\pm$0.0 & \nodata & \nodata & \nodata & Konno+2016 (their Schechter)\tablenotemark{c}\\ 2.2 & 42.4$\pm$0.1 & -2.9$\pm$0.1 & \nodata  & 39.7$\pm$0.1 & -2.0(fixed) & 40.0$\pm$0.1 & 39.5$\pm$0.1\tablenotemark{d} & Konno+2016 (Schechter + fPL)\\ 2.2 & 42.4$\pm$0.1 & -2.8$\pm$0.1 & \nodata  & 39.7$\pm$0.1 & -2.0$\pm$0.5 & 38.5$\pm$21.4 & 39.5$\pm$0.1\tablenotemark{d} & Konno+2016 (Schechter + PL)\\ 2.2 & 42.0$\pm$0.4 & -2.6$\pm$0.4 & 0.4$\pm$0.2  & 39.7$\pm$0.1 & -2.0(fixed) & 40.0$\pm$0.1 & 39.5$\pm$0.1\tablenotemark{d} &  Konno+2016 (Saunders + fPL)\\ 2.2 & 41.8$\pm$0.3 & -2.4$\pm$0.3 & 0.5$\pm$0.2  & 39.8$\pm$0.1 & -1.8$\pm$0.4 & 27.9$\pm$17.7 & 39.4$\pm$0.2\tablenotemark{d} &  Konno+2016 (Saunders + PL) \enddata   \tablecomments{} \tablenotetext{a}{\rm{Upper} integration limit set to the maximum observered $z=0.3$ Ly$\alpha$ luminosity of log $L=43.7$.} \tablenotetext{b}{\rm{Identified} AGN removed prior to Schechter function fit (see \citealt{wold14} for details).} \tablenotetext{c}{\rm{Best} fit Schechter function as listed in Table 5 of \citet{konno16}, symmetric errors estimated from same table.} \tablenotetext{d}{\rm{Upper} integration limit set to the maximum observered $z=2.2$ Ly$\alpha$ luminosity of log $L=44.4$.} \label{evotable} \end{deluxetable*}

At higher redshifts the LAEs that make up the bright-end tail of
the Ly$\alpha$ LF are typically \citep[but not always e.g.,][]{matthee15,hu16}
attributed to AGNs due to their bright counterparts in X-ray, UV and/or
radio imaging data \citep[e.g.,][]{ouchi08,konno16}. However, high-redshift
LAEs with faint luminosities are generally assumed to be star-formers.
With our low-redshift survey that has identified AGNs in multiple
ways (see Section \ref{agn}), we have shown that AGNs are also present
at lower Ly$\alpha$ luminosities. With this in mind, we developed
a procedure that does not require AGN identification -- yet can accurately
recover the SF luminosity density -- by simultaneously fitting the
SF and AGN Ly$\alpha$ LFs. In Section \ref{evo}, we use this procedure
to help avoid potential systematic errors in the study of the evolution
of the SF luminosity density. 

In Figure \ref{comboLF} (a) and (b), we show the combined $z\sim0.3$
SF+AGN Ly$\alpha$ LF (here the LF is computed for all LAEs regardless
of their SF or AGN classification). In Figure \ref{comboLF}(a), we
simultaneously fit a Schechter function and power-law and find log
$\rho_{Ly\alpha,SF}^{obs}=38.3\pm0.1$ (integrating over the luminosity
range of log $L=41.2$ to infinity) and log $\rho_{Ly\alpha,AGN}^{obs}=38.2\pm0.2$
erg s$^{-1}$Mpc$^{-3}$(integrating over the survey's luminosity
range of log $L=41.2$ to $43.7$) which is consistent with our best
estimate based on the isolated SF and AGN Ly$\alpha$ LFs. We have
fixed the power-law slope to the best-fit value of $-2.0$ (see Row
3 of Table \ref{schechter}) because the AGN LF is hard to constrain
at the faint end where the SF+AGN Ly$\alpha$ LF is dominated by star-forming
galaxies. We find that allowing the power-law slope to be a free parameter
(grey curves) does not alter our luminosity density measurements beyond
our 1$\sigma$ error bars (see Table \ref{schechter}). 

Observational and theoretical studies \citep[e.g., ][]{gunawardhana15,salim12},
have suggested that LFs of SFR tracers are better fit by a Saunders
function \citep{saunders90}:
\begin{equation}
\Phi_{SF}(L)dL=\phi^{\star}\left(\frac{L}{L^{\star}}\right)^{\alpha}e^{-{\rm {log}^{2}(1+L/L^{\star})}/(2\sigma^{2})}d\left(\frac{L}{L^{\star}}\right).
\end{equation}
This function is similar to a Schechter function, except beyond $L^{\star}$
the function declines in a Gaussian manner rather than exponentially.
The increased occurrence of AGNs at higher Ly$\alpha$ luminosities
makes the exact shape of the Ly$\alpha$ SF LF beyond $L^{\star}$
difficult to constrain, and we fit our combined SF$+$AGN LF data
points with a Saunders function to access any effect on our computed
luminosity densities. In Figure \ref{comboLF}(b), we simultaneously
fit a Saunders function and power-law and find  log $\rho_{Ly\alpha,SF}^{obs}=38.4\pm0.2$
and log $\rho_{Ly\alpha,AGN}^{obs}=38.1\pm0.3$ erg s$^{-1}$Mpc$^{-3}$.
We find that the measured luminosity densities are not significantly
altered by the choice of Schechter or Saunders function.

The agreement between isolated SF LF fits (see Figure \ref{LF}) and
simultaneous SF+AGN fits (see Figure \ref{comboLF}) suggests that
we can accurately recover SF luminosity densities even without AGN
classification.  The best-fit LF parameters and observed Ly$\alpha$
luminosity densities are summarized in Table \ref{schechter}.

\section{Ly$\alpha$ LF Evolution}

\label{evo}

\noindent 

\begin{figure*}[!]
\includegraphics[bb=95bp 75bp 720bp 520bp,clip,width=8.5cm]{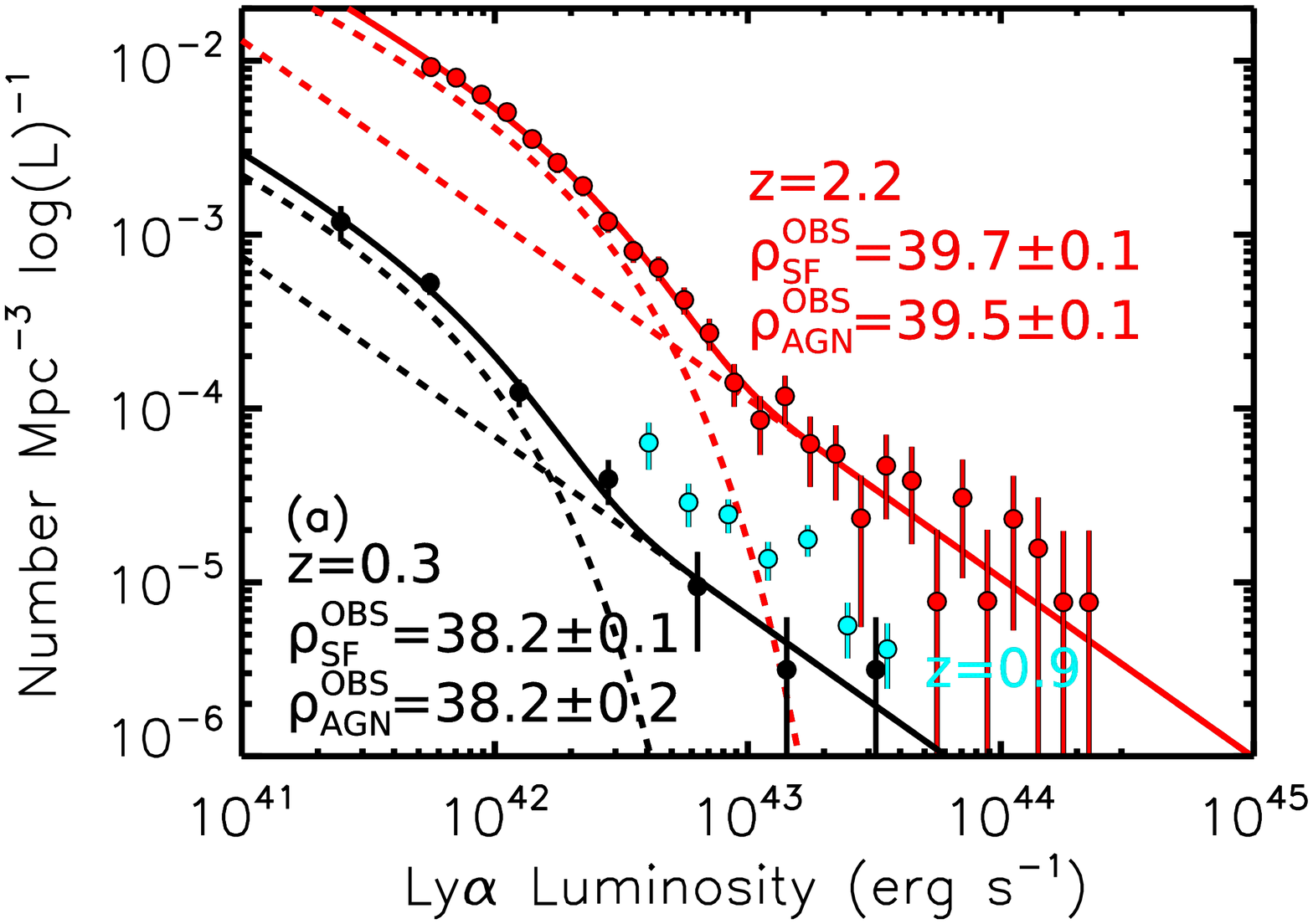}\includegraphics[bb=95bp 75bp 720bp 520bp,clip,width=8.5cm]{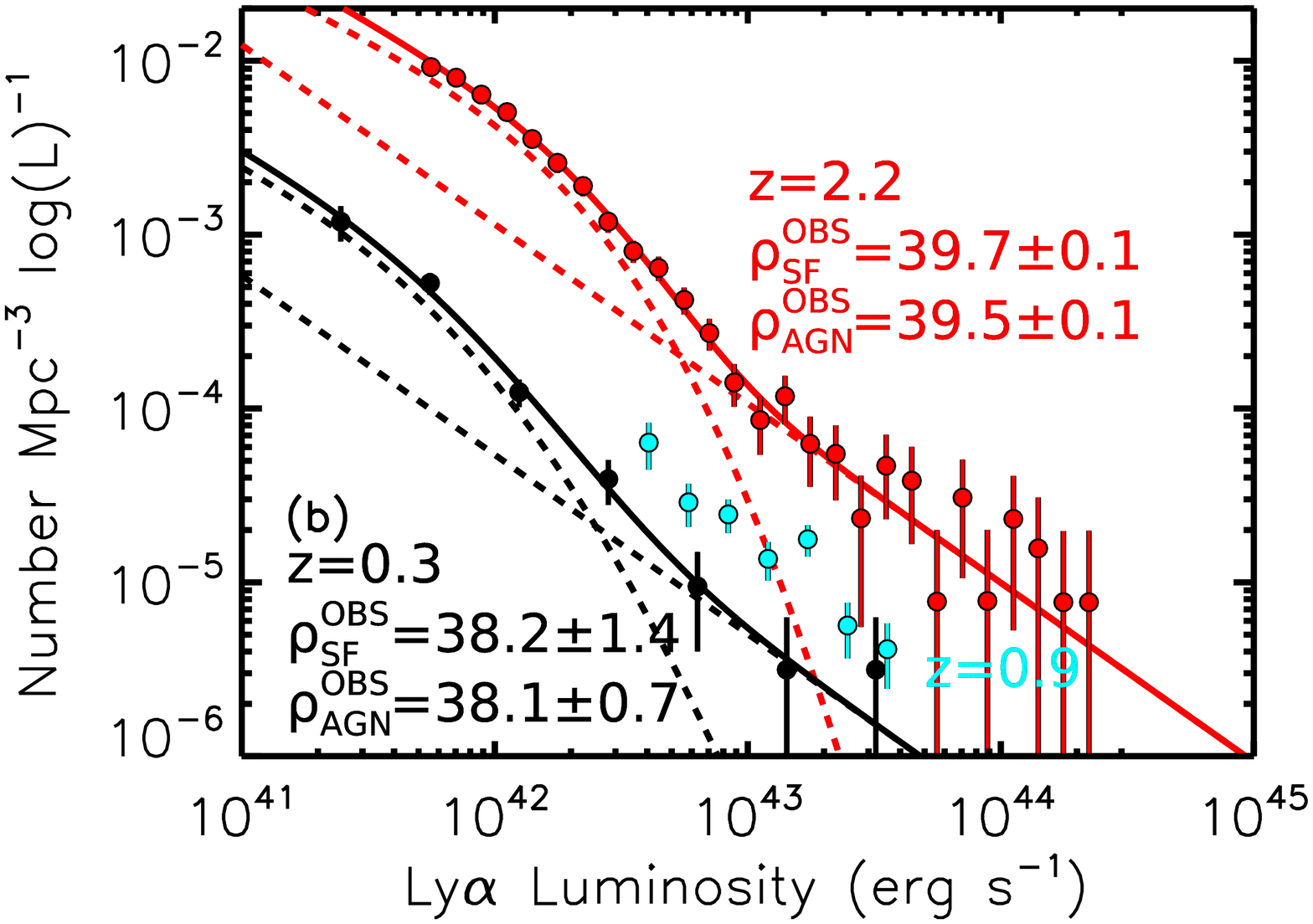}

\caption{\textbf{(a)} Evolution of the\textbf{\textit{ }}combined SF and AGN
Ly$\alpha$ LFs from $z\sim0.3$ to $z=2.2$ with best-fit\textbf{
}Schechter function $+$ power-law. The black points show our $z\sim0.3$
LF data. The cyan points show $z\sim0.9$ LF data from \citet{wold14}.
The red points show the $z=2.2$ LF data from \citet{konno16}.\textbf{
(b)} Same as Figure \ref{evoLF}(a), but with the data fit by a Saunders
function $+$ power-law. The listed log luminosity densities and corresponding
best-fit parameters are summarized in Table \ref{evotable}.}

\label{evoLF}

{\footnotesize (A color version of this figure is available in the online journal.)}
\end{figure*}

In Figure \ref{evoLF}, we show the evolution of the SF $+$ AGN Ly$\alpha$
LF for $z\sim0.3$, $0.9$, and $2.2$ (black, cyan, and red data
points, respectively). The $z=2.2$ data are from the deep Subaru
narrowband survey presented by \citet{konno16}. This sample contains
a total of 3,137 LAEs covering a Ly$\alpha$ luminosity range of log
$L_{{\rm {Ly}\alpha}}=41.7$ - $44.4$ erg s$^{-1}$. The $z\sim0.9$
data are from the archival \emph{GALEX} NUV LAE survey presented by
\citet{wold14}. This sample contains 60 SF LAEs covering a Ly$\alpha$
luminosity range of log $L_{{\rm {Ly}\alpha}}=42.5$ - $43.4$ erg
s$^{-1}$ and a redshift range of $z=0.67-1.16$. 
\begin{figure}[!th]
\includegraphics[bb=95bp 75bp 720bp 520bp,clip,width=8.5cm]{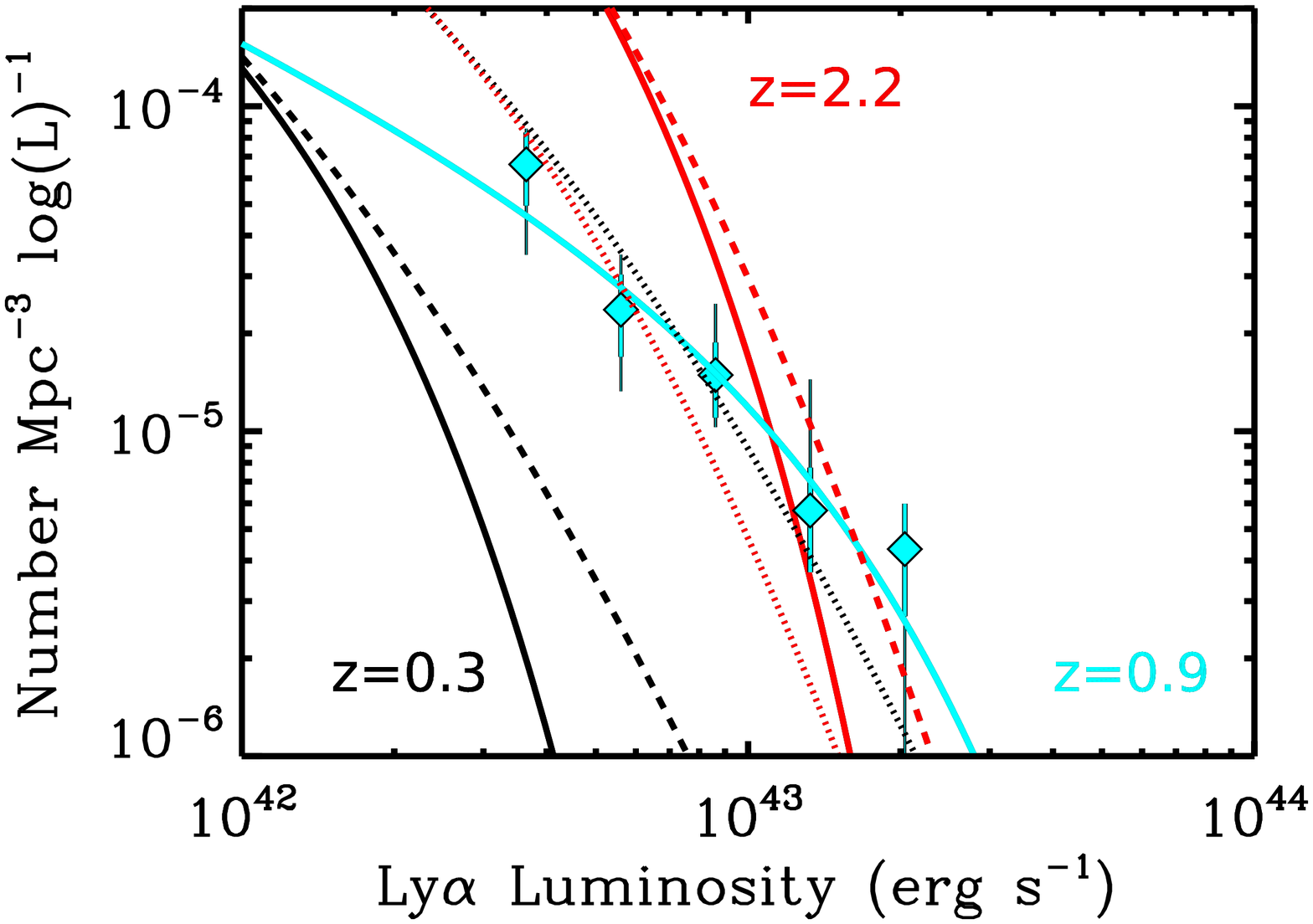}

\caption{The $z\sim0.9$ Ly$\alpha$ SF LF compared to the $z\sim0.3$ and
$z=2.2$ Ly$\alpha$ SF LFs. The cyan data points indicate the $z\sim0.9$
SF LF with thick Poisson error bars. To assess the effect of any unidentified
AGNs, we also show thin error bars that indicate the 1$\sigma$ Poisson
error obtained by restricting the $z\sim0.9$ survey area to regions
with deep X-ray data. All SF LAEs within this sub-sample are not identified
as AGNs in any way (see \citealt{wold14} for details) and are known
to have X-ray luminosities below $10^{42}$ erg s$^{-1}$. The best-fit
$z\sim0.3$, $0.9$, and $2.2$ Schechter functions are indicated
by black, cyan, and red solid curves, respectively. The best-fit $z\sim0.3$
and $2.2$ Saunders functions are indicated by black and red dashed
curves, respectively. The $z\sim0.3$ Saunders function with $L^{\star}$
increased by $0.45$ dex is indicated by the black dotted curve. The
$z=2.2$ Saunders function with $\phi^{\star}$ decreased by $0.8$
dex is indicated by the red dotted curve. }

\label{nlf}

{\footnotesize (A color version of this figure is available in the online journal.)}
\end{figure}

For the $z\sim0.3$ and $2.2$ data, we apply our fitting technique
developed in the previous section to obtain self-consistently measured
luminosity densities. In this section, we are most interested in a
direct comparison of our results to \citet{konno16}. Thus, for the
luminosity density calculations we adopt Konno et al.'s lower integration
limit of log $L=41.41$ erg s$^{-1}$ that corresponds to $(0.03)L_{Ly\alpha,z=3}^{\star}$
\citep{ouchi08}. For the $\rho_{Ly\alpha,AGN}^{obs}$ computations,
we set the upper integration limit to the survey's maximum observed
Ly$\alpha$ luminosity. For the $z=0.3$ and $z=2.2$ survey, this
corresponds to log $L=$$43.7$ and $44.4$ erg s$^{-1}$, respectively
. 

In Figure \ref{evoLF}(a), we simultaneously fit a Schechter function
+ a fixed slope power-law (listed as fPL in Table \ref{evotable})
to the $z=0.3$ and $z=2.2$ data. It is not clear that our fixed-slope
assumption is accurate but given the limited luminosity range over
which AGN density dominates over SF galaxies, we adopt this convention.
We find that allowing the AGN power-law slope to be a free parameter
does not significantly alter our results (see Table \ref{evotable}).
Using this simultaneous fitting method, we find a factor of 30 increase
in the SF luminosity density from $z\sim0.3$ to $z=2.2$. Over the
same redshift range, \citet{konno16} found a more dramatic factor
of $\sim100$ increase in SF luminosity density. The steeper drop
found by Konno et al.\ is caused by their use of the Cowie et al.\
$z=0.3$ LF. We find a $z=2.2$ log $\rho_{Ly\alpha,SF}^{obs}$ of
$39.7\pm0.1$ erg s$^{-1}$Mpc$^{-3}$ based on our fit to the data
from \citet{konno16}. Our $z=2.2$ measurement is consistent with
the value reported in \citet[log $\rho\sim39.8$ erg s$^{-1}$ Mpc$^{-3}$;][]{konno16}.
Our simultaneous fits favor a lower $L^{\star}$ (by 0.3 dex) and
higher $\phi^{\star}$ (by 0.3 dex) at $z=2.2$ than found by \citet[][their result labeled `Best estimate']{konno16}.
Given the number of assumptions in our fitting method, we do not consider
our $L^{\star}$ and $\phi^{\star}$ results to supersede the results
of Konno et al. However, if $z=2.2$ AGNs contribute to the overall
LAE population in a manner similar to our low-redshift sample, then
these proposed offsets may prove to be real. 

Comparing our computed $z=2.2$ SF and AGN luminosity densities, we
estimate an AGN contribution of $\sim40\%$ to the total observed
Ly$\alpha$ luminosity density with a lower-limit estimate of $\sim20\%$
. These results are comparable to our previously computed $z\sim0.3$
AGN contribution estimates of $39\%$ with a lower-limit estimate
of $17\%$  (See Section \ref{sec_lf}). Even if all the low-luminosity
AGNs found in our $z=0.3$ sample disappear at $z=2.2$, a significant
AGN contribution to the total luminosity density is still expected.
For example, integrating the bright-end tail of the AGN LF from $(2.5)L_{{\rm {Ly}\alpha}}^{\star}(z=2.2)=10^{42.8}$
to the observed maximum Ly$\alpha$ luminosity of $10^{44.4}$ erg
s$^{-1}$, we find that $\rho_{Ly\alpha,AGN}=$39.1 erg s$^{-1}$Mpc$^{-3}$
which corresponds to an AGN contribution of $\sim20\%$ to the total
observed Ly$\alpha$ luminosity density. These results suggest tentatively
that the SF and AGN luminosity densities coevolve from $z=0.3$ to
$2.2$ such that star-forming galaxies and AGNs contribute roughly
equally to the observed Ly$\alpha$ light.

As in Section 5 and Figure \ref{comboLF}, we also consider a Saunders
function fit to the data. In Figure\ref{evoLF}(b), we simultaneously
fit a Saunders function + power-law to the $z=0.3$ and $z=2.2$ data.
We find that this alteration does not significantly change the measured
luminosity densities. Our results are summarized in Table \ref{evotable}.
Our main conclusion from these results is that the drop in SF luminosity
density from $z=2.2$ to $0.3$ is not as large as some studies have
previously claimed (though still very large). Additionally, the contribution
of AGNs to the total observed Ly$\alpha$ luminosity density at $z\sim0.3$
is comparable to the contribution from SF galaxies and this trend
appears to continue out to $z=2.2$.  

Although the $z\sim0.9$ data lack sufficient luminosity range to
allow us to simultaneously fit the $z\sim0.9$ SF+AGN LF, in Figure
\ref{nlf} we reproduce the $z\sim0.9$ SF LF from \citet[][cyan curve]{wold14}
to show how the $z\sim0.3$ and $z=2.2$ SF LFs compare, solid black
and red curves, respectively. Taken at face value, the intersection
of the best-fit $z\sim0.9$ Schechter function with the best-fit $z=2.2$
Schechter function implies that SF LAEs more luminous than $\sim2\times10^{43}$
erg s$^{-1}$ are more common at $z\sim0.9$ than at $z=2.2$. Over
the same redshift range, a similar behavior is not observed in H$\alpha$
LFs \citep{sobral13}, and a discordant H$\alpha$ / Ly$\alpha$ LF
evolution is not naively expected because to first order Ly$\alpha$
emitting galaxies will be a subset drawn from H$\alpha$ emitting
galaxies modulo the escape fraction. 

We note that the rate of decline of the Ly$\alpha$ SF LF at high
luminosities is difficult to measure due to the increasing AGN contribution,
and we suspect that the inferred Ly$\alpha$ LF evolution can be explained
by a non-exponential decline in the the bright end of the LF coupled
with the attempted Schechter function fit to a very limited and bright
$z\sim0.9$ Ly$\alpha$ luminosity range. As shown in Figure \ref{nlf},
the $z\sim0.9$ SF LF data-points (cyan diamonds with Poisson error
bars) are relatively flat, and they are not well fit by an exponentially
declining function. However, we find that our best-fit $z\sim0.3$
Saunders function with an $L^{\star}$ boost of 0.45 dex (black dotted
curve) or our best-fit $z=2.2$ Saunders function with a $\phi^{\star}$
decline by $0.8$ dex (red dotted curve) provide a reasonable fit
to the $z\sim0.9$ SF LF. This implies a $z\sim0.9$ log $\rho_{Ly\alpha,SF}$
of approximately $38.9$ erg s$^{-1}$Mpc$^{-3}$ which is $0.5$
dex higher than the SF luminosity density computed from the best-fit
Schechter function (see Table \ref{evotable}). We suggest the the
previous estimate for $\rho_{Ly\alpha,SF}$ at $z\sim0.9$ is likely
biased low, but we cannot completely rule out alternative explanations
such as discordant H$\alpha$ / Ly$\alpha$ LF evolution or undiagnosed
AGNs in the $z\sim0.9$ LAE sample. 
\begin{figure}[!th]
\includegraphics[bb=95bp 150bp 720bp 510bp,clip,width=8.5cm]{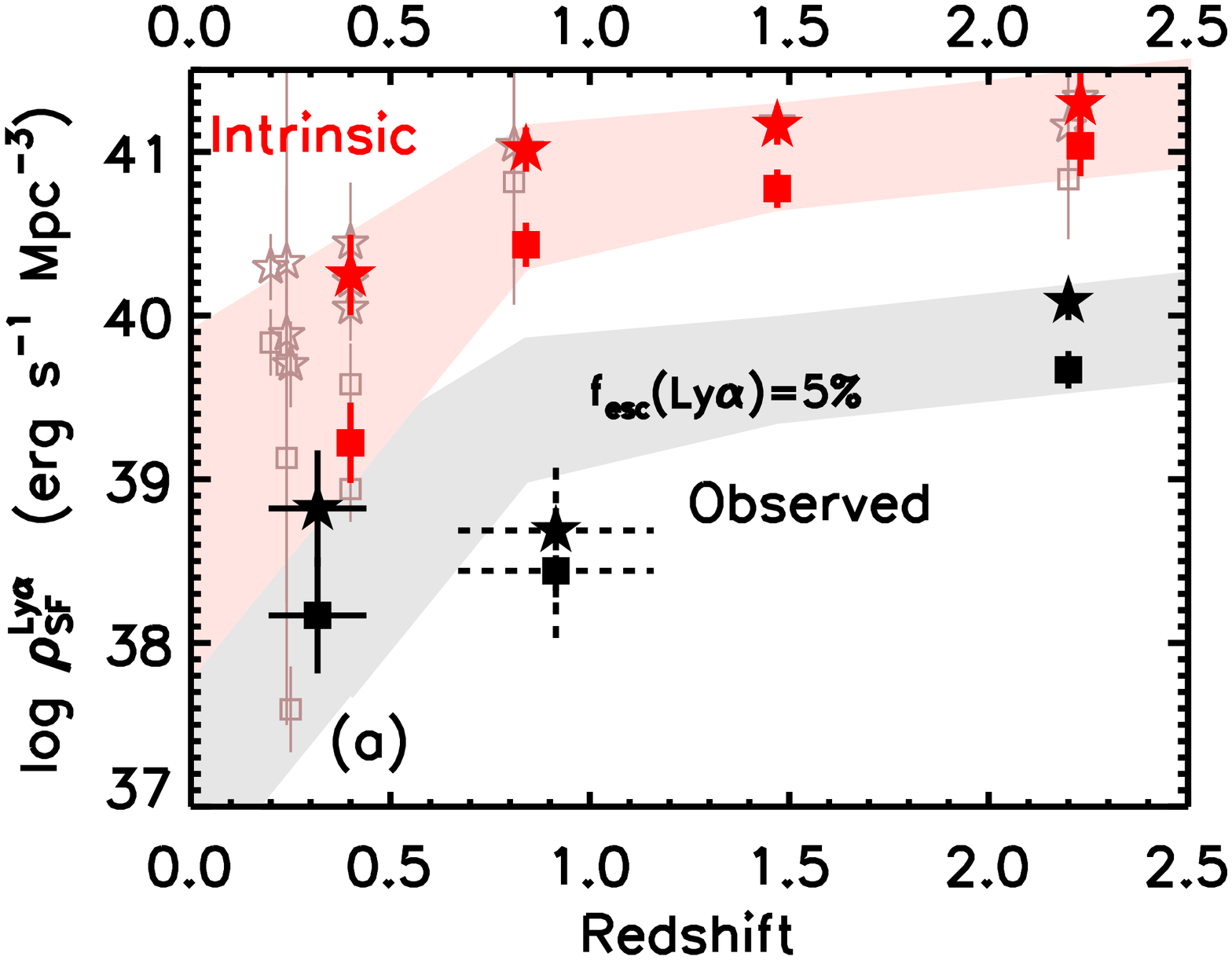}

\includegraphics[bb=95bp 75bp 720bp 520bp,clip,width=8.5cm]{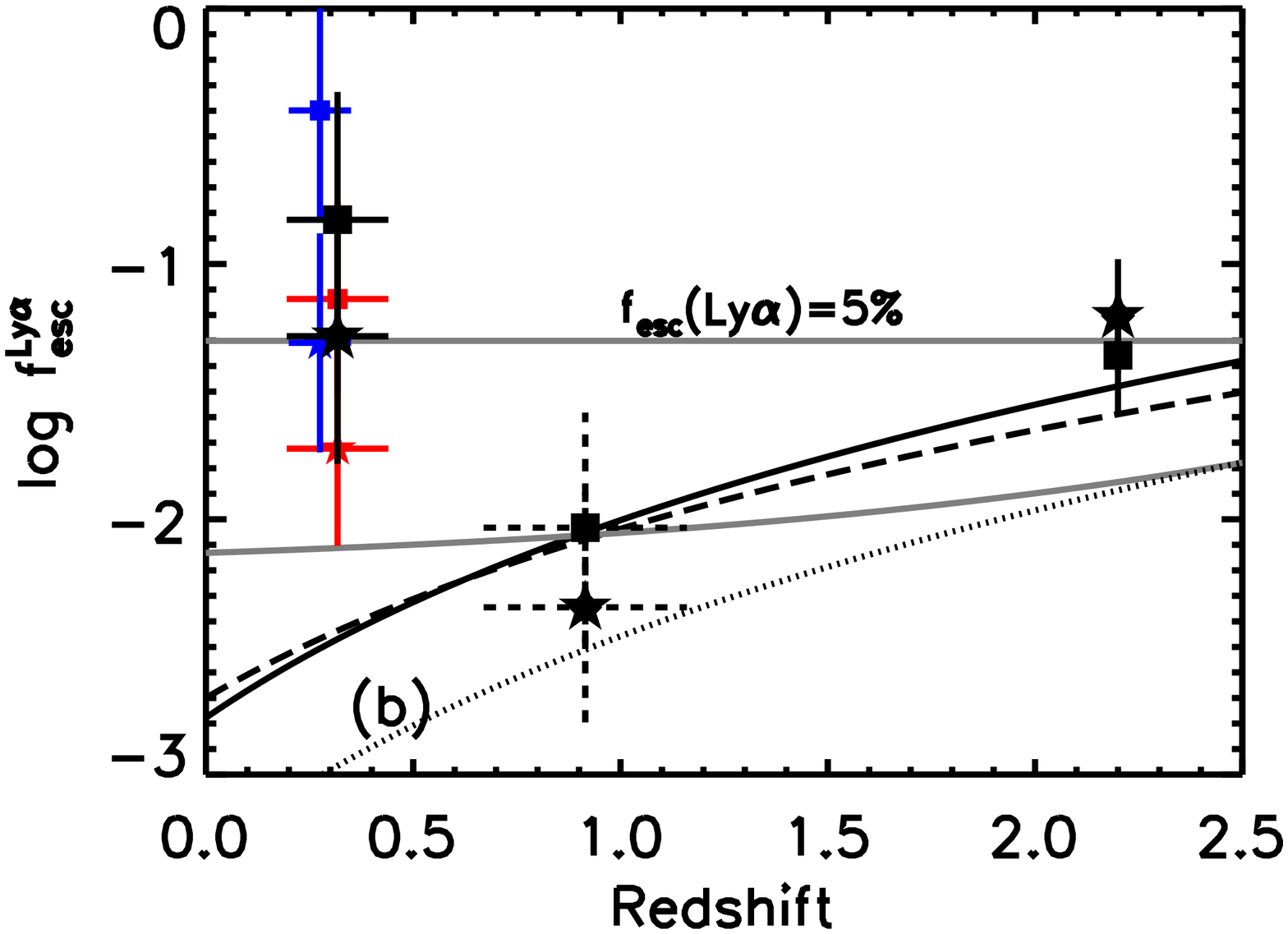}

\caption{\textbf{(a)} The evolution of the observed (black symbols) and intrinsic
(red symbols) Ly$\alpha$ luminosity densities from $0.3<z<2.2$.
Partial luminosity densities are shown with square symbols, while
total luminosity densities are shown with star symbols. The red shaded
region maps out the intrinsic Ly$\alpha$ luminosity density evolution
implied by our fixed $\alpha=-1.6$ fits to Sobral et al.'s H$\alpha$
LFs. The grey region shows this same region offset by -1.3 dex indicating
a Ly$\alpha$ escape fraction of $5\%$. The open shaded square and
star symbols show partial and total intrinsic Ly$\alpha$ luminosity
densities computed from alternative dust-corrected H$\alpha$ LFs
(\citealt{ly07,ly11,shioya08,lee12,drake13,stroe15} and total luminosity
densities from \citealt{sobral13}). \textbf{(b)} The volumetric Ly$\alpha$
escape fraction computed from the ratio of the observed and intrinsic
Ly$\alpha$ luminosity densities. The black star and square symbols
show our measured Ly$\alpha$ escape fractions with full and partial
integration limits, respectively. The red and blue data points show
escape fractions computed from the Ly$\alpha$ studies of \citet{cowie10}
and \citet{deharveng08}, respectively. The solid, dashed, and dotted
black curves are from \citet{hayes11}, \citet{blanc11}, and \citet{konno16}
, respectively, and show their best fit power law to $f_{esc}({\rm {Ly}}\alpha)$
data. The lower gray curve shows the best fit transition curve from
\citet{blanc11}. The gray horizontal line shows a constant escape
fraction of $5\%$. The calculated points and the selected curves
have not been corrected for IGM absorption, which should be small
for all presented $z<2.2$ data. The $z=0.9$ Ly$\alpha$ luminosity
density and escape fraction error bars are dashed because these data
points are less secure as discussed in Section \ref{evo} and Figure
\ref{nlf}.}

\label{ld_evo}

{\footnotesize (A color version of this figure is available in the online journal.)}
\end{figure}

\section{Luminosity Density and Ly$\alpha$ Escape Fraction Evolution}

In Figure \ref{ld_evo} (a), we show the observed and intrinsic Ly$\alpha$
luminosity density evolution from $z=0.3$ to $2.2$. The observed
Ly$\alpha$ luminosity densities are computed from the best-fit Schechter
function parameters reported in our Table \ref{evotable}. The intrinsic
Ly$\alpha$ luminosity densities are computed from dust-corrected
H$\alpha$ LFs from the High-redshift(Z) Emission Line Survey \citep[HiZELS;][]{sobral13}.
HiZELS is a series of narrow-band surveys that produced self-consistent
H$\alpha$ LFs at $z=0.4$, $0.84$, $1.47$, and $2.23$. A constant
$A_{{\rm {H}\alpha}}=1$ magnitude of dust extinction correction is
applied for all four redshifts, where 
\begin{equation}
L_{{\rm {H}\alpha}}=10^{0.4A_{{\rm {H}\alpha}}}L_{{\rm {H}\alpha}}^{{\rm {uncorr}}}.
\end{equation}
For consistency with our study, we chose to independently fit the
dust-corrected H$\alpha$ LF data (Sobral et al.'s Table 4) with Schechter
functions rather than directly using Sobral et al.'s best-fit parameters.
Sobral et al.\ found the faint-end slope of the H$\alpha$ luminosity
function to be $\alpha=-1.60\pm0.08$ with no significant evolution
from $z=0.4$ to $2.2$. Thus, we assume a constant $\alpha=-1.6$
for our Schechter functions fits. We find that altering the faint-end
to a constant $\alpha=-1.75$, which is consistent with our Ly$\alpha$
LF fits, does not significantly change our results. As prescribed
by Sobral et al., we make a 10 to 15$\%$ correction to our H$\alpha$
luminosity densities to account for any AGN contribution (see their
Section 4.1). To convert from H$\alpha$ to intrinsic Ly$\alpha$
luminosity, we assume the typical case B recombination ratio of 8.7.
In Figure \ref{ld_evo} (a), we also show intrinsic Ly$\alpha$ luminosity
densities from other dust-corrected H$\alpha$ LFs obtained from the
literature. For consistency across both H$\alpha$ and Ly$\alpha$
surveys, we estimate the luminosity density errors by adding in quadrature
the reported 1$\sigma$ errors in the best fit Schechter function
parameters $L^{\star}$ and $\phi^{\star}$.

As in the previous section, for the Ly$\alpha$ luminosity density
calculations we adopt Konno et al.'s lower integration limit of $L=10^{41.41}$
erg s$^{-1}$ which corresponds to $10\%$ of our best fit $L_{{\rm {Ly}\alpha}}^{\star}$
at $z=2.2$ (see Table \ref{evotable}). We adopt a consistent H$\alpha$
lower integration limit of $(0.10)L_{{\rm {H}\alpha}}^{\star}(z=2.23)=10^{41.9}$
erg s$^{-1}$. Thus, the intrinsic Ly$\alpha$ luminosity densities
(square red symbols) are computed with integration limits from $(0.10)L_{{\rm {H}\alpha}}^{\star}(z=2.23)=10^{41.9}$
erg s$^{-1}$ to infinity, and the observed Ly$\alpha$ luminosity
densities (square black symbols) are computed with integration limits
from $(0.10)L_{{\rm {Ly}\alpha}}^{\star}(z=2.2)=10^{41.41}$ erg s$^{-1}$
(see Table \ref{evotable}) to infinity. 

While our adopted lower integration limits are roughly consistent
with the values used by previous studies \citep[e.g., see][]{hayes11,blanc11,konno16},
these values are somewhat arbitrary and a source of systematic uncertainty.
In particular, we find that the convention of fixing the lower integration
limit to a percentage of $L_{{\rm }}^{\star}$ at high-redshift can
contribute to large variations in the computed luminosity densities
at low-redshift. Moving from $z=2.2$ to $0$, $L_{{\rm }}^{\star}$
declines rapidly and if the chosen integration limit approaches $L_{{\rm }}^{\star}$,
then relatively small differences in the best-fit $L_{{\rm }}^{\star}$
between studies can result in very different luminosity density measures.
Integrating down to zero removes this effect but makes our calculation
more sensitive to the assumed faint-end slope, but given reasonable
$\alpha$ values  we do not expect our total luminosity density calculations
to be altered by more than a factor of 2. Consistent with this expectation,
we find that the variation seen between H$\alpha$ studies in partial
luminosity densities (red square symbols) is significantly larger
than the variation in total luminosity densities (red star symbols).
Given these issues, we consider our total luminosity density measurements
to be more reliable when evaluating evolutionary trends, and unless
otherwise noted, we use total luminosity densities in the following
discussion.

In Figure \ref{ld_evo} (a), the red shaded region maps out the intrinsic
Ly$\alpha$ luminosity density evolution. The grey region shows this
same region offset by -1.3 dex, which corresponds to a Ly$\alpha$
escape fraction of $5\%$. We find that the decline in observed Ly$\alpha$
luminosity density from $z=2.2$ to $z=0.3$ may simply mirror the
decline seen in the intrinsic Ly$\alpha$ luminosity density and hence
the H$\alpha$ luminosity density. At $z=0.9$, the observed Ly$\alpha$
luminosity density may dip relative to the intrinsic Ly$\alpha$ luminosity
density, but this data point is less secure because the luminosity
data covers a smaller dynamical range and is limited to the bright
end of the LF (see Figure \ref{evoLF}). 

The volumetric Ly$\alpha$ escape fraction is a measure of the fraction
of Ly$\alpha$ photons that escape from the survey volume. It is defined
as the ratio of the observed and intrinsic Ly$\alpha$ luminosity
densities:

\begin{equation}
f_{esc}^{{\rm {Ly}\alpha}}=\frac{\rho_{Ly\alpha,SF}^{obs}}{\rho_{Ly\alpha,SF}^{int}}=\frac{\rho_{Ly\alpha,SF}^{obs}}{8.7\times\rho_{H\alpha,SF}^{int}}.
\end{equation}

Many groups have studied the redshift evolution of this quantity and
concluded that the volumetric Ly$\alpha$ escape fraction increases
rapidly with redshift until $z=7$ at which point the escape fraction
drops, which is typically attributed to the increasing opacity of
the IGM and the onset of reionization \citep{hayes11,blanc11,konno16}.
For this study, we are concerned with the previously claimed rapid
decline in $f_{esc}^{{\rm {Ly}\alpha}}$ at low redshifts where the
intervening IGM will have a negligible effect. Various explanations
for the low-redshift decline have been proposed including increasing
dust content and increasing neutral column density of star-forming
galaxies. We can now better constrain the escape fraction by including
our robust low-redshift constraint and by making a more direct comparison
to H$\alpha$ results that are now self-consistently measured out
to a redshift of $z=2.23$. Beyond this redshift, H$\alpha$ surveys
are not currently available and the intrinsic Ly$\alpha$ luminosities
must be estimated from the dust-corrected UV luminosity functions,
which require large corrections for extinction and are dependent on
the assumed initial mass function, metallicity, and star formation
history. We also have the advantage of obtaining all of our H$\alpha$
constraints from a single study \citep{sobral13}. This ensures consistency
in the employed data reduction and LF incompleteness corrections.
We have also assumed the same standard $H_{0}=70$ km s$^{-1}$ Mpc$^{-1}$,
$\Omega_{M}=0.3$, and $\Omega_{\Lambda}=0.7$ cosmology.

In Figure \ref{ld_evo} (b), we show the evolution of the volumetric
Ly$\alpha$ escape fraction inferred from the intrinsic and observed
Ly$\alpha$ luminosity densities presented in Figure \ref{ld_evo}
(a). Comparing our computed $z=0.3$ and $z=2.2$ Ly$\alpha$ escape
fractions, we find results that are consistent with a relatively constant
$f_{esc}^{{\rm {Ly}\alpha}}$ with a value of $\sim10^{-1.3}$ or
$\sim5\%$.  The $z=0.9$ data-point may suggest a dip in the escape
fraction, but as discussed in Section \ref{evo} this data point is
less secure. Although our 1$\sigma$ error bars are quite large, we
find that the existing low-redshift observational constraints do
not provide convincing evidence for rapidly declining $f_{esc}^{{\rm {Ly}\alpha}}$
with decreasing redshift at $0.3<z<2.2$. We emphasize that our results
are not inconsistent with an evolving $f_{esc}^{{\rm {Ly}\alpha}}$
at $z>2$. For example, \citet{blanc11} found that the overall $0.3<z<7.7$
evolution of $f_{esc}^{{\rm {Ly}\alpha}}$ can be described by a function
that levels off at both low-redshift (with $f_{esc}^{{\rm {Ly}\alpha}}(z=0.3)\sim0.01$)
and high-redshift (with $f_{esc}^{{\rm {Ly}\alpha}}(z\sim6)\sim0.80$)
with a transition between these two extremes at $z\sim4$. This transitional
function was motivated by the observed evolution of dust extinction
derived from the UV slope of continuum selected galaxies \citep{bouwens09}.
Blanc et al.'s best-fit transitional function is shown as a grey curve
in Figure \ref{ld_evo} (b). While our study favors a higher normalization
at low redshifts, the relatively constant $f_{esc}^{{\rm {Ly}\alpha}}$
from $z=0.3$ to $z=2.2$ is consistent with our results.

We find that the difference between our study and previous results
\citep[e.g.,][]{hayes11,konno16} suggesting a factor of 10 decline
in $f_{esc}^{{\rm {Ly}\alpha}}$ from $z=2.2-0.3$ cannot be solely
attributed to the numerator, our new $z=0.3$ Ly$\alpha$ LF measurement.
Our LF when compared to Cowie et al.'s Ly$\alpha$ LF can only account
for a factor of $3$ boost to the $f_{esc}^{{\rm {Ly}\alpha}}(z=0.3)$.
To investigate whether the denominator, our $\rho_{Ly\alpha,SF}^{int}(z=0.3)$
measure, is reliable, we compiled measurements from low-redshift dust-corrected
H$\alpha$ LFs from the literature (see Figure \ref{ld_evo} (a)).
Overall, we find that our utilized intrinsic luminosity density is
not an outlier when compared to other measurements. If we adopt the
highest $\rho_{Ly\alpha,SF}^{int}(z=0.4)=10^{40.48}$erg s$^{-1}$Mpc$^{-3}$
estimate, which is roughly consistent with the UV derived value used
by \citet[]{konno16}, we can reduce our $f_{esc}^{{\rm {Ly}\alpha}}(z=0.3)$
measure by an additional factor of 2. We find that other small differences
between $f_{esc}^{{\rm {Ly}\alpha}}$ studies are explained by the
assumed integration limits, Ly$\alpha$ EW cuts, and best-fit Schechter
parameters. 

Overall we consider our results that show a relatively constant $f_{esc}^{{\rm {Ly}\alpha}}=0.05$
from $z=0.3-2.2$ to be more reliable because our study has the advantage
of a robust $\rho_{Ly\alpha,SF}^{obs}(z=0.3)$ estimate and uniform
H$\alpha$ constraints. At the very least, we have shown that the
existing low-redshift observational constraints do not provide clear-cut
evidence for rapidly evolving $f_{esc}^{{\rm {Ly}\alpha}}$ at $z<2$.
\begin{figure}[!th]
\includegraphics[bb=95bp 70bp 720bp 510bp,clip,width=8.5cm]{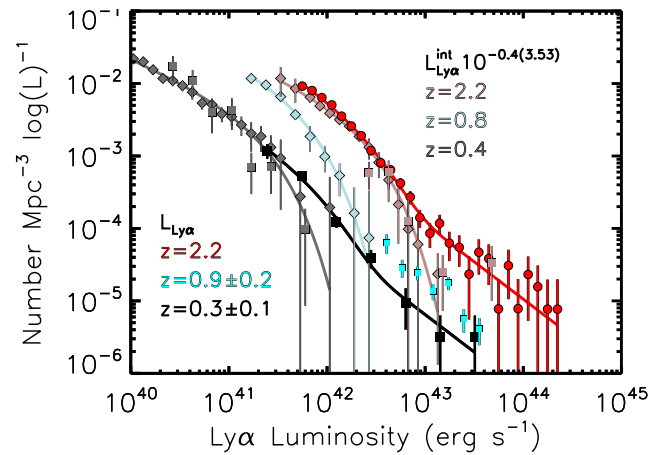}

\caption{The observed Ly$\alpha$ luminosity functions (circle symbols) compared
to the intrinsic Ly$\alpha$ luminosity functions (shaded diamond
and square symbols from \citealt{sobral13} and \citealt{matthee17},
respectively) with a constant $A_{{\rm {Ly}\alpha}}=3.53$ magnitude
of dust extinction applied. Our best-fit H$\alpha$ Schechter functions
and our best-fit Ly$\alpha$ Schechter plus power-law functions are
shown over the extent of their data-points.}

\label{lf_evo}

{\footnotesize (A color version of this figure is available in the online journal.)}
\end{figure}

A constant $f_{esc}^{{\rm {Ly}\alpha}}=0.05$ measurement is roughly
consistent with expectations given the assumed constant $A_{{\rm {H}\alpha}}=1$
magnitude of dust extinction. \citet{sobral13} argue that past H$\alpha$
studies typically find $A_{{\rm {H}\alpha}}=1\pm0.2$ with no clear
redshift evolution. For these reasons, a simple 1 magnitude of H$\alpha$
extinction is corrected for in Sobral et al.\ and in this study.
Assuming a \citet{calzetti00} dust law and $R_{V}=4.1$, a magnitude
of dust extinction at $\lambda6563$ implies $A_{{\rm {Ly}\alpha}}=3.53$
mag. In Figure \ref{lf_evo}, we show the intrinsic Ly$\alpha$ LFs
at $z=0.4$, $0.8$, and $2.2$ with a constant 3.53 magnitudes of
extinction applied. This is equivalent to multiplying the intrinsic
Ly$\alpha$ $L$$^{\star}$ values by a factor of $0.04$ and is consistent
with our proposed constant $\sim5\%$ Ly$\alpha$ escape fraction.
Given this very simple assumption the agreement between observed Ly$\alpha$
LFs and intrinsic Ly$\alpha$ LFs with extinction applied is encouraging.
Particularly, the agreement is notable for the two narrow-band $z=2.2$
LFs where the bright-end AGN tail becomes dominant at L$_{{\rm {Ly}\alpha}}\gtrsim2\times10^{43}$erg
s$^{-1}$ in both cases. This scenario implies that within the LAE
population the average Ly$\alpha$ photon encounters the same amount
of dust opacity as H$\alpha$ photons. This has been previously suggested
by studies that examined the relation between the Ly$\alpha$ escape
fraction and the dust extinction for samples of LAEs (e.g., \citealt[][]{cowie11,blanc11}).
If dust extinction is the main driver of Ly$\alpha$ escape, then
this may also help to explain the non-evolution of the Ly$\alpha$
EW scale length since similar to the Ly$\alpha$ escape fraction the
EW is also governed by H{\small{}I} scattering and dust absorption,
but complicating the interpretation, EWs will also depend on the star
formation history and metallicity of the host galaxy. While more
complex scenarios cannot be ruled out, we find that the simplest explanation
for the lack of evolution observed in the Ly$\alpha$ escape fraction
(and perhaps the EW scale length) from $z=0.3$ to $2.2$ is a relatively
constant dust extinction over this same redshift range.

\section{Summary}

Previous studies have suggested that at low redshifts high-EW LAEs
become less prevalent and that the amount of Ly$\alpha$ emission
able to escape (as measured by $f_{esc}^{{\rm {Ly}\alpha}}$) declines
rapidly. A number of explanations for these trends have been suggested
including increasing dust content, increasing neutral column density,
and/or increasing metallicity of star-forming galaxies at lower redshifts.
In this paper we presented the first local sample of LAEs selected
based solely on their Ly$\alpha$ emission and showed that the dramatic
decline previously suggested in the Ly$\alpha$ EW distribution scale
length and volumetric Ly$\alpha$ escape fraction from $z=2.2$ to
$0.3$ becomes less convincing when local LAEs are selected in manner
similar to high-redshift LAEs. Our results are consistent with these
quantities not evolving, despite the intrinsic Ly$\alpha$ luminosity
(as probed by $L_{{\rm {H}\alpha}}^{\star}$) plummeting by an order
of magnitude from $z=2.2$ to $0.4$ \citep{sobral13}. This may imply
that the physical conditions that allow strong Ly$\alpha$ emission
are present at both low and high redshifts, or that changing conditions
conspire make no apparent evolutionary trend. We show that the current
Ly$\alpha$ and H$\alpha$ LFs are surprisingly consistent with a
simple scenario in which dust extinction is relatively constant and
is the main driver of Ly$\alpha$ escape. Finally, our work finds
that AGNs contribute significantly to the total Ly$\alpha$ luminosity
density, and we find evidence that this holds true out to a redshift
of $z=2.2$. We emphasize that larger and more sensitive LAE surveys
are needed to further constrain the evolution of the EW distribution
scale length and volumetric Ly$\alpha$ escape fraction. The limited
facilities currently available in the ultraviolet prevent significant
improvement below a redshift of $z\sim2$. However, the HETDEX survey
which will detect close to one million LAEs will resolve whether these
quantities evolve from a redshift of $z=2-3.5$.

\acknowledgements{We would like to thank the referee for their critical reading of
the paper and useful suggestions for improving it. The authors wish
to thank David Sobral for insightful discussions. This work was supported
by a NASA Keck PI Data Award, administered by the NASA Exoplanet Science
Institute. Data presented herein were obtained at the W. M. Keck Observatory
from telescope time allocated to the National Aeronautics and Space
Administration through the agency's scientific partnership with the
California Institute of Technology and the University of California.
The Observatory was made possible by the generous financial support
of the W. M. Keck Foundation. The authors wish to recognize and acknowledge
the very significant cultural role and reverence that the summit of
Mauna Kea has always had within the indigenous Hawaiian community.
We are most fortunate to have the opportunity to conduct observations
from this mountain. IGBW and SLF acknowledge support from the NSF
AAG award AST-1518183. IGBW, AJB, and LLC acknowledge support from
the Munich Institute for Astro- and Particle Physics (MIAPP) of the
Deutsche Forschungsgemeinschaft (DFG) cluster of excellence \textquotedblleft Origin
and Structure of the Universe''. AJB acknowledges support from NASA
ADAP grant NNX14AJ66G, the John Simon Guggenheim Memorial Foundation,
and the Trustees of the William F. Vilas Estate. }

\bibliographystyle{apj} 
\bibliography{wold}
\clearpage \begin{landscape} 

\begin{deluxetable}{ccccccccccccccc} \tabletypesize{\scriptsize} \setlength{\tabcolsep}{0.02in} \tablecolumns{15} \tablewidth{0pc} \tablecaption{Emission-line Sample: CDFS-00} \tablehead{ \colhead{Num} & \colhead{Name}    &  \colhead{R.A.} &   \colhead{Decl.} & \colhead{FUV} & \colhead{NUV} & \colhead{$z^{c}_{\rm{galex}}$} & \colhead{log $L$(Ly$\alpha$)} & \colhead{EW$_{\rm{r}}$(Ly$\alpha$)}  & \colhead{R.A.(opt)} &  \colhead{Decl.(opt)} & \colhead{Offset} & \colhead{$z_{\rm{opt}}$} & \colhead{log $L^{d}_{2-8~\rm{keV}}$} & \colhead{Class}\\ \colhead{} & \colhead{} & \colhead{(J2000.0)}   & \colhead{(J2000.0)} & \colhead{(AB)} & \colhead{(AB)} & \colhead{} & \colhead{(erg s$^{-1}$)} & \colhead{(\AA)} & \colhead{(J2000.0)}   & \colhead{(J2000.0)} & \colhead{(arcsec)} & \colhead{} & \colhead{(erg s$^{-1}$)} & \colhead{}\\ \colhead{(1)} & \colhead{(2)} & \colhead{(3)} & \colhead{(4)} & \colhead{(5)} & \colhead{(6)} & \colhead{(7)}  & \colhead{(8)}  & \colhead{(9)} & \colhead{(10)} & \colhead{(11)} & \colhead{(12)} & \colhead{(13)} & \colhead{(14)} & \colhead{(15)}} \startdata       1 & GALEX033102-275130\tablenotemark{p} &  52.760130 & -27.858466 & 21.97 & 21.50 & 0.340\tablenotemark{2} & 41.95 & 24$\pm$3 &  52.760235 & -27.858490 & 0.3 & 0.335\tablenotemark{e} & \nodata\phm{f} & \nodata\\       2 & GALEX033108-274214\phn &  52.785524 & -27.704094 & 23.03 & 23.44 & 0.230\tablenotemark{2} & 41.58 & 92$\pm$37 &  52.785574 & -27.704395 & 1.1 & \nodata\phm{d} & \nodata\phm{f} & \nodata\\       3 & GALEX033111-275258\tablenotemark{p} &  52.797118 & -27.882930 & 21.24 & 20.62 & 0.267\tablenotemark{1} & 41.54 & 8$\pm$2 &  52.797794 & -27.882475 & 2.7 & 0.265\tablenotemark{e} & \nodata\phm{f} & w\\       4 & GALEX033112-274801\tablenotemark{p} &  52.800331 & -27.800321 & 22.24 & 21.96 & 0.260\tablenotemark{2} & 41.43 & 18$\pm$6 &  52.800362 & -27.800318 & 0.1 & 0.258\tablenotemark{e} & \nodata\phm{f} & \nodata\\       5 & GALEX033115-274953\phn &  52.814216 & -27.831617 & 21.20 & 21.52 & 0.184\tablenotemark{2} & 42.17 & 132$\pm$53 &  52.814224 & -27.831604 & 0.1 & 0.182\tablenotemark{f} & -999\phm{f} & \nodata\\       6 & GALEX033120-275449\phn &  52.837457 & -27.913616 & 23.62 & 24.26 & 0.289\tablenotemark{2} & 41.97 & 346$\pm$79 &  52.836938 & -27.913108 & 2.5 & 0.290\tablenotemark{f} & -999\phm{f} & \nodata\\       7 & GALEX033129-273449\phn &  52.873386 & -27.580340 & 21.47 & 21.07 & 0.303\tablenotemark{2} & 41.48 & 7$\pm$2 &  52.872960 & -27.580422 & 1.4 & 0.297\tablenotemark{f} & -999\phm{f} & \nodata\\       8 & GALEX033132-273803\phn &  52.884050 & -27.634405 & 21.95 & 21.73 & 0.253\tablenotemark{3} &   &   &  52.884105 & -27.634381 & 0.2 & (0.088)\tablenotemark{f} & -999\phm{f} & \nodata\\       9 & GALEX033132-275007\phn &  52.885785 & -27.835416 & 21.02 & 21.01 & 0.183\tablenotemark{2} & 41.69 & 26$\pm$3 &  52.885924 & -27.835377 & 0.5 & 0.180\tablenotemark{f} & -999\phm{f} & \nodata\\      10 & GALEX033137-273844\phn &  52.906576 & -27.645563 & 23.24 & 22.80 & 0.198\tablenotemark{3} & 41.36 & 97$\pm$17 &  52.907163 & -27.645389 & 2.0 & 0.195\tablenotemark{f} & 41.4\phm{f} & \nodata\\      11 & GALEX033143-281145\tablenotemark{p} &  52.930008 & -28.196035 & 22.14 & 21.91 & 0.243\tablenotemark{3} & 41.41 & 19$\pm$7 &  52.930058 & -28.196028 & 0.2 & 0.242\tablenotemark{e} & \nodata\phm{f} & \nodata\\      12 & GALEX033145-281038\phn &  52.938978 & -28.177395 & 22.05 & 23.60 & 0.218\tablenotemark{3} & 41.88 & 85$\pm$9 &  52.942211 & -28.180876 & 16.2 & 0.216\tablenotemark{e} & \nodata\phm{f} & \nodata\\      13 & GALEX033145-274615\phn &  52.940221 & -27.770865 & 24.83 & 25.64 & 0.249\tablenotemark{3} & 41.27 & 359$\pm$301 &  52.939929 & -27.770922 & 1.0 & 0.247\tablenotemark{h} & -999\phm{f} & a\\      14 & GALEX033147-280812\phn &  52.946718 & -28.136678 & 22.98 & 23.14 & 0.286\tablenotemark{1} & 41.96 & 129$\pm$54 &  52.946629 & -28.136681 & 0.3 & 0.283\tablenotemark{e} & \nodata\phm{f} & \nodata\\      15 & GALEX033148-273946\phn &  52.950147 & -27.663035 & 23.44 & 23.53 & 0.257\tablenotemark{3} & 41.60 & 108$\pm$47 &  52.950225 & -27.662988 & 0.3 & 0.259\tablenotemark{f} & -999\phm{f} & \nodata\\      16 & GALEX033150-280811\phn &  52.958425 & -28.136512 & 22.74 & 22.13 & 0.219\tablenotemark{2} & 42.24 & 1110$\pm$381 &  52.957666 & -28.135944 & 3.2 & 0.219\tablenotemark{g} & \nodata\phm{f} & n\\      17 & GALEX033150-281120\tablenotemark{p} &  52.962210 & -28.189010 & 21.01 & 20.88 & 0.216\tablenotemark{1} & 41.74 & 19$\pm$6 &  52.962204 & -28.189028 & 0.1 & 0.213\tablenotemark{e} & \nodata\phm{f} & \nodata\\      18 & GALEX033154-281419\tablenotemark{p} &  52.976513 & -28.238641 & 21.75 & 21.64 & 0.282\tablenotemark{1} & 42.21 & 67$\pm$23 &  52.976501 & -28.238640 & 0.0 & 0.280\tablenotemark{e} & \nodata\phm{f} & \nodata\\      19 & GALEX033154-281409\tablenotemark{p} &  52.978447 & -28.235871 & 21.91 & 21.27 & 0.319\tablenotemark{1} & 41.94 & 26$\pm$9 &  52.978458 & -28.235889 & 0.1 & 0.316\tablenotemark{e} & \nodata\phm{f} & \nodata\\      20 & GALEX033155-281245\phn &  52.979571 & -28.212720 & 22.17 & 22.04 & 0.329\tablenotemark{2} & 42.05 & 42$\pm$14 &  52.979557 & -28.212870 & 0.5 & 0.326\tablenotemark{e} & \nodata\phm{f} & \nodata\\      21 & GALEX033159-280951\tablenotemark{p} &  52.999330 & -28.164434 & 21.25 & 21.15 & 0.239\tablenotemark{1} & 41.72 & 18$\pm$6 &  52.999458 & -28.164722 & 1.1 & 0.236\tablenotemark{e} & \nodata\phm{f} & \nodata\\      22 & GALEX033200-281057\tablenotemark{p} &  53.001047 & -28.182500 & 22.46 & 21.97 & 0.279\tablenotemark{2} & 41.81 & 47$\pm$6 &  53.002308 & -28.182331 & 4.0 & 0.278\tablenotemark{e} & \nodata\phm{f} & \nodata\\      23 & GALEX033202-281112\phn &  53.010433 & -28.186867 & 23.37 & 24.22 & 0.258\tablenotemark{2} & 41.65 & 105$\pm$44 &  53.010105 & -28.187319 & 1.9 & 0.257\tablenotemark{e} & \nodata\phm{f} & \nodata\\      24 & GALEX033204-280429\phn &  53.018110 & -28.074785 & 22.72 & 22.32 & 0.282\tablenotemark{2} & 41.67 & 42$\pm$15 &  53.018246 & -28.075474 & 2.5 & 0.280\tablenotemark{e} & 42.6\phm{f} & xn\\      25 & GALEX033211-280911\tablenotemark{p} &  53.049368 & -28.153184 & 21.09 & 20.49 & 0.239\tablenotemark{2} & 41.44 & 7$\pm$2 &  53.049267 & -28.152910 & 1.0 & 0.237\tablenotemark{e} & \nodata\phm{f} & \nodata\\      26 & GALEX033211-280130\tablenotemark{p} &  53.049749 & -28.025048 & 21.54 & 21.51 & 0.218\tablenotemark{1} & 41.88 & 52$\pm$18 &  53.049732 & -28.025000 & 0.2 & 0.215\tablenotemark{e} & -999\phm{f} & \nodata\\      27 & GALEX033213-280405\phn &  53.056197 & -28.068263 & 23.80 & 22.55 & 0.298\tablenotemark{3} & 41.71 & 133$\pm$25 &  53.056889 & -28.068632 & 2.6 & 0.302\tablenotemark{e} & -999\phm{f} & a\\      28 & GALEX033214-273102\phn &  53.059095 & -27.517397 & 22.92 & 22.53 & 0.298\tablenotemark{2} & 41.70 & 52$\pm$9 &  53.059971 & -27.517139 & 2.9 & 0.287\tablenotemark{e} & \nodata\phm{f} & n\\      29 & GALEX033214-281111\tablenotemark{p} &  53.061569 & -28.186519 & 21.71 & 21.18 & 0.263\tablenotemark{1} & 41.71 & 21$\pm$2 &  53.061661 & -28.186567 & 0.3 & 0.261\tablenotemark{e} & \nodata\phm{f} & n\\      30 & GALEX033216-281308\phn &  53.067724 & -28.219051 & 22.51 & 22.05 & 0.278\tablenotemark{3} & 41.84 & 55$\pm$7 &  53.068790 & -28.219053 & 3.4 & 0.277\tablenotemark{e} & \nodata\phm{f} & \nodata\\      31 & GALEX033218-281320\tablenotemark{p} &  53.078007 & -28.222441 & 21.54 & 21.16 & 0.283\tablenotemark{1} & 42.37 & 82$\pm$29 &  53.078033 & -28.222441 & 0.1 & 0.279\tablenotemark{e} & \nodata\phm{f} & \nodata\\      32 & GALEX033219-274122\phn &  53.082533 & -27.689716 & 23.52 & 23.56 & 0.229\tablenotemark{2} & 41.44 & 108$\pm$19 &  53.082588 & -27.689594 & 0.5 & 0.227\tablenotemark{e} & 41.0\tablenotemark{j}\phm{f} & b\\      33 & GALEX033221-273044\phn &  53.088036 & -27.512391 & 22.75 & 22.83 & 0.250\tablenotemark{2} & 41.62 & 58$\pm$21 &  53.088074 & -27.512411 & 0.1 & 0.248\tablenotemark{e} & \nodata\phm{f} & \nodata\\      34 & GALEX033221-275602\phn &  53.088287 & -27.934010 & 22.95 & 22.75 & 0.239\tablenotemark{3} & 41.31 & 34$\pm$13 &  53.088452 & -27.934212 & 0.9 & 0.237\tablenotemark{f} & -999\phm{f} & \nodata\\      35 & GALEX033221-273528\phn &  53.091494 & -27.591373 & 23.62 & 24.11 & 0.245\tablenotemark{2} & 41.61 & 197$\pm$105 &  53.091785 & -27.591337 & 0.9 & 0.242\tablenotemark{e} & -999\phm{f} & \nodata\\      36 & GALEX033225-275857\phn &  53.107151 & -27.982616 & 23.25 & 23.55 & 0.279\tablenotemark{2} & 41.86 & 158$\pm$70 &  53.107237 & -27.982728 & 0.5 & 0.278\tablenotemark{h} & -999\phm{f} & \nodata\\      37 & GALEX033225-272956\phn &  53.107671 & -27.499077 & 23.05 & 22.82 & 0.248\tablenotemark{2} & 41.48 & 55$\pm$21 &  53.107868 & -27.499100 & 0.6 & 0.245\tablenotemark{e} & \nodata\phm{f} & \nodata\\      38 & GALEX033232-275705\phn &  53.134436 & -27.951597 & 22.55 & 22.58 & 0.233\tablenotemark{2} & 41.39 & 29$\pm$10 &  53.134488 & -27.951649 & 0.2 & 0.233\tablenotemark{e} & -999\phm{f} & \nodata\\      39 & GALEX033234-274707\phn &  53.144657 & -27.785525 & 23.55 & 23.38 & 0.248\tablenotemark{3} & 41.25 & 51$\pm$21 &  53.144683 & -27.785447 & 0.3 & 0.247\tablenotemark{e} & 39.4\tablenotemark{k}\phm{f} & \nodata\\      40 & GALEX033235-273630\phn &  53.148298 & -27.608605 & 22.92 & 22.53 & 0.370\tablenotemark{3} &   &   &  53.146029 & -27.607698 & 7.9 & (0.251)\tablenotemark{f} & -999\phm{f} & \nodata\\      41 & GALEX033236-281038\tablenotemark{p} &  53.154145 & -28.177407 & 21.32 & 21.01 & 0.207\tablenotemark{1} & 41.75 & 27$\pm$3 &  53.155334 & -28.177498 & 3.8 & 0.204\tablenotemark{e} & \nodata\phm{f} & \nodata\\      42 & GALEX033238-272946\phn &  53.159236 & -27.496269 & 23.21 & 23.47 & 0.214\tablenotemark{2} & 41.20 & 45$\pm$18 &  53.158962 & -27.495813 & 1.9 & 0.211\tablenotemark{e} & \nodata\phm{f} & \nodata\\      43 & GALEX033241-273620\phn &  53.171766 & -27.605734 & 22.32 & 22.37 & 0.260\tablenotemark{1} & 41.88 & 65$\pm$23 &  53.171760 & -27.605750 & 0.1 & 0.257\tablenotemark{f} & -999\phm{f} & \nodata\\      44 & GALEX033241-281125\tablenotemark{p} &  53.174262 & -28.190311 & 20.03 & 20.26 & 0.207\tablenotemark{1} & 42.29 & 30$\pm$10 &  53.174255 & -28.190306 & 0.0 & 0.204\tablenotemark{e} & \nodata\phm{f} & \nodata\\      45 & GALEX033244-275139\phn &  53.184739 & -27.860948 & 22.75 & 22.26 & 0.280\tablenotemark{2} & 41.69 & 51$\pm$8 &  53.184493 & -27.861438 & 1.9 & 0.273\tablenotemark{h} & 40.8\tablenotemark{j}\phm{f} & n\\      46 & GALEX033244-274514\phn &  53.185187 & -27.754043 & 23.61 & 24.25 & 0.263\tablenotemark{2} & 41.62 & 145$\pm$66 &  53.185311 & -27.753926 & 0.6 & 0.260\tablenotemark{h} & -999\phm{f} & \nodata\\      47 & GALEX033246-274714\tablenotemark{p} &  53.195124 & -27.787351 & 21.97 & 22.12 & 0.228\tablenotemark{1} & 41.85 & 56$\pm$3 &  53.195644 & -27.787776 & 2.3 & 0.226\tablenotemark{e} & -999\phm{f} & \nodata\\ \enddata \tablecomments{ }\end{deluxetable}
\clearpage \addtocounter{table}{-1} \begin{deluxetable}{ccccccccccccccc} \tabletypesize{\scriptsize} \setlength{\tabcolsep}{0.02in} \tablecolumns{15} \tablewidth{0pc} \tablecaption{Emission-line Sample: CDFS-00 (Continued)} \tablehead{ \colhead{Num} & \colhead{Name}    &  \colhead{R.A.} &   \colhead{Decl.} & \colhead{FUV} & \colhead{NUV} & \colhead{$z^{c}_{\rm{galex}}$} & \colhead{log $L$(Ly$\alpha$)} & \colhead{EW$_{\rm{r}}$(Ly$\alpha$)}  & \colhead{R.A.(opt)} &  \colhead{Decl.(opt)} & \colhead{Offset} & \colhead{$z_{\rm{opt}}$} & \colhead{log $L^{d}_{2-8~\rm{keV}}$} & \colhead{Class}\\ \colhead{} & \colhead{} & \colhead{(J2000.0)}   & \colhead{(J2000.0)} & \colhead{(AB)} & \colhead{(AB)} & \colhead{} & \colhead{(erg s$^{-1}$)} & \colhead{(\AA)} & \colhead{(J2000.0)}   & \colhead{(J2000.0)} & \colhead{(arcsec)} & \colhead{} & \colhead{(erg s$^{-1}$)} & \colhead{}\\ \colhead{(1)} & \colhead{(2)} & \colhead{(3)} & \colhead{(4)} & \colhead{(5)} & \colhead{(6)} & \colhead{(7)}  & \colhead{(8)}  & \colhead{(9)} & \colhead{(10)} & \colhead{(11)} & \colhead{(12)} & \colhead{(13)} & \colhead{(14)} & \colhead{(15)}} \startdata      48 & GALEX033248-274550\phn &  53.201922 & -27.764088 & 22.87 & 22.96 & 0.215\tablenotemark{2} & 41.26 & 38$\pm$14 &  53.201833 & -27.764037 & 0.3 & 0.214\tablenotemark{e} & 39.6\tablenotemark{k}\phm{f} & \nodata\\      49 & GALEX033249-273243\tablenotemark{p} &  53.208065 & -27.545501 & 21.91 & 21.81 & 0.221\tablenotemark{1} & 41.64 & 34$\pm$4 &  53.208122 & -27.545031 & 1.7 & 0.219\tablenotemark{e} & \nodata\phm{f} & \nodata\\      50 & GALEX033251-280305\tablenotemark{p} &  53.213904 & -28.051582 & 22.30 & 21.88 & 0.213\tablenotemark{1} & 41.77 & 84$\pm$8 &  53.214363 & -28.051258 & 1.9 & 0.214\tablenotemark{e} & 41.0\tablenotemark{i} & a\\      51 & GALEX033253-274834\tablenotemark{p} &  53.221815 & -27.809538 & 21.57 & 21.77 & 0.229\tablenotemark{1} & 42.16 & 95$\pm$37 &  53.221824 & -27.809313 & 0.8 & 0.226\tablenotemark{e} & -999\phm{f} & \nodata\\      52 & GALEX033253-280704\phn &  53.223911 & -28.117843 & 22.15 & 22.14 & 0.305\tablenotemark{1} & 41.82 & 29$\pm$10 &  53.223999 & -28.117916 & 0.4 & 0.297\tablenotemark{e} & \nodata\phm{f} & \nodata\\      53 & GALEX033307-274433\tablenotemark{p} &  53.279518 & -27.742550 & 20.79 & 20.27 & 0.220\tablenotemark{1} & 41.53 & 10$\pm$1 &  53.280520 & -27.742386 & 3.2 & 0.218\tablenotemark{e} & 40.6\tablenotemark{k}\phm{f} & \nodata\\      54 & GALEX033321-273339\tablenotemark{p} &  53.340413 & -27.560853 & 22.03 & 21.93 & 0.278\tablenotemark{1} & 42.13 & 76$\pm$27 &  53.340530 & -27.560711 & 0.6 & 0.276\tablenotemark{e} & -999\phm{f} & \nodata\\      55 & GALEX033333-275645\tablenotemark{p} &  53.389526 & -27.945835 & 21.80 & 21.38 & 0.429\tablenotemark{2} & 42.12 & 15$\pm$6 &  53.389565 & -27.946243 & 1.5 & 0.422\tablenotemark{f} & -999\phm{f} & \nodata\\      56 & GALEX033334-281127\phn &  53.392972 & -28.191032 & 23.41 & 22.94 & 0.340\tablenotemark{2} & 42.06 & 157$\pm$68 &  53.393167 & -28.191083 & 0.6 & 0.338\tablenotemark{g} & \nodata\phm{f} & \nodata\\      57 & GALEX033343-280108\phn &  53.431270 & -28.019159 & 23.28 & 22.09 & 0.355\tablenotemark{3} &   &   &  53.431339 & -28.020054 & 3.2 & (0.656)\tablenotemark{e} & \nodata\phm{f} & \nodata\\      58 & GALEX033346-274736\phn &  53.444531 & -27.793457 & 23.36 & 22.97 & 0.363\tablenotemark{1} & 42.43 & 361$\pm$54 &  53.445130 & -27.793115 & 2.3 & \nodata\phm{d} & \nodata\phm{f} & \nodata\\      59 & GALEX033351-273559\phn &  53.464008 & -27.599818 & 21.15 & 20.99 & 0.184\tablenotemark{3} & 41.28 & 10$\pm$4 &  53.463917 & -27.599806 & 0.3 & 0.181\tablenotemark{g} & \nodata\phm{f} & \nodata\\      60 & GALEX033357-274910\tablenotemark{p} &  53.489190 & -27.819597 & 21.96 & 21.65 & 0.242\tablenotemark{3} & 41.43 & 16$\pm$6 &  53.489895 & -27.819418 & 2.3 & 0.242\tablenotemark{e} & \nodata\phm{f} & \nodata\\      61 & GALEX033359-275759\tablenotemark{p} &  53.496427 & -27.966480 & 22.16 & 21.72 & 0.370\tablenotemark{3} & 42.21 & 44$\pm$16 &  53.496417 & -27.966500 & 0.1 & 0.358\tablenotemark{e} & \nodata\phm{f} & n\\      62 & GALEX033413-275246\tablenotemark{p} &  53.555083 & -27.879595 & 21.68 & 21.66 & 0.235\tablenotemark{2} & 41.43 & 14$\pm$5 &  53.554714 & -27.880039 & 2.0 & 0.233\tablenotemark{e} & \nodata\phm{f} & \nodata\\ \enddata \tablecomments{ \tablenotetext{c}{\rm{Confidence} in LAE candidate} \tablenotetext{d}{\rm{X-ray} data from the extended survey (Lehmer et al.\ 2005)} \tablenotetext{e}{\rm{This} paper's DEIMOS spectra} \tablenotetext{f}{\rm{Spectroscopic redshifts} from Cooper et al.\ (2012)} \tablenotetext{g}{\rm{Spectroscopic redshifts} from Mao et al.\ (2012)} \tablenotetext{h}{\rm{Spectra} from Le F{\`e}vre et al.\ (2013)} \tablenotetext{i}{\rm{X-ray} luminosity computed from the soft 0.5-2 keV flux band as catalogued in the extended Lehmer et al.\ (2005) survey} \tablenotetext{j}{\rm{X-ray} luminosity computed from the hard 2-7 keV flux band as catalogued in the deeper Luo et al.\ (2017) survey} \tablenotetext{k}{\rm{X-ray} luminosity computed from the soft 0.5-2 keV flux band as catalogued in the deeper Luo et al.\ (2017) survey} \tablenotetext{p}{\rm{Pipeline} LAE candidate identified in Cowie et al.\ (2010) and Cowie et al.\ (2011)} \tablenotetext{a}{\rm{Absorber} classification based on optical spectra} \tablenotetext{b}{\rm{BLAGN} classification based on optical spectra} \tablenotetext{n}{\rm{BPT} AGN classification based on optical spectra} \tablenotetext{u}{\rm{AGN} classification based on UV spectra} \tablenotetext{w}{\rm{AGN} classification based on \em{WISE} imaging data} \tablenotetext{x}{\rm{AGN} classification based on X-ray imaging data} }\label{cdfs} \end{deluxetable}

\clearpage

\begin{deluxetable}{ccccccccccccccc} \tabletypesize{\scriptsize} \setlength{\tabcolsep}{0.02in} \tablecolumns{15} \tablewidth{0pc} \tablecaption{Emission-line Sample: GROTH-00} \tablehead{ \colhead{Num} & \colhead{Name}    &  \colhead{R.A.} &   \colhead{Decl.} & \colhead{FUV} & \colhead{NUV} & \colhead{$z^{c}_{\rm{galex}}$} & \colhead{log $L$(Ly$\alpha$)} & \colhead{EW$_{\rm{r}}$(Ly$\alpha$)}  & \colhead{R.A.(opt)} &  \colhead{Decl.(opt)} & \colhead{Offset} & \colhead{$z_{\rm{opt}}$} & \colhead{log $L^{d}_{2-10~\rm{keV}}$} & \colhead{Class}\\ \colhead{} & \colhead{} & \colhead{(J2000.0)}   & \colhead{(J2000.0)} & \colhead{(AB)} & \colhead{(AB)} & \colhead{} & \colhead{(erg s$^{-1}$)} & \colhead{(\AA)} & \colhead{(J2000.0)}   & \colhead{(J2000.0)} & \colhead{(arcsec)} & \colhead{} & \colhead{(erg s$^{-1}$)} & \colhead{}\\ \colhead{(1)} & \colhead{(2)} & \colhead{(3)} & \colhead{(4)} & \colhead{(5)} & \colhead{(6)} & \colhead{(7)}  & \colhead{(8)}  & \colhead{(9)} & \colhead{(10)} & \colhead{(11)} & \colhead{(12)} & \colhead{(13)} & \colhead{(14)} & \colhead{(15)}} \startdata       1 & GALEX141731+524610\phn & 214.383060 &  52.769511 & 22.51 & 22.54 & 0.214\tablenotemark{2} & 41.73 & 99$\pm$36 & 214.381970 &  52.768856 & 3.3 & 0.213\tablenotemark{h} & \nodata & \nodata\\       2 & GALEX141745+524618\tablenotemark{p} & 214.438460 &  52.771876 & 22.08 & 21.84 & 0.246\tablenotemark{1} & 41.63 & 30$\pm$4 & 214.437670 &  52.771730 & 1.8 & 0.244\tablenotemark{e} & -999 & \nodata\\       3 & GALEX141752+524316\phn & 214.466990 &  52.721363 & 23.11 & 23.19 & 0.273\tablenotemark{2} & 41.58 & 58$\pm$20 & 214.466800 &  52.721325 & 0.4 & 0.267\tablenotemark{h} & -999 & \nodata\\       4 & GALEX141758+523811\phn & 214.495610 &  52.636401 & 23.33 & 23.21 & 0.285\tablenotemark{2} & 41.50 & 50$\pm$18 & 214.496290 &  52.636562 & 1.6 & 0.284\tablenotemark{h} & -999 & \nodata\\       5 & GALEX141800+524401\phn & 214.501170 &  52.733693 & 22.42 & 22.44 & 0.251\tablenotemark{1} & 41.75 & 57$\pm$18 & 214.501820 &  52.733604 & 1.5 & 0.249\tablenotemark{h} & -999 & \nodata\\       6 & GALEX141805+524507\tablenotemark{p} & 214.521650 &  52.752176 & 21.50 & 21.32 & 0.245\tablenotemark{1} & 41.71 & 20$\pm$6 & 214.521420 &  52.751970 & 0.9 & 0.244\tablenotemark{e} & -999 & \nodata\\       7 & GALEX141810+524659\phn & 214.541760 &  52.783332 & 24.08 & 22.28 & 0.360\tablenotemark{3} & 42.19 & 542$\pm$195 & 214.540000 &  52.784660 & 6.1 & 0.356\tablenotemark{e} & -999 & n\\       8 & GALEX141821+525725\tablenotemark{p} & 214.588890 &  52.957118 & 20.87 & 20.53 & 0.384\tablenotemark{1} & 42.90 & 50$\pm$14 & 214.589200 &  52.957214 & 0.8 & 0.379\tablenotemark{h} & \nodata & ubw\\       9 & GALEX141833+530540\phn & 214.640710 &  53.094582 & 22.56 & 22.77 & 0.291\tablenotemark{2} & 41.92 & 66$\pm$21 & 214.640470 &  53.094540 & 0.5 & 0.287\tablenotemark{h} & \nodata & \nodata\\      10 & GALEX141845+525659\phn & 214.687990 &  52.949806 & 21.93 & 22.22 & 0.354\tablenotemark{1} & 42.48 & 66$\pm$20 & 214.687867 &  52.949653 & 0.6 & 0.350\tablenotemark{e} & -999 & \nodata\\      11 & GALEX141851+522459\phn & 214.714340 &  52.416608 & 23.82 & 23.94 & 0.303\tablenotemark{2} & 41.65 & 118$\pm$48 & 214.714342 &  52.416469 & 0.5 & 0.300\tablenotemark{e} & \nodata & \nodata\\      12 & GALEX141854+530747\tablenotemark{p} & 214.729040 &  53.129873 & 21.72 & 22.10 & 0.205\tablenotemark{2} & 41.73 & 44$\pm$13 & 214.728740 &  53.129887 & 0.6 & 0.203\tablenotemark{e} & \nodata & \nodata\\      13 & GALEX141855+525935\tablenotemark{p} & 214.732990 &  52.993122 & 21.92 & 21.47 & 0.288\tablenotemark{2} & 41.79 & 24$\pm$3 & 214.732960 &  52.992176 & 3.4 & 0.287\tablenotemark{e} & -999 & \nodata\\      14 & GALEX141859+522329\phn & 214.749790 &  52.391640 & 23.03 & 23.15 & 0.250\tablenotemark{2} & 42.02 & 438$\pm$78 & 214.749813 &  52.391408 & 0.8 & 0.249\tablenotemark{e} & \nodata & a\\      15 & GALEX141914+522326\tablenotemark{p} & 214.812240 &  52.390660 & 22.18 & 21.86 & 0.256\tablenotemark{2} & 41.49 & 21$\pm$7 & 214.811371 &  52.390653 & 1.9 & 0.252\tablenotemark{e} & \nodata & \nodata\\      16 & GALEX141915+524825\phn & 214.815100 &  52.807083 & 22.32 & 21.88 & 0.278\tablenotemark{3} & 41.57 & 21$\pm$7 & 214.813220 &  52.806420 & 4.7 & 0.284\tablenotemark{f} & -999 & \nodata\\      17 & GALEX141915+530246\phn & 214.815250 &  53.046206 & 23.19 & 23.33 & 0.207\tablenotemark{3} & 41.41 & 97$\pm$37 & 214.816610 &  53.046950 & 4.0 & 0.203\tablenotemark{f} & \nodata & \nodata\\      18 & GALEX141920+530244\phn & 214.836310 &  53.045636 & 24.81 & 24.48 & 0.270\tablenotemark{3} & 41.56 & 1100$\pm$917 & 214.836400 &  53.046524 & 3.2 & 0.271\tablenotemark{h} & -999 & n\\      19 & GALEX141925+522333\phn & 214.854710 &  52.392550 & 23.72 & 24.14 & 0.243\tablenotemark{2} & 41.44 & 122$\pm$51 & 214.854700 &  52.392375 & 0.6 & 0.239\tablenotemark{e} & \nodata & \nodata\\      20 & GALEX141934+525659\phn & 214.894880 &  52.949932 & 24.13 & 24.58 & 0.328\tablenotemark{2} & 41.81 & 234$\pm$114 & 214.895029 &  52.949811 & 0.5 & 0.325\tablenotemark{e} & -999 & \nodata\\      21 & GALEX141937+523024\tablenotemark{p} & 214.907460 &  52.506826 & 21.75 & 21.18 & 0.282\tablenotemark{2} & 41.46 & 10$\pm$3 & 214.907440 &  52.506832 & 0.0 & 0.282\tablenotemark{e} & \nodata & \nodata\\      22 & GALEX141938+523049\phn & 214.912120 &  52.513875 & 23.33 & 23.84 & 0.250\tablenotemark{2} & 41.67 & 137$\pm$57 & 214.912925 &  52.513550 & 2.1 & 0.248\tablenotemark{e} & \nodata & \nodata\\      23 & GALEX141946+525942\phn & 214.943380 &  52.995262 & 22.78 & 21.53 & 0.275\tablenotemark{2} &   &   & 214.943219 &  52.995189 & 0.4 & (STAR)\phm{d} & \nodata & \nodata\\      24 & GALEX141947+522304\phn & 214.948280 &  52.384671 & 22.72 & 22.47 & 0.270\tablenotemark{3} & 41.61 & 42$\pm$7 & 214.949570 &  52.385818 & 5.0 & 0.267\tablenotemark{h} & \nodata & \nodata\\      25 & GALEX141951+524210\phn & 214.964510 &  52.702953 & 23.80 & 23.48 & 0.241\tablenotemark{3} &   &   & 214.964523 &  52.701321 & 5.9 & (0.549)\tablenotemark{e} & 43.2 & x\\      26 & GALEX141959+524243\phn & 214.998650 &  52.712043 & 22.62 & 22.62 & 0.241\tablenotemark{2} & 41.64 & 59$\pm$21 & 214.999179 &  52.712519 & 2.1 & 0.240\tablenotemark{e} & 42.4 & xb\\      27 & GALEX142010+524029\phn & 215.042870 &  52.674995 & 22.15 & 21.15 & 0.319\tablenotemark{3} &   &   & 215.042847 &  52.674995 & 0.1 & (0.549)\tablenotemark{e} & \nodata & \nodata\\      28 & GALEX142010+524231\phn & 215.045110 &  52.708868 & 22.20 & 22.04 & 0.239\tablenotemark{2} & 41.38 & 18$\pm$3 & 215.046310 &  52.707977 & 4.1 & 0.239\tablenotemark{h} & \nodata & n\\      29 & GALEX142011+522906\phn & 215.048720 &  52.485110 & 24.30 & 24.49 & 0.231\tablenotemark{3} & 41.29 & 204$\pm$108 & 215.048221 &  52.485156 & 1.1 & 0.229\tablenotemark{e} & \nodata & \nodata\\      30 & GALEX142011+524122\phn & 215.049060 &  52.689652 & 23.14 & 23.22 & 0.224\tablenotemark{2} & 41.35 & 131$\pm$55 & 215.049200 &  52.689378 & 1.0 & 0.221\tablenotemark{e} & \nodata & \nodata\\      31 & GALEX142013+525357\phn & 215.055240 &  52.899426 & 23.84 & 22.89 & 0.359\tablenotemark{3} & 42.11 & 271$\pm$120 & 215.056592 &  52.900261 & 4.2 & 0.349\tablenotemark{e} & -999 & \nodata\\      32 & GALEX142013+524652\phn & 215.056810 &  52.781112 & 22.33 & 22.17 & 0.266\tablenotemark{2} & 41.55 & 26$\pm$8 & 215.056690 &  52.780621 & 1.8 & 0.262\tablenotemark{h} & -999 & \nodata\\      33 & GALEX142031+524757\tablenotemark{p} & 215.132760 &  52.799350 & 21.10 & 20.73 & 0.253\tablenotemark{1} & 41.84 & 17$\pm$2 & 215.133040 &  52.799362 & 0.6 & 0.253\tablenotemark{e} & \nodata & \nodata\\      34 & GALEX142043+524306\tablenotemark{p} & 215.180380 &  52.718352 & 21.38 & 21.32 & 0.251\tablenotemark{1} & 41.62 & 14$\pm$4 & 215.180354 &  52.718847 & 1.8 & 0.247\tablenotemark{e} & \nodata & \nodata\\      35 & GALEX142043+523612\tablenotemark{p} & 215.181420 &  52.603344 & 21.16 & 20.90 & 0.341\tablenotemark{1} & 42.68 & 68$\pm$21 & 215.181370 &  52.603191 & 0.6 & 0.337\tablenotemark{e} & \nodata & ubw\\      36 & GALEX142044+525006\tablenotemark{p} & 215.186110 &  52.835135 & 21.60 & 21.45 & 0.255\tablenotemark{1} & 41.70 & 20$\pm$6 & 215.186130 &  52.835140 & 0.0 & 0.251\tablenotemark{e} & -999 & \nodata\\      37 & GALEX142048+522917\phn & 215.203490 &  52.488241 & 23.16 & 22.45 & 0.239\tablenotemark{2} &   &   & 215.203380 &  52.488094 & 0.6 & (0.169)\tablenotemark{h} & \nodata & \nodata\\      38 & GALEX142048+525152\phn & 215.203960 &  52.864454 & 22.31 & 21.82 & 0.213\tablenotemark{3} &   &   & 215.203950 &  52.864460 & 0.0 & (0.290)\tablenotemark{h} & -999 & \nodata\\      39 & GALEX142051+524331\phn & 215.212510 &  52.725413 & 22.70 & 22.71 & 0.277\tablenotemark{2} & 41.55 & 32$\pm$10 & 215.212283 &  52.725239 & 0.8 & 0.275\tablenotemark{e} & \nodata & \nodata\\      40 & GALEX142102+525410\phn & 215.259780 &  52.902805 & 25.13 & 23.54 & 0.188\tablenotemark{3} &   &   & 215.265590 &  52.903996 & 13.3 & (0.351)\tablenotemark{h} & -999 & \nodata\\      41 & GALEX142121+523805\tablenotemark{p} & 215.341500 &  52.634802 & 21.77 & 21.63 & 0.284\tablenotemark{1} & 42.01 & 38$\pm$11 & 215.340940 &  52.634773 & 1.2 & 0.281\tablenotemark{h} & \nodata & b\\      42 & GALEX142124+523919\tablenotemark{p} & 215.352620 &  52.655506 & 22.07 & 21.81 & 0.258\tablenotemark{2} & 41.60 & 24$\pm$7 & 215.352620 &  52.655499 & 0.0 & 0.258\tablenotemark{e} & \nodata & \nodata\\      43 & GALEX142127+524022\phn & 215.363800 &  52.673039 & 24.63 & 23.60 & 0.281\tablenotemark{3} &   &   & 215.363710 &  52.672962 & 0.3 & (0.468)\tablenotemark{h} & \nodata & \nodata\\      44 & GALEX142130+525304\phn & 215.375990 &  52.884719 & 22.10 & 22.04 & 0.195\tablenotemark{3} & 41.42 & 33$\pm$10 & 215.376190 &  52.884827 & 0.6 & 0.193\tablenotemark{h} & \nodata & \nodata\\      45 & GALEX142135+523139\tablenotemark{p} & 215.399700 &  52.527553 & 20.21 & 20.09 & 0.250\tablenotemark{1} & 42.44 & 39$\pm$12 & 215.399540 &  52.527500 & 0.4 & 0.249\tablenotemark{e} & \nodata & ubw\\      46 & GALEX142140+523512\phn & 215.417840 &  52.586910 & 22.86 & 23.00 & 0.249\tablenotemark{2} & 41.56 & 57$\pm$19 & 215.417750 &  52.586611 & 1.1 & 0.247\tablenotemark{e} & \nodata & \nodata\\      47 & GALEX142151+524950\tablenotemark{p} & 215.462830 &  52.830705 & 21.83 & 21.64 & 0.201\tablenotemark{1} & 41.39 & 21$\pm$7 & 215.462970 &  52.830853 & 0.6 & 0.202\tablenotemark{e} & \nodata & b\\ \enddata \tablecomments{ }\end{deluxetable} \clearpage \addtocounter{table}{-1} \begin{deluxetable}{ccccccccccccccc} \tabletypesize{\scriptsize} \setlength{\tabcolsep}{0.02in} \tablecolumns{15} \tablewidth{0pc} \tablecaption{Emission-line Sample: GROTH-00 (Continued)} \tablehead{ \colhead{Num} & \colhead{Name}    &  \colhead{R.A.} &   \colhead{Decl.} & \colhead{FUV} & \colhead{NUV} & \colhead{$z^{c}_{\rm{galex}}$} & \colhead{log $L$(Ly$\alpha$)} & \colhead{EW$_{\rm{r}}$(Ly$\alpha$)}  & \colhead{R.A.(opt)} &  \colhead{Decl.(opt)} & \colhead{Offset} & \colhead{$z_{\rm{opt}}$} & \colhead{log $L^{d}_{2-10~\rm{keV}}$} & \colhead{Class}\\ \colhead{} & \colhead{} & \colhead{(J2000.0)}   & \colhead{(J2000.0)} & \colhead{(AB)} & \colhead{(AB)} & \colhead{} & \colhead{(erg s$^{-1}$)} & \colhead{(\AA)} & \colhead{(J2000.0)}   & \colhead{(J2000.0)} & \colhead{(arcsec)} & \colhead{} & \colhead{(erg s$^{-1}$)} & \colhead{}\\ \colhead{(1)} & \colhead{(2)} & \colhead{(3)} & \colhead{(4)} & \colhead{(5)} & \colhead{(6)} & \colhead{(7)}  & \colhead{(8)}  & \colhead{(9)} & \colhead{(10)} & \colhead{(11)} & \colhead{(12)} & \colhead{(13)} & \colhead{(14)} & \colhead{(15)}} \startdata      48 & GALEX142157+524545\phn & 215.488030 &  52.762516 & 23.34 & 22.51 & 0.294\tablenotemark{3} &   &   & 215.488860 &  52.763329 & 3.4 & (0.381)\tablenotemark{h} & \nodata & \nodata\\      49 & GALEX142206+523658\phn & 215.525010 &  52.616296 & 23.33 & 23.79 & 0.330\tablenotemark{2} & 42.11 & 214$\pm$95 & 215.525729 &  52.617000 & 3.0 & 0.325\tablenotemark{e} & \nodata & \nodata\\      50 & GALEX142208+525225\tablenotemark{p} & 215.533960 &  52.873800 & 22.49 & 21.37 & 0.307\tablenotemark{1} & 42.05 & 76$\pm$25 & 215.533500 &  52.873810 & 1.0 & 0.302\tablenotemark{e} & \nodata & w\\      51 & GALEX142244+524805\phn & 215.686910 &  52.801616 & 23.52 & 23.28 & 0.357\tablenotemark{3} & 42.09 & 158$\pm$57 & 215.685780 &  52.801151 & 3.0 & 0.354\tablenotemark{h} & \nodata & b\\ \enddata \tablecomments{ \tablenotetext{c}{\rm{Confidence} in LAE candidate} \tablenotetext{d}{\rm{X-ray} data from Laird et al.\ (2009)} \tablenotetext{e}{\rm{This} paper's DEIMOS spectra} \tablenotetext{f}{\rm{Spectroscopic redshifts} from Matthews et al.\ (2013)} \tablenotetext{h}{\rm{This} paper's WIYN spectra} \tablenotetext{p}{\rm{Pipeline} LAE candidate identified in Cowie et al.\ (2010) and Cowie et al.\ (2011)} \tablenotetext{a}{\rm{Absorber} classification based on optical spectra} \tablenotetext{b}{\rm{BLAGN} classification based on optical spectra} \tablenotetext{n}{\rm{BPT} AGN classification based on optical spectra} \tablenotetext{u}{\rm{AGN} classification based on UV spectra} \tablenotetext{w}{\rm{AGN} classification based on \em{WISE} imaging data} \tablenotetext{x}{\rm{AGN} classification based on X-ray imaging data} }\label{groth} \end{deluxetable}

\clearpage

\begin{deluxetable}{ccccccccccccccc} \tabletypesize{\scriptsize} \setlength{\tabcolsep}{0.02in} \tablecolumns{15} \tablewidth{0pc} \tablecaption{Emission-line Sample: NGPDWS-00} \tablehead{ \colhead{Num} & \colhead{Name}    &  \colhead{R.A.} &   \colhead{Decl.} & \colhead{FUV} & \colhead{NUV} & \colhead{$z^{c}_{\rm{galex}}$} & \colhead{log $L$(Ly$\alpha$)} & \colhead{EW$_{\rm{r}}$(Ly$\alpha$)}  & \colhead{R.A.(opt)} &  \colhead{Decl.(opt)} & \colhead{Offset} & \colhead{$z_{\rm{opt}}$} & \colhead{log $L^{d}_{2-7~\rm{keV}}$} & \colhead{Class}\\ \colhead{} & \colhead{} & \colhead{(J2000.0)}   & \colhead{(J2000.0)} & \colhead{(AB)} & \colhead{(AB)} & \colhead{} & \colhead{(erg s$^{-1}$)} & \colhead{(\AA)} & \colhead{(J2000.0)}   & \colhead{(J2000.0)} & \colhead{(arcsec)} & \colhead{} & \colhead{(erg s$^{-1}$)} & \colhead{}\\ \colhead{(1)} & \colhead{(2)} & \colhead{(3)} & \colhead{(4)} & \colhead{(5)} & \colhead{(6)} & \colhead{(7)}  & \colhead{(8)}  & \colhead{(9)} & \colhead{(10)} & \colhead{(11)} & \colhead{(12)} & \colhead{(13)} & \colhead{(14)} & \colhead{(15)}} \startdata       1 & GALEX143443+351052\phn & 218.679860 &  35.181172 & 22.13 & 22.20 & 0.265\tablenotemark{3} & 41.52 & 19$\pm$8 & 218.679520 &  35.181183 & 1.0 & 0.262\tablenotemark{g} & -999 & \nodata\\       2 & GALEX143446+351703\phn & 218.695440 &  35.284394 & 21.41 & 20.92 & 0.191\tablenotemark{3} & 41.64 & 35$\pm$4 & 218.695370 &  35.283958 & 1.6 & 0.190\tablenotemark{g} & 41.5 & n\\       3 & GALEX143519+345241\phn & 218.829270 &  34.878325 & 23.96 & 23.66 & 0.234\tablenotemark{3} &   &   & 218.826907 &  34.877567 & 7.5 & no $z$\tablenotemark{g} & -999 & \nodata\\       4 & GALEX143533+352741\phn & 218.890140 &  35.461598 & 24.41 & 23.58 & 0.402\tablenotemark{3} &   &   & 218.890710 &  35.461303 & 2.0 & no $z$\tablenotemark{g} & -999 & \nodata\\       5 & GALEX143544+350020\phn & 218.935920 &  35.005605 & 22.43 & 22.43 & 0.343\tablenotemark{3} & 42.07 & 49$\pm$18 & 218.935930 &  35.005558 & 0.2 & 0.339\tablenotemark{g} & -999 & \nodata\\       6 & GALEX143554+351910\phn & 218.976530 &  35.319503 & 20.80 & 20.55 & 0.317\tablenotemark{3} &   &   & 218.977102 &  35.319122 & 2.2 & (0.552)\tablenotemark{f} & 42.2 & xw\\       7 & GALEX143556+345006\phn & 218.984970 &  34.835058 & 22.04 & 21.64 & 0.290\tablenotemark{3} & 42.05 & 52$\pm$19 & 218.984820 &  34.835007 & 0.5 & 0.289\tablenotemark{g} & -999 & \nodata\\       8 & GALEX143605+352729\phn & 219.023510 &  35.458240 & 21.84 & 21.21 & 0.252\tablenotemark{1} & 41.52 & 16$\pm$7 & 219.023280 &  35.458237 & 0.7 & 0.251\tablenotemark{g} & -999 & \nodata\\       9 & GALEX143609+352242\phn & 219.041500 &  35.378367 & 21.99 & 21.85 & 0.217\tablenotemark{3} &   &   & 219.041391 &  35.378250 & 0.5 & no $z$\tablenotemark{g} & -999 & \nodata\\      10 & GALEX143613+344813\phn & 219.056080 &  34.803716 & 23.15 & 24.28 & 0.205\tablenotemark{2} & 41.78 & 438$\pm$157 & 219.055690 &  34.804249 & 2.2 & 0.204\tablenotemark{g} & -999 & a\\      11 & GALEX143618+345630\phn & 219.077080 &  34.941691 & 21.57 & 21.89 & 0.271\tablenotemark{2} & 42.24 & 62$\pm$23 & 219.077000 &  34.941654 & 0.3 & 0.269\tablenotemark{g} & -999 & \nodata\\      12 & GALEX143622+345632\tablenotemark{p} & 219.091790 &  34.942329 & 21.27 & 21.21 & 0.271\tablenotemark{1} & 42.01 & 25$\pm$9 & 219.092209 &  34.942055 & 1.6 & 0.269\tablenotemark{e} & -999 & \nodata\\      13 & GALEX143624+345938\tablenotemark{p} & 219.100460 &  34.993919 & 21.54 & 21.50 & 0.214\tablenotemark{1} & 41.77 & 35$\pm$13 & 219.100418 &  34.993500 & 1.5 & 0.213\tablenotemark{e} & -999 & \nodata\\      14 & GALEX143636+345033\tablenotemark{p} & 219.152880 &  34.842770 & 20.91 & 20.63 & 0.283\tablenotemark{1} & 42.33 & 36$\pm$12 & 219.152878 &  34.842777 & 0.0 & 0.280\tablenotemark{e} & -999 & \nodata\\      15 & GALEX143647+352606\phn & 219.197900 &  35.435101 & 20.78 & 20.94 & 0.180\tablenotemark{1} & 42.23 & 84$\pm$31 & 219.197952 &  35.435272 & 0.6 & 0.178\tablenotemark{e} & -999 & \nodata\\      16 & GALEX143647+351032\phn & 219.198950 &  35.175702 & 20.98 & 20.46 & 0.253\tablenotemark{1} & 41.76 & 13$\pm$2 & 219.198930 &  35.175804 & 0.4 & 0.251\tablenotemark{g} & -999 & \nodata\\      17 & GALEX143733+352212\tablenotemark{p} & 219.388110 &  35.370157 & 21.11 & 20.66 & 0.249\tablenotemark{1} & 41.84 & 19$\pm$2 & 219.388123 &  35.370167 & 0.1 & 0.243\tablenotemark{e} & -999 & \nodata\\      18 & GALEX143738+352232\phn & 219.409870 &  35.375633 & 22.39 & 22.00 & 0.239\tablenotemark{1} & 41.84 & 82$\pm$34 & 219.410080 &  35.375557 & 0.7 & 0.237\tablenotemark{g} & 41.4 & n\\      19 & GALEX143745+352825\tablenotemark{p} & 219.437740 &  35.473771 & 21.90 & 21.54 & 0.396\tablenotemark{1} & 42.61 & 71$\pm$25 & 219.437520 &  35.473343 & 1.7 & 0.394\tablenotemark{g} & 42.1 & xub\\      20 & GALEX143805+345849\tablenotemark{p} & 219.523550 &  34.980417 & 19.50 & 19.38 & 0.430\tablenotemark{3} & 43.04 & 15$\pm$5 & 219.523575 &  34.980476 & 0.2 & 0.425\tablenotemark{e} & 43.1 & xubw\\      21 & GALEX143806+351351\phn & 219.528710 &  35.230987 & 25.20 & 23.85 & 0.334\tablenotemark{3} &   &   & 219.529090 &  35.231754 & 3.0 & no $z$\tablenotemark{g} & -999 & \nodata\\      22 & GALEX143808+352111\phn & 219.533370 &  35.353059 & 25.16 & 24.97 & 0.354\tablenotemark{3} &   &   & 219.534508 &  35.353594 & 3.9 & no $z$\tablenotemark{g} & -999 & \nodata\\ \enddata \tablecomments{Same format as Table \ref{cdfs} \tablenotetext{c}{\rm{Confidence} in LAE candidate} \tablenotetext{d}{\rm{X-ray} data from Kenter et al.\ (2005)} \tablenotetext{e}{\rm{This} paper's DEIMOS spectra} \tablenotetext{f}{\rm{Spectroscopic redshifts} from Kochanek et al.\ (2012)} \tablenotetext{g}{\rm{This} paper's WIYN spectra} \tablenotetext{p}{\rm{Pipeline} LAE candidate identified in Cowie et al.\ (2010) and Cowie et al.\ (2011)} \tablenotetext{a}{\rm{Absorber} classification based on optical spectra} \tablenotetext{b}{\rm{BLAGN} classification based on optical spectra} \tablenotetext{n}{\rm{BPT} AGN classification based on optical spectra} \tablenotetext{u}{\rm{AGN} classification based on UV spectra} \tablenotetext{w}{\rm{AGN} classification based on \em{WISE} imaging data} \tablenotetext{x}{\rm{AGN} classification based on X-ray imaging data} }\label{ngpdws} \end{deluxetable}

\clearpage

\begin{deluxetable}{ccccccccccccccc} \tabletypesize{\scriptsize} \setlength{\tabcolsep}{0.02in} \tablecolumns{15} \tablewidth{0pc} \tablecaption{Emission-line Sample: COSMOS-00} \tablehead{ \colhead{Num} & \colhead{Name}    &  \colhead{R.A.} &   \colhead{Decl.} & \colhead{FUV} & \colhead{NUV} & \colhead{$z^{c}_{\rm{galex}}$} & \colhead{log $L$(Ly$\alpha$)} & \colhead{EW$_{\rm{r}}$(Ly$\alpha$)}  & \colhead{R.A.(opt)} &  \colhead{Decl.(opt)} & \colhead{Offset} & \colhead{$z_{\rm{opt}}$} & \colhead{log $L^{d}_{2-10~\rm{keV}}$} & \colhead{Class}\\ \colhead{} & \colhead{} & \colhead{(J2000.0)}   & \colhead{(J2000.0)} & \colhead{(AB)} & \colhead{(AB)} & \colhead{} & \colhead{(erg s$^{-1}$)} & \colhead{(\AA)} & \colhead{(J2000.0)}   & \colhead{(J2000.0)} & \colhead{(arcsec)} & \colhead{} & \colhead{(erg s$^{-1}$)} & \colhead{}\\ \colhead{(1)} & \colhead{(2)} & \colhead{(3)} & \colhead{(4)} & \colhead{(5)} & \colhead{(6)} & \colhead{(7)}  & \colhead{(8)}  & \colhead{(9)} & \colhead{(10)} & \colhead{(11)} & \colhead{(12)} & \colhead{(13)} & \colhead{(14)} & \colhead{(15)}} \startdata       1 & GALEX095902+021906\tablenotemark{p} & 149.761580 &   2.318473 & 20.36 & 20.01 & 0.350\tablenotemark{1} & 43.00 & 69$\pm$19 & 149.761505 &   2.318429 & 0.3 & 0.345\tablenotemark{g} & 44.2 & xuw\\       2 & GALEX095909+022154\phn & 149.790540 &   2.365217 & 24.12 & 23.02 & 0.416\tablenotemark{3} &   &   & 149.791031 &   2.365119 & 1.8 & (0.452)\tablenotemark{i} & -999 & \nodata\\       3 & GALEX095910+020732\phn & 149.793100 &   2.125737 & 21.87 & 21.95 & 0.357\tablenotemark{1} & 42.65 & 103$\pm$30 & 149.792969 &   2.125635 & 0.6 & 0.353\tablenotemark{g} & 42.7 & xu\\       4 & GALEX095913+021841\phn & 149.804440 &   2.311585 & 22.39 & 22.50 & 0.245\tablenotemark{2} & 41.92 & 82$\pm$28 & 149.804291 &   2.311690 & 0.7 & 0.250\tablenotemark{j} & -999 & \nodata\\       5 & GALEX095920+021431\phn & 149.837270 &   2.241971 & 21.45 & 21.70 & 0.307\tablenotemark{2} & 42.25 & 36$\pm$10 & 149.837006 &   2.241960 & 1.0 & 0.303\tablenotemark{h} & -999 & \nodata\\       6 & GALEX095921+020906\phn & 149.840630 &   2.151739 & 22.07 & 22.06 & 0.360\tablenotemark{2} & 42.41 & 66$\pm$11 & 149.840350 &   2.151680 & 1.0 & 0.355\tablenotemark{i} & -999 & \nodata\\       7 & GALEX095924+021447\phn & 149.852770 &   2.246435 & 22.16 & 21.61 & 0.348\tablenotemark{2} & 42.14 & 39$\pm$12 & 149.853820 &   2.245720 & 4.6 & 0.345\tablenotemark{i} & -999 & \nodata\\       8 & GALEX095928+015935\phn & 149.868860 &   1.993090 & 23.25 & 22.29 & 0.255\tablenotemark{3} &   &   & 149.868546 &   1.992900 & 1.3 & (1.166)\tablenotemark{h} & 43.6 & x\\       9 & GALEX095929+020849\phn & 149.874530 &   2.147199 & 22.61 & 22.76 & 0.202\tablenotemark{3} & 41.60 & 61$\pm$21 & 149.872147 &   2.148790 & 10.3 & 0.220\tablenotemark{j} & -999 & \nodata\\      10 & GALEX095939+022838\phn & 149.915380 &   2.477289 & 22.38 & 22.70 & 0.253\tablenotemark{2} & 41.96 & 90$\pm$17 & 149.915110 &   2.477260 & 1.0 & 0.250\tablenotemark{f} & -999 & \nodata\\      11 & GALEX095940+015122\tablenotemark{p} & 149.917560 &   1.856112 & 21.30 & 21.13 & 0.251\tablenotemark{1} & 41.93 & 25$\pm$7 & 149.917760 &   1.855986 & 0.9 & 0.251\tablenotemark{e} & -999 & \nodata\\      12 & GALEX095943+020503\phn & 149.930820 &   2.084299 & 21.38 & 21.62 & 0.189\tablenotemark{1} & 41.72 & 34$\pm$10 & 149.930600 &   2.083960 & 1.5 & 0.186\tablenotemark{f} & -999 & \nodata\\      13 & GALEX095943+021022\phn & 149.932750 &   2.172934 & 22.03 & 21.86 & 0.263\tablenotemark{2} & 41.70 & 25$\pm$8 & 149.932632 &   2.173122 & 0.8 & 0.262\tablenotemark{i} & -999 & \nodata\\      14 & GALEX095946+020226\phn & 149.942960 &   2.040575 & 23.85 & 24.69 & 0.394\tablenotemark{3} &   &   & 149.942825 &   2.040343 & 1.0 & (0.208)\tablenotemark{i} & -999 & \nodata\\      15 & GALEX095950+022501\phn & 149.960410 &   2.416999 & 22.83 & 23.21 & 0.242\tablenotemark{2} & 41.80 & 113$\pm$42 & 149.960379 &   2.416972 & 0.1 & 0.240\tablenotemark{e} & -999 & \nodata\\      16 & GALEX095955+023111\phn & 149.982740 &   2.519945 & 23.97 & 24.09 & 0.422\tablenotemark{3} &   &   & 149.983307 &   2.519970 & 2.0 & no $z$\tablenotemark{j} & -999 & \nodata\\      17 & GALEX095958+014929\phn & 149.995580 &   1.824732 & 22.86 & 22.99 & 0.221\tablenotemark{2} & 41.66 & 102$\pm$37 & 149.995400 &   1.824658 & 0.7 & 0.219\tablenotemark{e} & -999 & \nodata\\      18 & GALEX100003+021137\phn & 150.012980 &   2.193794 & 22.57 & 22.50 & 0.238\tablenotemark{2} & 41.91 & 126$\pm$48 & 150.012758 &   2.193483 & 1.4 & 0.238\tablenotemark{e} & -999 & \nodata\\      19 & GALEX100006+015524\phn & 150.026660 &   1.923501 & 22.85 & 22.94 & 0.265\tablenotemark{2} &   &   & 150.026581 &   1.923215 & 1.1 & (0.206)\tablenotemark{i} & -999 & \nodata\\      20 & GALEX100006+022224\phn & 150.027990 &   2.373421 & 21.28 & 22.00 & 0.219\tablenotemark{3} & 41.62 & 15$\pm$3 & 150.028450 &   2.375111 & 6.3 & 0.223\tablenotemark{e} & -999 & \nodata\\      21 & GALEX100010+015453\phn & 150.043040 &   1.914787 & 23.94 & 24.34 & 0.269\tablenotemark{2} & 41.69 & 239$\pm$129 & 150.042969 &   1.914580 & 0.8 & 0.266\tablenotemark{j} & -999 & a\\      22 & GALEX100017+022417\phn & 150.074090 &   2.404956 & 24.85 & 25.45 & 0.288\tablenotemark{3} &   &   & 150.073914 &   2.405210 & 1.1 & no $z$\tablenotemark{j} & -999 & \nodata\\      23 & GALEX100019+020438\phn & 150.080640 &   2.077489 & 23.64 & 23.30 & 0.268\tablenotemark{2} & 41.69 & 153$\pm$71 & 150.080279 &   2.077317 & 1.4 & 0.265\tablenotemark{e} & -999 & \nodata\\      24 & GALEX100025+015853\tablenotemark{p} & 150.105520 &   1.981567 & 19.16 & 19.07 & 0.379\tablenotemark{1} & 43.68 & 80$\pm$21 & 150.105220 &   1.981140 & 1.9 & 0.372\tablenotemark{e} & 44.1 & xubw\\      25 & GALEX100027+015705\tablenotemark{p} & 150.115670 &   1.951455 & 20.56 & 20.71 & 0.268\tablenotemark{1} & 42.45 & 38$\pm$10 & 150.115650 &   1.951172 & 1.0 & 0.265\tablenotemark{e} & 42.0 & \nodata\\      26 & GALEX100032+015721\phn & 150.134570 &   1.955847 & 22.85 & 22.79 & 0.389\tablenotemark{3} & 42.37 & 99$\pm$33 & 150.134384 &   1.955910 & 0.7 & 0.384\tablenotemark{i} & -999 & \nodata\\      27 & GALEX100035+020113\tablenotemark{p} & 150.148970 &   2.020404 & 20.87 & 20.73 & 0.269\tablenotemark{1} & 41.98 & 16$\pm$4 & 150.148880 &   2.020370 & 0.3 & 0.266\tablenotemark{e} & -999 & \nodata\\      28 & GALEX100040+021833\phn & 150.169970 &   2.309275 & 23.09 & 24.02 & 0.266\tablenotemark{1} & 41.89 & 151$\pm$58 & 150.169879 &   2.308947 & 1.2 & 0.262\tablenotemark{e} & -999 & \nodata\\      29 & GALEX100043+020637\tablenotemark{p} & 150.179780 &   2.110421 & 21.76 & 20.78 & 0.367\tablenotemark{1} & 42.61 & 76$\pm$22 & 150.179779 &   2.110381 & 0.1 & 0.360\tablenotemark{e} & 43.3 & xub\\      30 & GALEX100047+020406\phn & 150.198650 &   2.068510 & 21.56 & 21.23 & 0.187\tablenotemark{1} & 41.90 & 64$\pm$7 & 150.198479 &   2.068497 & 0.6 & 0.185\tablenotemark{e} & -999 & n\\      31 & GALEX100055+015636\phn & 150.232640 &   1.943342 & 21.86 & 23.44 & 0.220\tablenotemark{1} & 42.28 & 238$\pm$40 & 150.232590 &   1.942600 & 2.7 & 0.219\tablenotemark{i} & 42.8 & xn\\ \enddata \tablecomments{ }\end{deluxetable} \clearpage \addtocounter{table}{-1} \begin{deluxetable}{ccccccccccccccc} \tabletypesize{\scriptsize} \setlength{\tabcolsep}{0.02in} \tablecolumns{15} \tablewidth{0pc} \tablecaption{Emission-line Sample: COSMOS-00 (Continued)} \tablehead{ \colhead{Num} & \colhead{Name}    &  \colhead{R.A.} &   \colhead{Decl.} & \colhead{FUV} & \colhead{NUV} & \colhead{$z^{c}_{\rm{galex}}$} & \colhead{log $L$(Ly$\alpha$)} & \colhead{EW$_{\rm{r}}$(Ly$\alpha$)}  & \colhead{R.A.(opt)} &  \colhead{Decl.(opt)} & \colhead{Offset} & \colhead{$z_{\rm{opt}}$} & \colhead{log $L^{d}_{2-10~\rm{keV}}$} & \colhead{Class}\\ \colhead{} & \colhead{} & \colhead{(J2000.0)}   & \colhead{(J2000.0)} & \colhead{(AB)} & \colhead{(AB)} & \colhead{} & \colhead{(erg s$^{-1}$)} & \colhead{(\AA)} & \colhead{(J2000.0)}   & \colhead{(J2000.0)} & \colhead{(arcsec)} & \colhead{} & \colhead{(erg s$^{-1}$)} & \colhead{}\\ \colhead{(1)} & \colhead{(2)} & \colhead{(3)} & \colhead{(4)} & \colhead{(5)} & \colhead{(6)} & \colhead{(7)}  & \colhead{(8)}  & \colhead{(9)} & \colhead{(10)} & \colhead{(11)} & \colhead{(12)} & \colhead{(13)} & \colhead{(14)} & \colhead{(15)}} \startdata      32 & GALEX100106+020502\phn & 150.277590 &   2.084125 & 22.24 & 21.94 & 0.284\tablenotemark{1} & 42.14 & 83$\pm$25 & 150.277400 &   2.084197 & 0.7 & 0.283\tablenotemark{e} & -999 & n\\      33 & GALEX100110+022049\tablenotemark{p} & 150.294790 &   2.346956 & 21.27 & 21.50 & 0.250\tablenotemark{1} & 42.11 & 41$\pm$11 & 150.294650 &   2.347014 & 0.5 & 0.248\tablenotemark{e} & -999 & \nodata\\      34 & GALEX100133+022500\phn & 150.388390 &   2.416912 & 22.92 & 22.32 & 0.349\tablenotemark{3} & 42.03 & 64$\pm$20 & 150.388214 &   2.417060 & 0.8 & 0.350\tablenotemark{j} & -999 & \nodata\\      35 & GALEX100140+020508\phn & 150.418990 &   2.085591 & 24.21 & 23.20 & 0.377\tablenotemark{3} &   &   & 150.418321 &   2.085153 & 2.9 & (0.425)\tablenotemark{e} & 43.4 & x\\      36 & GALEX100143+020437\phn & 150.431150 &   2.077187 & 24.40 & 24.01 & 0.239\tablenotemark{3} &   &   & 150.430496 &   2.076470 & 3.5 & no $z$\tablenotemark{j} & -999 & \nodata\\      37 & GALEX100150+022451\phn & 150.459720 &   2.414426 & 22.42 & 21.66 & 0.416\tablenotemark{3} &   &   & 150.459564 &   2.414530 & 0.7 & (0.267)\tablenotemark{j} & -999 & \nodata\\      38 & GALEX100152+021158\phn & 150.467490 &   2.199598 & 21.93 & 19.93 & 0.277\tablenotemark{1} &   &   & 150.467380 &   2.199554 & 0.4 & (STAR) & \nodata & \nodata\\ \enddata \tablecomments{Same format as Table \ref{cdfs} \tablenotetext{c}{\rm{Confidence} in LAE candidate} \tablenotetext{d}{\rm{X-ray} data from Civano et al.\ (2016)} \tablenotetext{e}{\rm{This} paper's DEIMOS spectra} \tablenotetext{f}{\rm{Spectroscopic redshifts} from Knobel et al.\ (2012)} \tablenotetext{g}{\rm{Spectroscopic redshifts} from Adelman-McCarthy et al.\ (2009)} \tablenotetext{h}{\rm{Spectroscopic redshifts} from Prescott et al.\ (2006)} \tablenotetext{i}{\rm{Spectra} from Lilly et al.\ (2007)} \tablenotetext{j}{\rm{This} paper's WIYN spectra} \tablenotetext{p}{\rm{Pipeline} LAE candidate identified in Cowie et al.\ (2010) and Cowie et al.\ (2011)} \tablenotetext{a}{\rm{Absorber} classification based on optical spectra} \tablenotetext{b}{\rm{BLAGN} classification based on optical spectra} \tablenotetext{n}{\rm{BPT} AGN classification based on optical spectra} \tablenotetext{u}{\rm{AGN} classification based on UV spectra} \tablenotetext{w}{\rm{AGN} classification based on \em{WISE} imaging data} \tablenotetext{x}{\rm{AGN} classification based on X-ray imaging data} }\label{cosmos} \end{deluxetable}

\clearpage

\end{landscape}

\end{document}